\documentclass[11pt, letterpaper]{article}

\usepackage[latin1]{inputenc}
\usepackage{graphicx}
\usepackage{amsfonts,amsmath,amsopn,amssymb,amsthm,bbm,latexsym,mathrsfs,pstricks,verbatim,caption,pst-node,stmaryrd}

\newcommand{\group}[1]{\left(#1\right)}

\newcommand{\qbin}[2]{\left[\begin{array}{c}#1\\#2\end{array}\right]}
\let\rw\rightarrow

\let\lf\lfloor
\let\rf\rfloor

\newcommand{\gs}{\text{gs}}
\let\t\times

\newcommand{\su}[1]{\widehat{su}(2)_{#1}}

\let\ga\gamma
\def\M{{\cal M}}

\newcommand{\mS}{\mathcal{S}}

\newcommand{\mk}{\mathcal{S}}
\newcommand{\mkp}[1]{\mathcal{S}^{(#1)}}

\newcommand{\ym}{\gamma}
\newcommand{\ymp}[1]{\ga_{#1}}

\newcommand{\wm}{\hat{w}}

\newcommand{\wt}{\tilde{w}}
\newcommand{\cm}{s}

\newcommand{\tl}{\mathcal{\bar{P}}}

\newcommand{\h}{\bar{h}}
\newcommand{\ytp}[1]{\bar{h}_{#1}}

\newcommand{\ct}{s}

\newcommand{\ak}{\mathcal{P}}

\newcommand{\ya}{h}
\newcommand{\yap}[1]{h_{#1}}

\newcommand{\wa}{w}

\newcommand{\ca}{s}

\newcommand{\ab}{b}
\newcommand{\abp}[2]{b_{#1}^{#2}}

\newcommand{\ppp}{p^+}
\newcommand{\ppm}{p^-}
\newcommand{\ppx}{p^\times}
\newcommand{\ppo}{p^0}
\newcommand{\ppi}{p^\vee}
\newcommand{\pp}[1]{p^{#1}}

\newcommand{\nnp}{n_+}
\newcommand{\nnm}{n_-}
\newcommand{\nnx}{n_\t}
\newcommand{\nno}{n_0}
\newcommand{\nni}{n_\vee}
\newcommand{\nn}[1]{n_{#1}}

\let\y\infty

\let\lrw\leftrightarrow

\title{\vskip60pt {Path representation of $\su{k}$ states I:}
\vskip1pt{Operators and particles for $k=1,2$}}
\author{\bf{Jo\"el Lamy-Poirier\footnote{Present address: Perimeter Institute for Theoretical Physics, Waterloo, Ontario, N2J 2Y5, Canada; email: jlamypoirier@perimeterinstitute.ca} 
and  Pierre
Mathieu} \\ 
\\
D\'epartement de physique, de g\'enie physique et d'optique,\\
Universit\'e Laval,
Qu\'ebec, Canada, G1K 7P4.\\
joel.lamy-poirier.1@ulaval.ca,
pmathieu@phy.ulaval.ca}

\begin{document}
\maketitle

\begin{abstract}
This is the first of two articles devoted to the analysis of the path description of the states in $\su{k}$ WZW models,
a representation well suited  for constructive derivations of the fermionic characters.
In this first article, the cases $k=1,2$ are  treated in detail,  emphasizing  a different description in each case (operators vs particles). For $k=1$, we first prove, as a side result, the equivalence of two known path representations for the finitized $\su{1}$ states by displaying an explicit bijection. An immediate offshoot is the gain of a new and simple weighting for the (Kyoto) path representation
that generalizes to level $k$. 
The bijection also suggests  
two operator constructions for the $\su{1}$ paths, a local and a nonlocal one, both interrelated. These are formal operators that map a path to another path, so that any path can be obtained by successive applications of these operators on a simple  reference (ground-state) path.
 The nonlocal operator description is the starting point for a direct and elementary derivation of the $\su{1}$ spinon character. The second part presents an extensive study of the $\su{2}$ paths from their particle point of view,  where the particles are defined as the path building blocks. The resulting generating functions appear to provide new (at least superficially) fermionic forms of the  characters. 
 In particular, a nice relationship between the sum of the $j=0,1$ characters at $k=2$ and the two ones at $k=1$ is unravelled. 

\end{abstract} 
\newpage
\section{Introduction}

\subsection{Background}
The organization of the states in conformal field theories is coded in the characters. These are readily fixed by the Virasoro representation theory. For the minimal models, an irreducible character is expressed as an infinite alternating sum (dubbed bosonic), reflecting the exclusion-inclusion procedure resulting from the subtraction of the singular vectors \cite{Roc} (see e.g., \cite{CFT}). Quite interestingly, there exists  alternative expressions for these irreducible characters. For instance, a minimal model can be viewed as the scaling limit of a solvable statistical model at criticality -- specifically, a RSOS model \cite{ABF,FB} -- and this implies, via the corner-transfer matrix \cite{Bax}, a representation of each (CFT) state by a one-dimensional configuration.  An irreducible character can thus be re-expressed as a one-dimensional configuration sum \cite{Mel2,Kyoto}. Equivalently, since a configuration defines a path, a character is a path generating function. This function in turn can be evaluated explicitly in terms of a positive multiple sum, called a fermionic character \cite{KKMM}.\footnote{The representation of the states by paths is actually the most effective method for obtaining such so-called fermionic forms, either   directly \cite{OleJS,JMpar} or by combinatorial recursive methods \cite{FLPW,Wel}. From an intrinsic conformal field theoretical perspective, fermionic characters encode a description of the states in terms of quasi-particles (whose  bases, to a large extent, remain to be worked out but for some examples see \cite{Fetal,JM3p}): the Hilbert space is build up by a filling process (by the action of creation operators) subject to exclusion rules. Quasi-particle bases are not unique; for instance, it is expected that there will be one canonical form for each integrable perturbation (classified in \cite{Zam}), so that a set of  quasi-particles would represent the massless limits of the off-critical massive particles appropriate to the perturbation under consideration -- see e.g., \cite{KKMM}. 
But the same is true for configuration sums: there should be as many different path representations as there are integrable perturbations. Indeed, a configuration sum is characteristic of an off-critical sector. Moreover, since paths have a natural particle description (where particles in this context refer to the basic building blocks in terms of which all paths can be decomposed -- see e.g., \cite{OleJS,PMjmp}), these provide the natural combinatorial representation of the CFT quasi-particles (see for instance \cite{JMpar}).}

For WZW models, the states are similarly organized in terms of the irreducible characters of the affine Lie algebras associated to the symmetry group \cite{KZ,CFT}. Considered as the scaled limit of a face model, these WZW states also have a combinatorial representation in terms of paths \cite{DJKMO1,DJKMO2,DJKMO3}.\footnote{There is a vast literature on the fermionic characters of affine Lie algebras obtained via some combinatorial description of the states; see for instance \cite{Hata} and references therein.}
In this context, quantum group symmetry provide an important simplification: the character being $q$-independent (where $q$ here stands for  the quantum parameter, so that the model is critical at $q=1$), it can be most simply studied at $q=0$, where the basis of states crystalizes to a simple form \cite{JMMO}.

For the simplest model, $\su{1}$, the  link between the path construction of \cite{DJKMO2} and the fermionic characters has been made in \cite{Mel}. Within the CFT framework, this character has been re-obtained from the construction of an explicit basis of states -- the spinon basis of \cite{BPS,BLS} --, a construction inspired by the correspondence between the  $\su{1}$ WZW and an integrable spin chain with long-range interaction \cite{Hetal}.\footnote{More recently, a similar basis was derived from an extended simple-current-type reformulation of the model \cite{MR}.}
The level-$k$ generalization of the spinon basis and the associated fermionic characters  have been conjectured in \cite{BLSb} 
and proved in \cite{Ara,NYa} 
and indirectly in \cite {SW}. 

\subsection{Objectives and organization of this work}

The main goal of the  present article and its sequel \cite{LPM}  is to work  out the fermionic $\su{k}$ characters directly from the (Kyoto) path representation \cite{DJKMO3,JMMO} by means of a variant of the constructive method pioneered in \cite{OleJS} in the context of unitary minimal models (and rooted in the work \cite{BreL}). The motivation for focussing on the constructive method is that it seems to be best suited to eventually unravel the physics behind the fermionic forms.
Here we concentrate on the cases $k=1,2$, postponing the analysis of the general case to the sequel article.

In the first step, we present a rather detailed analysis of the $k=1$ case. The interest of working out 
extensively this special case is twofold. On the one hand, the states pertaining to a specific sector at level 1
have been described by new SOS-type paths  in \cite{FPW}, a representation rather different from  that given in \cite{DJKMO2,DJKMO3,JMMO,Mel}. Our first objective is  to obtain a bijective correspondence between these two types of paths. This bijection has an immediate spin-off: it provides an alternative and much simpler weighting of the $\su{1}$  Kyoto paths, an expression that turns out to remain valid to all levels \cite{LPM}.
On the other hand, we develop a formal operator representation of the paths
that is instrumental in our constructive approach to the generating function, both at level 1 and for the general case worked out in the sequel article. Our analysis of the $\su{1}$ paths is the subject of Sect. \ref{Seck1}, whose detailed  structure is presented in its introducing paragraphs.

In our study of the $k=1$ paths, we point out that the operator construction can equally well be phrased in terms of a particle decomposition of the paths. In this context, the particle-operator relationship is trivial. Such a particle reinterpretation is not completely immediate in the general case. However, for $k=2$, there is a natural particle description of the paths   that is worth making explicit. The analysis of $\mkp{2}$ paths from this point of view -- presented in Sect. \ref{cask2} -- is quite interesting and (if we can say so) rather nice. For instance, the paths pertaining to the $j=0,1$ modules are naturally decomposed in terms of four particles. However, we unravel a transmutation process which reduces their number to the a priori expected three
  (this expectation being based on the representation of the $\su{2}$ algebra in terms of three free fermions \cite{GodO}). The resulting characters are (apparently) new fermionic expressions, two of which ($j=0,1$) are readily reduced to  known expressions by manipulations of $q$-identities. We also  point out the interesting formula, eq. (\ref{caun2}), relating the sum of the $j=0,1$ characters to that of the two $k=1$ modules.


\section{Path descriptions of the $\su{1}$ states}\label{Seck1}

The general objective of this section is to present an exhaustive analysis of the path descriptions of the
states in integrable $\su{1}$ modules. Two quite different path representations are known, the one given in \cite{JMMO,Mel}, in which the path height can take two possible values, and the one introduced recently in \cite{FPW}, where the path heights can rather take any positive values. 
These two classes of paths are respectively denoted as $\mk$ (defined in Sect. \ref{Sspa} ) and $\tl$ (defined in Sect. \ref{Spbpa}). Note that our definition of $\tl$ paths generalizes the one given in \cite{FPW}, the latter description being recovered with special boundary conditions. Our generalization covers states in all sectors (to be defined shortly), the two $\su{1}$ modules and nails down a prescription to lift the infinite degeneracy inherent to this description.

Our first specific aim is to demonstrate the equivalence of these two representations by displaying a bijection between them. This is the subject of Sect. \ref{Sbisp}. The key tool for establishing this bijection is the construction of yet a third class of paths, denoted $\ak$ and introduced  in Sect. \ref{Sdefak}. The bijection $\mk\lrw\tl$ is proved via these intermediate paths as follows: $\mk\lrw\ak\lrw \tl$. The second and simplest step is proved in Sect. \ref{bijt}. It is clear from this analysis  that $\ak$ paths are sort of regularized versions of the $\tl$ paths (and this is how we came to this notion). While the latter have a deformed-type (unitary) RSOS weight function \cite{ABF, OleJS} -- the deformation being rooted in the presence of horizontal edges -- the former have a genuine RSOS weighting. The first step, establishing $\mk\lrw\ak$, is worked out in two parts. The bijection is displayed in Sect. \ref{bisp} and  the demonstration that it preserves the weight is given in Sect. \ref{wpbeta}.

The demonstration of the equivalence $\mk\lrw\tl$ is a result of independent interest. However, in relation with our main concern for the rest of this section, that is, the combinatorial analysis of the infinite $\mk$ paths, this equivalence obtained via the $\ak$ paths equips us with an alternative and much simpler weight function for the $\mk$ paths, namely, the $\ak$-weight expression rewritten in terms of  the $\mk$ path data. 

For those readers not  interested in the bijection, which is irrelevant for the rest  of the article, Sect. \ref{Spbpa} and \ref{Sbisp} can be skipped. It suffices to read the last two paragraphs of Sect. \ref{Suup}.

With the goal of working out constructively the fermionic generating function for $\mk$ paths, namely, the $\su{1}$ fermionic characters, we present in Sect. \ref{b1} a nonlocal operator construction of these paths. This is inspired by a similar construction introduced in \cite{JM} (see also \cite{BMW}). Here it is defined somewhat differently, directly in terms of an infinite product of local operators introduced previously in Sect. \ref{wpbeta}.
In this description, a path is constructed from the ground-state path (representing the vacuum state) by an ordered  sequence of operators. In Sect. \ref{pacara}, the path characteristics are readily re-expressed in terms of the operator indices.
This operator construction is naturally dressed with a particle interpretation  in Sect. \ref{oppart}.

Finally, in Sect. \ref{SGF}, the  $\su{1}$ fermionic characters for the $j=0,1/2$ modules are built up following the standard strategy \cite{OleJS}. In this way, the spinon formula  \cite{BPS,BLS} (see also \cite{MR})  are recovered.

\subsection{$\mk$ paths}\label{Sspa}

A finite $\mk$ path $\ga$ is a  sequence of numbers $(\ymp{1},\ymp{2},\cdots,\ymp{L},\ymp{L+1})$, $\ymp{l} \in \{-1,1\}$, with adjacent points linked by edges \cite{Mel, JMMO}.\footnote{We use a sightly different definition of the paths than the one in \cite{DJKMO1,JMMO} and in \cite{Dav,Mel}, where $\ga_l\in\{0,1\}$. The relation with the definition used in \cite{Dav,Mel} is a simple rescaling $(0,1)\rw(-1,1)$, while compared with the definition used in \cite{DJKMO1,JMMO}, the path is not only scaled but it is also reverted: $(0,1)\rw(1,-1)$. Our path definition is similar to that in \cite{Ara}, called there a spin configuration.} The edges may be  NE, SE  or H (horizontal), i.e., $|\ymp{l}-\ymp{l+1}|=0,2$. With the last point fixed according to  $\ymp{L+1}=-\ymp{L}$, the paths are divided in two classes depending on the value of $\ymp{L+1   }$. This value fixes the parameter $j\in\{0,\tfrac12\}$ of the corresponding $\su{1}$ module:
\begin{align}
	j=\frac{1}{4}(1+(-1)^{L+1}\ymp{L+1}).
\end{align}
It will prove convenient to define the intermediate numbers $j_l$:
\begin{align}\label{jlun}
	j_l&=\frac{1}{4}(1+(-1)^l\ymp{l}),\qquad 
	j\equiv j_{{L+1}}.
\end{align}
We will denote by $\mS_j$ the set of paths for which {$\ymp{L+1}=(-1)^{L}(1-4j)$}.

The unnormalized weight $w^\circ(\ga)$ of a path $\ga$ is defined  as follows:
\begin{align}\label{poids}
	w^\circ(\ga) = 
	\frac12\sum_{l=1}^L\, l(1-\min(\ymp{l},-\ymp{l+1})).
\end{align}
Since each term under the sum involves two adjacent vertices, $w(\ga)$ can be viewed as a sum over the weight of the edges of $\ga$. 
An edge linking $\ga_l$ to $\ga_{l+1}$ contributes to $l$ in (\ref{poids}), unless it is a SE edge, which does not contribute. \footnote{This level-1 unnormalized weight  $w^\circ(\ga)$ is often presented in the form $\sum_l 
lH(\ga_l,\ga_{l+1})$, where $H$ is called the energy function. In terms of   our variables $\ga_l$, it is defined by the matrix (cf. \cite[eqs (2.30-(2.4)]{Ara}):
\begin{align*}\label{H3}
H(\ga_l,\ga_{l+1})=H_{\ga_l,\ga_{l+1}}=\begin{pmatrix}1&0\\ 1&1\end{pmatrix},
\end{align*}
where rows and columns are ordered as $(1,-1)$ (i.e., $H(1,-1)=0$). (Compare also with the $su(2)$ specialization of the $H$ function in \cite {DJKMO1}, with paths vertically reverted.) Note that the relevant weight expression in the relative one defined in (\ref{poidsre}). Therefore, a modification  of $H(\ga_l,\ga_{l+1})$ by a constant $a$ does not affect this relative weight. This explain the different normalizations found in the literature (in particular $a=-1$ in \cite{Dav,Mel}). The advantage of the one chosen here is to make the contribution of each edge to $w^\circ(\ga)$ positive.} 


Given a value of $j$, the path with minimal weight (referred to as the ground state) is the one with maximal number of SE edges, packed as close as possible to the right.
 For $j=0$, this path is unique
\begin{equation}\label{gs0}
	\ga^{\gs(0)}=(1,-1,1,-1,\cdots),
\end{equation}
but for $j=1/2$ it is twice degenerate:
\begin{align}\label{gsgs'}
	\ga^{\gs(\frac12)}=(-1,1,-1,1,\cdots,),
	\qquad \qquad
	\ga^{\gs'(\frac12)}=(1,1,-1,1,\cdots).
\end{align}
We will refer to the first of these -- the purely oscillating path -- as being the true $j=1/2$ ground state.
The two paths in (\ref{gsgs'}) belong to different sectors, a notion defined below. For the two genuine ground-states $\ga^{\gs(j)}$, one has:
\begin{equation}
	\label{gsj1} \ga^{\gs(j)}_l = (-1)^{l-2j+1}\qquad \text{and} \quad j_l=j\quad \forall \,l.
\end{equation}
Having characterized the ground-state paths, we note that the definition (\ref{jlun}) of $j_l(\gamma)$ can be made more natural, being simply the normalized difference between $\ymp{l}$ and $\ym_{l}^{\gs(0)}$, that is,
\begin{align}
j_l&=\frac{1}{4}(1+(-1)^l\ymp{l})
   =\frac{1}{4}|\ymp{l}-(-1)^{l+1}|
   =\frac{1}{4}|\ymp{l}-\ym_{l}^{\gs(0)}|
   .\label{defdejlun}
\end{align}
We stress that this holds true for  $\mS_0$ as well as $\mS_{\frac12}$ paths.  

The weight $\wm$ of a path $\ga\in\mS_j$ is defined to be the difference between its unnormalized weight minus that of the corresponding ground state:
\begin{align}\label{poidsre}
	\wm(\ga)=w^\circ(\ga)-w^\circ(\ga^{\gs(j)}) .
\end{align}

In addition to its weight, a $\mS_j$ path is characterized by its sector, denoted by $s$; it is defined as
\begin{align}
	\cm(\ga) &= -j+\frac12\sum_{l=1}^L(\ymp{l}-\ymp{l}^{\gs(j)}).\label{cml}
\end{align}
The additive term $-j$  fixes the normalization $s(\ga^{\gs(j)})=-j$, so that  $s=\mp\tfrac12$ for $\ym^{\gs(\frac12)}$ and $\ym^{\gs'(\frac12)}$ respectively. 
Since $|\ymp{l}-\ymp{l}^{\gs(j)}|=0,2$, it follows from (\ref{cml}) that $s+j\in \mathbb {Z}.$

\begin{figure}[ht]\
\caption{Example of a $\mk_{1/2}$ path, with $L=11$ (solid line). This path is superimposed  on the path $\ga^{\gs(\frac12)}$ in dotted lines. }
\label{exemplem1}
\begin{center}
\begin{pspicture}(1,-0.5)(8,0.5)
{\psset{yunit=10pt,xunit=20pt,linewidth=.8pt}
\psline{->}(1,-1.5)(1,1.75) \psline{->}(1,0)(12,0)
\rput(2,-0.4){{\scriptsize $2$}}
\rput(4,-0.4){{\scriptsize $4$}}
\rput(6,-0.4){{\scriptsize $6$}}
\rput(8,-0.4){{\scriptsize $8$}}
\rput(10,-0.4){{\scriptsize $10$}}
\rput(12,-0.4){{\scriptsize $12$}}

\rput[Br](0.8,-1){{\scriptsize $-1$}}
\rput[Br](0.8,1){{\scriptsize $1$}}

\psset{linestyle=solid}
\psline{-}(1,-0.1)(1,0.1)\psline{-}(2,-0.1)(2,0.1)
\psline{-}(3,-0.1)(3,0.1)\psline{-}(4,-0.1)(4,0.1)
\psline{-}(5,-0.1)(5,0.1)\psline{-}(6,-0.1)(6,0.1)
\psline{-}(7,-0.1)(7,0.1)\psline{-}(8,-0.1)(8,0.1)
\psline{-}(9,-0.1)(9,0.1)\psline{-}(10,-0.1)(10,0.1)
\psline{-}(11,-0.1)(11,0.1)\psline{-}(12,-0.1)(12,0.1)

\psline{-}(1,-1)(1.1,-1)
\psline{-}(1,1)(1.1,1)

\psline{-}(1,1)(2,1)
\psline{-}(2,1)(3,-1)
\psline{-}(3,-1)(4,-1)
\psline{-}(4,-1)(5,1)
\psline{-}(5,1)(6,1)
\psline{-}(6,1)(7,1)
\psline{-}(7,1)(8,-1)
\psline{-}(8,-1)(9,-1)
\psline{-}(9,-1)(10,1)
\psline{-}(10,1)(11,-1)
\psline{-}(11,-1)(12,1)

\psset{linestyle=dotted}

\psline{-}(1,-1)(2,1)
\psline{-}(3,-1)(4,1)
\psline{-}(4,1)(5,-1)

\psline{-}(5,-1)(6,1)
\psline{-}(6,1)(7,-1)

\psline{-}(7,-1)(8,1)
\psline{-}(8,1)(9,-1)
}

\end{pspicture}
\end{center}
\end{figure}
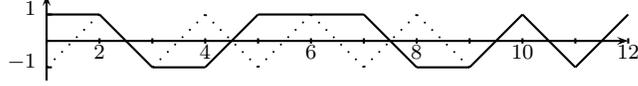

A typical $\mS$ path is displayed in Fig. \ref{exemplem1}. Its characteristics are readily evaluated:
\begin{align}
	j&={j_{12}=\frac{1}{4}(1+(-1)^{12}\ymp{12})=\frac{1}{2}}\nonumber\\
	\cm&=-j+\frac12(2+0+0-2+2+0+2-2+0+0+0)=\frac12\nonumber\\
	\wm &= (1+3+4+5+6+8+9+11)-(1+3+5+7+9+11)=11.
\end{align}
In the second line, the numbers in parenthesis are the values of $\ga_l-\ga_l^{\gs(\frac12)}$. In the third line, the numbers in the first and second parentheses are the initial positions of  all the edges but the SE ones for the displayed path and the ground-state path $\ym^{\gs(\frac12)}$ respectively.

In  the limit $L\rw\infty$, the weight $\wm$ of a path $\ga\in\mS_j$ is expressed as an indefinite (non-convergent) series unless $\ga$ has a purely oscillating tail matching that of the ground state with the corresponding value of $j$, in which case the series truncates to a finite sum. These  (finite-weight) infinite $\mS_j$ paths are those that represent the states of the  $\su{1}$ irreducible module labeled by $j$: their generating function is  the affine irreducible character \cite{Mel, DJKMO1, JMMO, DJKMO3}:
\begin{equation}\label{cara}
	\chi_j^{(1)}(z;q)=\sum_{\substack{\ga\in \mS_j\\ j_{l\gg 1}=j}}q^{\wm(\ga)}z^{s(\ga)},
\end{equation}
where the upper index $(1)$ stands for the level.

\subsection{$\tl$ paths}
\label{Spbpa}

A completely different sort of path representation of the $\su{1}$ states has been  introduced in \cite{FPW}.
These paths are akin to the RSOS paths of \cite{ABF} but with the upper height constraint pushed to infinity --- so that the height satisfies $h_l\geq 1$. A further difference is the possibility of having adjacent vertices of height 1, that is, H edges are allowed on the lower boundary of the defining strip.\footnote{These are called $T_L$ paths in \cite{FPW} since the constraints between successive heights are coded in the adjacency matrix of that tadpole graph.} The set of these paths will be denoted $\tl$, where the bar reminds of the possibility of H edges. A $\tl$ path $\h$ is thus defined by the sequence of positive integers $(\h_0,\h_1,\cdots, \h_L,\h_{L+1})$ subject to the constraints:
\begin{align}
	&|\h_{l+1}-\h_l|=1\quad \text{for}\quad \h_l>1,\qquad
	|\h_{l+1}-\h_l|=0,1\quad\text{if}\quad \h_l=1.
\end{align}
The weight $\wt$ of a $\tl$ path is defined to be
\begin{equation}\label{wdep}
	\wt(\h)=-\frac12(\ct+j)+\sum_{l=1}^L\,\wt_l,
\end{equation}
where the sum runs over the weight of all vertices (in contrast to a sum over edges  in $\mS$ paths) and the contribution of each vertex depends upon its nature \cite{FPW}:
\begin{equation}
\wt_l=\left\{
	\begin{array}{ll}
			\frac{l}{2} & \; \text{if}\; (\ytp{l+1}-\ytp{l})=(\ytp{l}-\ytp{l-1}),\quad\text{i.e., for straight-up/down/right vertices}, \\
		0           &\; \text{if}\; (\ytp{l+1}-\ytp{l})=-(\ytp{l}-\ytp{l-1})\; \text{and}\; h_l>1 ,\quad\text{i.e., for peaks/valleys }, \\
		\frac{l}{4} &\; \text{if}\;  \{\ytp{l+1},\ytp{l},\ytp{l-1}\}=\{1,1,2\} \mbox{ or }\{2,1,1\},\quad\text{i.e.,  for bends.} \\
	\end{array}
\right.\label{wtl}
\end{equation}
(Note that a straight-right vertex (or a plateau) is  in-between two H edges: $\ytp{l-1}=\ytp{l}=\ytp{l+1}=1$.) These weights can be written compactly as
\begin{equation}
\wt_l=\frac{l}4(2-|\ytp{l+1}+\ytp{l-1}-2\ytp{l}|).
\end{equation}
Except for $\{\ytp{l+1},\ytp{l},\ytp{l-1}\}=\{1,1,1\}$, this formula reduces to the simpler form
\begin{equation}
\wt_{l}=\frac{l}{4}|\ytp{l+1}-\ytp{l-1}| \qquad (\text{when}\;\{\ytp{l+1},\ytp{l},\ytp{l-1}\}\neq\{1,1,1\}).
\end{equation}
This is exactly the expression for the weight of the vertices in unitary RSOS models \cite{ABF,OleJS} (H edges being forbidden there).

 The correction term $-\frac12(\ct+j)$ in (\ref{wdep}) (absent in \cite{FPW}, where $\ct=j=0$) fixes a convenient normalization that will render the bijection between $\tl$ and $\mS$ paths weight preserving. However, for
 $\tl$ paths, there is no natural notion of module parameter $j$ and sector $s$. Rather, like the analogous finite RSOS paths, classes of $\tl$ paths are determined by the boundary conditions $\ytp{0}$, $\ytp{L}$ and $\ytp{L+1}$ \cite{ABF,OleJS,FLPW}. It turns out that, via the bijection to be spelled out below, these boundary values fix $j$ and $s$ as follows:
\begin{align}
	j&={\frac{1}{4}(1-(-1)^{L}(\ytp{L+1}-\ytp{L}))\label{jt}}\\
	2\ct&= \text{ min}(\ytp{L},\ytp{L+1})-\ytp{0}.
	\label{ct}
\end{align} Observe that the sector, which is a bulk property for $\mk$ paths, becomes a boundary condition for $\tl$ paths. These relations can be inverted\footnote{To invert the relation (\ref{ct}), we use its equivalent form:
$2s=\ytp{L}-\ytp{0}-[L+2j]_2$.} to obtain the boundary conditions that are determined by $j$ and $\ct$ :
\begin{align}\label{hLt}
	\ytp{L} &= \ytp{0}+2\ct+[L+2j]_2,\nonumber\\
	\ytp{L+1} &= \ytp{0}+2\ct +[L+1+2j]_2,
\end{align}
where the symbol $[x]_2$ is defined a
\begin{align}
	[x]_2 \equiv x \mbox{ mod } 2=\frac12(1-(-1)^x)~\in\{0,1\}, \label{nota}
\end{align}
a notation used extensively here.
Clearly, the values of $j$ and $s$ only fix the differences $\ytp{L+1}-\ytp{0}$ and $\ytp{L}-\ytp{0}$: out of the three boundary terms, one remains free. Let us take $\ytp{0}$ to be this free parameter. Its value can be any non-negative integer, the only  constraint being that the corresponding values of  $\ytp{L+1}$ and $\ytp{L}$ (fixed by $j,s$ and $\ytp{0}$) be $\geq1$. This forces $\ytp{0}\geq\ytp{0}^{\min}$, where $\ytp{0}^{\min}$ is determined from the condition : 
{\begin{align}\label{3min}
	1&=\min(\ytp{0}^{\min},\ytp{L}^{\min},\ytp{L+1}^{\min})\nonumber\\
	&=\ytp{0}^{\min}+\min(0,[L+2j]_2+2\ct,[L+1+2j]_2+2\ct)\nonumber\\
	&=\ytp{0}^{\min}+\min(0,2\ct).
\end{align}
The value of $\ytp{0}^{\min}$ is thus given by
\begin{align}\label{homin}
	\ytp{0}^{\min}=1+\max(0,-2\ct).
\end{align}
Setting $\ytp{0}=\ytp{0}^{\min}$ is a natural way -- and the one chosen here -- to lift the infinite degeneracy of the boundary values specified by $j$ and $s$.\footnote{The equivalence of the paths for different values of $\ytp{0}$ is simple to demonstrate and this proof will be omitted.} 
 From now on, $\tl$ paths will thus be understood to have 
$\ytp{0}=1+\max(0,-2\ct)$, that is, $\min(\ytp{0},\ytp{L},\ytp{L+1})=1$.

From the boundary conditions, we can find the ground-state paths for both values of $j$. Their heights take the forms
\begin{align}
	\ytp{l}^{\text{gs}(j)}&=1+[l+2j]_2.
\end{align}
For $j=1/2$, the second ground state is $\ytp{l\neq0}^{\text{gs'}(\frac12)}=2+[l+1]_2$, starting at $\ytp{0}^{\text{gs'}(\frac12)}=1$. Note that it is not necessary for $\tl$ paths to have the same tail as their ground state; they can differ by a vertical translation, depending on the sector.

Finally, the parameters $j$ and $s$ are not totally arbitrary, being linked by a parity constraint:
$s+j\in \mathbb Z$.
This implies that not all boundary conditions can be used for those $\tl$ paths that will eventually describe $\su{1}$  states. Indeed, the condition on $[2j+2s]_2=0$ induces a restriction on the parity of $\ytp{L}-\ytp{0}$ :
\begin{align}
[\ytp{L}-\ytp{0}]_2=[L+2j+2\ct]_2=[L]_2\label{condtlpair}.
\end{align}
This in turn implies that the number of H edges can only be even. Indeed, notice that the number of non-H edges is given by the number of $|\ytp{L}-\ytp{0}|$ NE or SE edges, accounting for the height difference between $\ytp{0}$ and $\ytp{L}$, plus an equal number of NE and SE edges (keeping the height constant), say $m$ of each. The number of remaining edges, those of H type, is $L-|\ytp{L}-\ytp{0}|-2m$, which, from eq. (\ref{condtlpair}), is seen to be even, as announced. (This is a technical point used in the bijection $\tl\leftrightarrow\mk$ via intermediate paths.)

These expressions for the boundary conditions in terms of $j,s$ are new; in the special case where $j=\ct=0$ and $L=0$ mod $2$, they match the ones given in \cite{FPW} (with $\ytp{0}=\ytp{0}^{\text{min}}=1$). For the example displayed in Fig. \ref{exempletl}, where $L=16$, eqs (\ref{ct}) and (\ref{jt}) give the values $s=0$ and $j=0$.

In the limit $L\rw\infty$, the weight is finite only if no vertex contributes to the weight from a finite $l$ onwards, meaning the path must be purely oscillating. From eq. (\ref{hLt}), the two heights delimiting the zigzag of the tail are $\ytp{0}+2s$ and $\ytp{0}+2s+1$; in addition, the path contain an even number of H edges.

\begin{figure}[ht]\
\caption{A $\tl$ path, with $L=16$, of weight 43, and with $j=s=0$.}
\label{exempletl}
\begin{center}
\begin{pspicture}(1,0)(7,2)
{\psset{yunit=15pt,xunit=15pt,linewidth=.8pt}
\psline{->}(0,0)(0,4) \psline{->}(0,0)(17,0)

\rput(2,-0.4){{\scriptsize $2$}}
\rput(4,-0.4){{\scriptsize $4$}}
\rput(6,-0.4){{\scriptsize $6$}}
\rput(8,-0.4){{\scriptsize $8$}}
\rput(10,-0.4){{\scriptsize $10$}}
\rput(12,-0.4){{\scriptsize $12$}}
\rput(14,-0.4){{\scriptsize $14$}}
\rput(16,-0.4){{\scriptsize $16$}}

\rput[Br](-0.2,0){{\scriptsize $1$}}
\rput[Br](-0.2,1){{\scriptsize $2$}}
\rput[Br](-0.2,2){{\scriptsize $3$}}
\rput[Br](-0.2,3){{\scriptsize $4$}}

\psset{linestyle=solid}
\psline{-}(1,0)(1,0.1)\psline{-}(2,0)(2,0.1)
\psline{-}(3,0)(3,0.1)\psline{-}(4,0)(4,0.1)
\psline{-}(5,0)(5,0.1)\psline{-}(6,0)(6,0.1)
\psline{-}(7,0)(7,0.1)\psline{-}(8,0)(8,0.1)
\psline{-}(9,0)(9,0.1)\psline{-}(10,0)(10,0.1)
\psline{-}(11,0)(11,0.1)\psline{-}(12,0)(12,0.1)
\psline{-}(13,0)(13,0.1)\psline{-}(14,0)(14,0.1)
\psline{-}(15,0)(15,0.1)\psline{-}(16,0)(16,0.1)

\psline{-}(0,1)(0.1,1)
\psline{-}(0,2)(0.1,2)
\psline{-}(0,3)(0.1,3)

\psline{-}(0,0)(1,1)
\psline{-}(1,1)(2,2)
\psline{-}(2,2)(3,3)
\psline{-}(3,3)(4,2)
\psline{-}(4,2)(5,1)
\psline{-}(5,1)(6,2)
\psline{-}(6,2)(7,1)
\psline{-}(7,1)(8,0)
\psline{-}(8,0)(9,0)
\psline{-}(9,0)(10,0)
\psline{-}(10,0)(11,1)
\psline{-}(11,1)(12,2)
\psline{-}(12,2)(13,1)
\psline{-}(13,1)(14,0)
\psline{-}(14,0)(15,0)
\psline{-}(15,0)(16,0)
\psline{-}(16,0)(17,1)

}
\end{pspicture}
\end{center}
\end{figure}
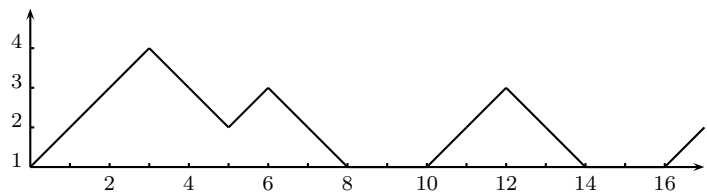

\subsection{Bijection $\tl\leftrightarrow\mk$ via $\ak$ paths}\label{Sbisp}

In this section, we present the bijection between $\tl$ and $\mk$ paths. This is our first result. This bijection will be established indirectly, by relating both paths to still another type of paths, denoted $\ak$. These are introduced in Sect. \ref{Sdefak}. Roughly,  a $\ak$ path is obtained from a $\tl$ path by replacing H edges by either SE or NE edges following a well-defined (and reversible) prescription. As a result, the path is extended to non-positive values of the vertical coordinates. The precise relationship is described in Sect. \ref{bijt}. On the other hand, the bijection between a $\ak$ and a $\mk$ path -- presented in Sect. \ref{bisp} -- corresponds to a simple relation between the edges of the former and the vertices of the latter: the SE and NE edges of the $\ak$ path  are mapped respectively to the vertex values $-1$ and $1$ of $\ga\in\mS$. The two bijections $\tl\leftrightarrow\ak$ and $\ak\leftrightarrow\mS$ are weight preserving. Once composed, they yield the desired bijection $\tl\leftrightarrow\mS$. 

Let us point out that in the demonstration of the weight-preserving character of the bijection $\ak\leftrightarrow\mS$, we introduce in Sect. \ref{wpbeta} an operator construction that relates any two paths (of the same type) with the same boundary conditions. (These operators are the building blocks of the nonlocal operators to be introduced in Sect. \ref{b1}.) 

\subsubsection{$\ak$ paths}\label{Sdefak}

A $\ak$ path $h$ is a sequence of integers $(h_0,h_1,\cdots, h_L,h_{L+1})$, with $\yap{0} = 0$ and $|h_{l+1}-h_l|=1$, with no further restriction on the height. We stress that the initial point is fixed and each $h_l$ can be non-positive. Fixing the initial point is a way to lift the infinite degeneracy of the paths determined by a given $j$ and $s$, exactly as for $\tl$ paths, when setting $\ytp{0}=\ytp{0}^{\min}$. The relation between the paths defined for different values of $h_0$ is trivial as these new paths will be seen   
to be invariant with respect to a vertical translation. 

The weight of a path $h$ is again the sum of the weight of all its vertices, which are defined as for $\tl$ paths in eq. (\ref{wtl}) -- except that there are no vertices with weight  $l/4$, H edges being forbidden. The weight of $h\in\ak$ is thus
\begin{align}
	\wa(h)=-\frac{1}{2}(\ca+j)+\sum_{l=1}^L \frac{l}{4}|\yap{l+1}-\yap{l-1}|.\label{wa}
\end{align}

Here again the values of the parameters $s$ and $j$ are not natural but they can be read off the extremal values 
$\yap{L}$ and $\yap{L+1}$ as (cf. Sect. \ref{bisp}) 
\begin{align}
	j&={\frac{1}{4}(1-(-1)^L(\yap{L+1}-\yap{L}))\label{ja}}\\
2\ca&= \text{ min}(\yap{L},\yap{L+1}).\label{ca}
\end{align}
These equations can be inverted to give the boundary conditions:
\begin{align}\label{hLa}
	\yap{L} &=  2s+[L+2j]_2 
	,\nonumber\\
	\yap{L+1} &=2s+ [L+1+2j]_2.  
\end{align}
As for $\tl$ paths, only half of the boundary conditions yield paths that can describe $\su{1}$ states:  since $[2j+2s]_2=0$, the parity of $h_L$ is linked to that of $L$.

The $\ak$ ground states are the same as for $\tl$ paths, except for a global translation:
\begin{align}
	h_{l}^{\text{gs}(j)}&=(-1)^{2j}[l]_2.
\end{align}
The second $j=1/2$ `ground state' is similar to the first, its vertices are $h_{l\neq0}^{\text{gs'}(\frac12)}=2-[l]_2$.

For the sample $\ak$ path displayed in Fig. \ref{exemplea1}, the values of $s$ and $j$ are easily found to be $s=j=0$; the weight is $w=\tfrac12(2+5+7+10+11+13)=24$.
 
In the limit $L\to \infty$, the situation  is analogous to that for $\tl$ paths: the tail oscillates between 
 $2s$ and $2s+1$.

\begin{figure}[ht]\
\caption{A $\ak$ path with {$L=14$}, $j=s=0$ and weight 24.} 
\label{exemplea1}
\begin{center}
\begin{pspicture}(1,-1.5)(7,1.5)
{\psset{yunit=15pt,xunit=15pt,linewidth=.8pt}
\psline{->}(0,-3)(0,3) \psline{->}(0,0)(16,0)
\rput(2,-0.4){{\scriptsize $2$}}
\rput(4,-0.4){{\scriptsize $4$}}
\rput(6,-0.4){{\scriptsize $6$}}
\rput(8,-0.4){{\scriptsize $8$}}
\rput(10,-0.4){{\scriptsize $10$}}
\rput(12,-0.4){{\scriptsize $12$}}
\rput(14,-0.4){{\scriptsize $14$}}

\rput[Br](-0.2,-2){{\scriptsize $-2$}}
\rput[Br](-0.2,-1){{\scriptsize $-1$}}
\rput[Br](-0.2,0){{\scriptsize $0$}}
\rput[Br](-0.2,1){{\scriptsize $1$}}
\rput[Br](-0.2,2){{\scriptsize $2$}}

\psset{linestyle=solid}
\psline{-}(1,-0.1)(1,0.1)\psline{-}(2,-0.1)(2,0.1)
\psline{-}(3,-0.1)(3,0.1)\psline{-}(4,-0.1)(4,0.1)
\psline{-}(5,-0.1)(5,0.1)\psline{-}(6,-0.1)(6,0.1)
\psline{-}(7,-0.1)(7,0.1)\psline{-}(8,-0.1)(8,0.1)
\psline{-}(9,-0.1)(9,0.1)\psline{-}(10,-0.1)(10,0.1)
\psline{-}(11,-0.1)(11,0.1)\psline{-}(12,-0.1)(12,0.1)
\psline{-}(13,-0.1)(13,0.1)\psline{-}(14,-0.1)(14,0.1)

\psline{-}(0,-2)(0.1,-2)
\psline{-}(0,-1)(0.1,-1)
\psline{-}(0,0)(0.1,0)
\psline{-}(0,1)(0.1,1)
\psline{-}(0,2)(0.1,2)

\psline{-}(0,0)(1,1)
\psline{-}(1,1)(2,0)
\psline{-}(2,0)(3,-1)
\psline{-}(3,-1)(4,0)
\psline{-}(4,0)(5,-1)
\psline{-}(5,-1)(6,-2)
\psline{-}(6,-2)(7,-1)
\psline{-}(7,-1)(8,0)
\psline{-}(8,0)(9,-1)
\psline{-}(9,-1)(10,0)
\psline{-}(10,0)(11,1)
\psline{-}(11,1)(12,2)
\psline{-}(12,2)(13,1)
\psline{-}(13,1)(14,0)
\psline{-}(14,0)(15,1)

}
\end{pspicture}
\end{center}
\end{figure}
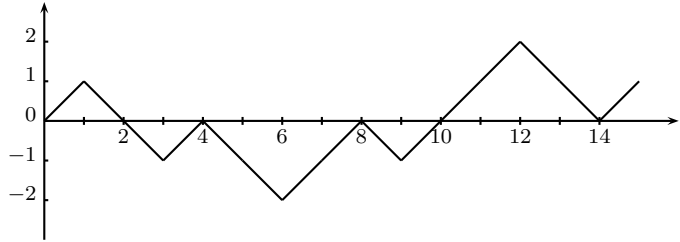

\subsubsection{Bijection $\mk\leftrightarrow\ak$}\label{bisp}

As already announced, the transformation from a $\mk$ to a $\ak$ path is a transformation from vertices to edges.
If we denote the edge linking the points {$\yap{l-1}$ and $\yap{l}$ as $(\yap{l-1},\yap{l})$, this is a rule that relates $\ga_l$ to  $(h_{l-1},h_{l})$}. This at once requires the $\ak$ path to have one more vertex than the $\mk$ one: the extra point is the fixed initial point $h_0$.
The map is then obtained by replacing the vertices equal to $-1$ (of the $\mk$ path) by SE edges and those equal to $1$ by NE ones.

To construct the path $h$ related to a path $\ga$, the initial step is to replace $\ymp{1}$ by the edge $(\yap{0}, \yap{1})$; with $h_0$ fixed (to the value 0), this determines $h_1$. The values of $h_l$ are thus fixed recursively from
\begin{align}
	\yap{l} = \yap{l-1} + \ymp{l},\label{bijection1}
\end{align}
The inverse transformation is immediate:
\begin{align}
	\ymp{l} = h_l - h_{l-1}.\label{bijection2}
\end{align}
This is manifestly a bijective transformation between $\ak$ and $\mk$ paths. \footnote{While this article was under preparation, T. Welsh pointed out that these $\ak$ paths have already been considered in \cite{DJKMO3,JMMO} -- see also \cite[Sect. 2.1--2.2]{FLOTW} and \cite{Ara} (with the only difference that instead of the initial point, it is the tail which is fixed, so that the $\ak$ paths there are constructed from right to left). From this perspective, the $\ak$ paths are dierct by-products of the $\mk$ paths. This bijection 
in itself could then be viewed as somewhat trivial, or at least very natural. However, when these $\ak$ paths are  ascribed a RSOS-type weighting modulo a non-trivial correction factor, demonstrating that this bijection is weight-preserving -- and this a key result here -- is not obvious. In contrast, in \cite{DJKMO3,JMMO,FLOTW,Ara}, these paths inherit their weight (called `energy' there) from the defining $\mk$ paths.}

As a simple illustration of the bijection, the $\ak$ path corresponding to the $\mk$ path of Fig. \ref{exemplem1} is pictured in Fig. \ref{exbijection1}.

\begin{figure}[ht]\
\caption{The $\ak$ path corresponding to the $\mk_{1/2}$ path of Fig. \ref{exemplem1} via the bijection (\ref{bijection1})} 
\label{exbijection1}
\begin{center}
\begin{pspicture}(0,0)(6,2)
{\psset{yunit=15pt,xunit=15pt,linewidth=.8pt}
\psline{->}(0,0)(0,4) \psline{->}(0,0)(12,0)
\rput(2,-0.4){{\scriptsize $2$}}
\rput(4,-0.4){{\scriptsize $4$}}
\rput(6,-0.4){{\scriptsize $6$}}
\rput(8,-0.4){{\scriptsize $8$}}
\rput(10,-0.4){{\scriptsize $10$}}

\rput[Br](-0.2,0){{\scriptsize $0$}}
\rput[Br](-0.2,1){{\scriptsize $1$}}
\rput[Br](-0.2,2){{\scriptsize $2$}}
\rput[Br](-0.2,3){{\scriptsize $3$}}

\psset{linestyle=solid}
\psline{-}(1,-0.1)(1,0.1)\psline{-}(2,-0.1)(2,0.1)
\psline{-}(3,-0.1)(3,0.1)\psline{-}(4,-0.1)(4,0.1)
\psline{-}(5,-0.1)(5,0.1)\psline{-}(6,-0.1)(6,0.1)
\psline{-}(7,-0.1)(7,0.1)\psline{-}(8,-0.1)(8,0.1)
\psline{-}(9,-0.1)(9,0.1)\psline{-}(10,-0.1)(10,0.1)
\psline{-}(11,-0.1)(11,0.1)

\psline{-}(0,0)(0.1,0)
\psline{-}(0,1)(0.1,1)
\psline{-}(0,2)(0.1,2)
\psline{-}(0,3)(0.1,3)

\psline{-}(0,0)(1,1)
\psline{-}(1,1)(2,2)
\psline{-}(2,2)(3,1)
\psline{-}(3,1)(4,0)
\psline{-}(4,0)(5,1)
\psline{-}(5,1)(6,2)
\psline{-}(6,2)(7,3)
\psline{-}(7,3)(8,2)
\psline{-}(8,2)(9,1)
\psline{-}(9,1)(10,2)
\psline{-}(10,2)(11,1)
\psline{-}(11,1)(12,2)

}
\end{pspicture}
\end{center}
\end{figure}

Using this bijection, we can now verify the expressions (\ref{ja}) and (\ref{ca}) giving the values of $\ca$ and $j$ for $\ak$ paths. For the expression of {$j_l$}, we have
{\begin{align}\label{jlenh}
	j_l&=\frac{1}{4}(1+(-1)^l\ga_{l})\nonumber\\
	&=\frac{1}{4}(1+(-1)^l(\yap{l}-\yap{l-1})).
\end{align}}
Using $j=j_{L+1}$, we recover the value (\ref{ja}). Similarly, for the sector $s$, starting from (\ref{cml}), we make the substitution (\ref{bijection2}), which transforms the sum into telescopic type, thereby readily evaluated:
\begin{align}
	\cm(\ga) &=-j+\frac12\sum_{l=1}^L(\ymp{l}-\ymp{l}^{\gs(j)})\nonumber\\
	&=-j+\frac12\sum_{l=1}^L(h_{l}-h_{l-1}-(h_{l}^{\gs(j)}-h_{l-1}^{\gs(j)}))\nonumber\\
	&=-j+\frac12(h_{L}-h_{0}-(h_{L}^{\gs(j)}-h_{0}^{\gs(j)}))\nonumber\\
	&=-j+\frac12(h_{L}-(-1)^{2j}[L]_2)\nonumber\\
		&= \frac12(h_{L}-[L+2j]_2) \nonumber\\
		&= \frac12 \text{min}\,(h_L,h_{L+1}), 
\end{align}
 where  in the last step, we used $h_{L+1}=h_L+(-1)^{L+2j}$ (cf. (\ref{ja})) and a case-by-case  comparison of the two sides of the equality. 
This is the announced expression (\ref{ca}).

\subsubsection{Weight-preserving character of $\mk\lrw\ak$: a local operator construction}\label{wpbeta}

We now verify the weight-preserving character of the bijection. We prove this by introducing a formal operator whose successive applications on a given path generate any other path with the same values of $j$ and $\cm$ (as long as the different application points are all strictly smaller than $L$). We thus define the local operator $\beta_l$ whose action on a path $h\in\ak$ reads:
\begin{align}\label{defbeta}
	\beta_l:&\left\{
	\begin{array}{ll}
		\yap{l}\rw\yap{l}+2&\mbox{if }\yap{l-1}=\yap{l+1}=\yap{l}+1,\\
		0&\mbox{otherwise.}
	\end{array}\right.
\end{align}
$\beta_l$ changes thus a valley at $l$ into  a peak at the same position; its action is illustrated in Fig. \ref{ex operation beta}. The inverse operation is
\begin{align}\label{defbetai}
	\beta_l^{-1}:&\left\{
	\begin{array}{ll}
		\yap{l}\rw\yap{l}-2&\mbox{if }\yap{l-1}=\yap{l+1}=\yap{l}-1,\\
		0&\mbox{otherwise.}
	\end{array}\right.
\end{align}
$\beta_l^{-1}$ changes a peak into a valley, as also shown in Fig. \ref{ex operation beta}. With $0<l<L$, the  boundary conditions are manifestly unaffected by the action  of $\beta_l$ or $\beta_l^{-1}$, so that $j$ and $s$ are not modified.

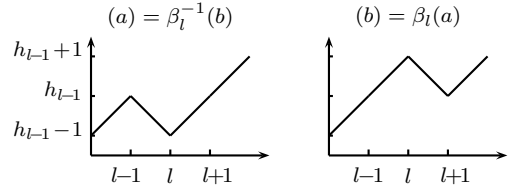
\begin{figure}[ht]\
\caption{Illustration of the operations $\beta$ and $\beta^{-1}$ on a portion of a $\ak$ path.} 
\label{ex operation beta}
\begin{center}
\begin{pspicture}(0,0)(5,1.7)
{\psset{yunit=15pt,xunit=15pt,linewidth=.8pt}

\psline{->}(0,-0.5)(0,2.5)
\rput[Br](-0.2,0){{\scriptsize $\yap{l\!-\!1}\!-\!1$}}
\rput[Br](-0.2,1){{\scriptsize $\yap{l\!-\!1}$}}
\rput[Br](-0.2,2){{\scriptsize $\yap{l\!-\!1}\!+\!1$}}
\psline{->}(0,-0.5)(4.5,-0.5)
\rput(0.89,-1){{\scriptsize $l\!\!-\!\!1$}}
\rput(2,-1){{\scriptsize $l$}}
\rput(3.2,-1){{\scriptsize $l\!\!+\!\!1$}}
\psline{-}(1,-0.5)(1,-0.35)
\psline{-}(2,-0.5)(2,-0.35)
\psline{-}(3,-0.5)(3,-0.35)
\psline{-}(0,0)(0.1,0)
\psline{-}(0,1)(0.1,1)
\psline{-}(0,2)(0.1,2)
\rput(2,3){{\scriptsize $(a)=\beta_l^{-1}(b)$}}

\psline{-}(0,0)(1,1)
\psline{-}(1,1)(2,0)
\psline{-}(2,0)(3,1)
\psline{-}(3,1)(4,2)

\psline{->}(6,-0.5)(6,2.5)
\psline{->}(6,-0.5)(10.5,-0.5)
\rput(7.,-1){{\scriptsize $l\!\!-\!\!1$}}
\rput(8,-1){{\scriptsize $l$}}
\rput(9.2,-1){{\scriptsize $l\!\!+\!\!1$}}
\psline{-}(7,-0.5)(7,-0.35)
\psline{-}(8,-0.5)(8,-0.35)
\psline{-}(9,-0.5)(9,-0.35)
\rput(8,3){{\scriptsize $(b)=\beta_l(a)$}}
\psline{-}(6,0)(6.1,0)
\psline{-}(6,1)(6.1,1)
\psline{-}(6,2)(6.1,2)

\psline{-}(6,0)(7,1)
\psline{-}(7,1)(8,2)
\psline{-}(8,2)(9,1)
\psline{-}(9,1)(10,2)
}
\end{pspicture}
\end{center}
\end{figure}

Before we plunge into the detailed analysis of the effect of these operators on the weight, we need to establish that any two paths with the same boundary conditions (or the same tail in the case of infinite paths) -- hence the same values of $j$ and $s$ -- can indeed by related to each other by a sequence of such operators. Consider then two paths $\ya,\ya'\in\ak$. We want to show that a sequence built out of $\beta_l$ and/or $\beta_l^{-1}$ on $h$ transforms it into $h'$. Note that the assumed equality of the boundary terms readily implies that $[h_l-h'_l]_2=0$, an obvious necessary requirement for the two paths to be $\beta$-related. The procedure will be to adjust the path $h$ step by step, by modifying its vertices successively from left to right. But a vertex can  be changed by a simple action of $\beta_l$ or $\beta_l^{-1}$ only if it  is a valley or a peak respectively. Otherwise, the procedure is as follows. Consider for definiteness a straight-down vertex at $l$ which we want to lift by two units. A direct action with $\beta_l$ is ruled out. But the trick is to pinpoint the next valley, say at $l+a$ and acts with $\beta_{l+a}$, thereby creating a valley at $l+a-1$.
We can then apply $\beta_{l+a-1}$, displacing the valley by one unit toward the left. Step by step, by means of the sequence $\beta_{l+1}...\beta_{l+a}\ya$, the valley is thereby moved at position $l$, where it is then changed into a peak by the action of $\beta_l$. This shows that the path $\ya'$ can be recursively constructed from $\ya$. 

We illustrate this procedure with an example, for which the paths $\ya$ and $\ya'$ are given by
\begin{align}
	\ya&=(0,1,0,-1,-2,-1,-2,-1,0,-1,0,-1,...),\nonumber\\
	\ya'&=(0,-1,0,1,2,1,0,1,0,-1,0,-1,...).
\end{align}
The procedure is as follow : 
\begin{align}
	\ya^{[1]}&=\beta_{1}^{-1}\ya=(0,-1,0,-1,-2,-1,-2,-1,0,-1,0,-1,...),\nonumber\\
	\ya^{[2]}&=\beta_{3}\beta_{4}\ya^{[1]}=(0,-1,0,1,0,-1,-2,-1,0,-1,0,-1,...),\nonumber\\
	\ya^{[3]}&=\beta_{4}\beta_{5}\beta_{6}\ya^{[2]}=(0,-1,0,1,2,1,0,-1,0,-1,0,-1,...),\nonumber\\
	\ya'&=\beta_{7}\ya^{[3]}=(0,-1,0,1,2,1,0,1,0,-1,0,-1,...).
\end{align}

Let us now see how these actions affect the weight. Let $\beta_lh\neq0$ and define
\begin{equation}
	\Delta\wa(\beta_l)\,{\equiv}\,\wa(\beta_l h)-\wa (h).
\end{equation}
Since a valley at $l$ is transformed into a peak, and that both types of vertices do not contribute to the weight, $\Delta\wa(\beta_l)$  only depends upon the way the two vertices at $l-1$ and $l+1$ are modified. Consider first the vertex at $l-1$.  $( \yap{l-2},\yap{l-1})$ is either a NE or a SE edge. In the former case, there is a peak at $l-1$ in $h$ (not contributing to the weight) which is transformed into a straight-up vertex in $\beta_lh$, with weight $(l-1)/2$. The weight increase is then $(l-1)/2$. If instead $( \yap{l-2},\yap{l-1})$ is a SE edge, the straight-down vertex is transformed into a valley and the weight decreases by $(l-1)/2$. The contribution to  $\Delta\wa(\beta_l)$ of the modified nature of the vertex at $l-1$ is thus $(l-1)(\yap{l-1}-\yap{l-2})/2$. A similar analysis of the vertex $l+1$ shows that the weight modification there is $(l+1)(\yap{l+1}-\yap{l+2})/2$. Adding up these two pieces, the net weight modification $\Delta\wa(\beta_l)$ resulting from the application of $\beta_l$ is then found to be:
\begin{align}
	\Delta\wa(\beta_l)=\frac{1}{2}(l-1)(\yap{l-1}-\yap{l-2})+\frac{1}{2}(l+1)(\yap{l+1}-\yap{l+2}).
\end{align}

We now perform a similar analysis for $\mk$ paths to show that the resulting weight modification $\Delta\wm(\beta_l)$ is equal to the above expression $\Delta\wa(\beta_l)$. The action of $\beta_l$ on a $\mk$ path is defined via the bijection (\ref{bijection2}): it consists in replacing a NE edge between $l$ and $l+1$ by a SE one as:
\begin{align}\label{actb}
	\beta_l:&\left\{
	\begin{array}{ll}
		(-1,1)\rw(1,-1) &\mbox{if }(\ymp{l},\ymp{l+1})=(-1,1)\\
		\; 0&\mbox{otherwise.}
	\end{array}\right.
\end{align}
The inverse action is
\begin{align}\label{actbi}
	\beta_l^{-1}:&\left\{
	\begin{array}{ll}
		(1,-1)\rw(-1,1) &\mbox{if }(\ymp{l},\ymp{l+1})=(1,-1)\\
		\; 0&\mbox{otherwise.}
	\end{array}\right.
\end{align}

The change $\Delta\wm(\beta_l)$ depends on the values of $\ymp{l-1}$ and $\ymp{l+2}$. Consider first the dependence upon $\ymp{l-1}$; if it is equal to $-1$, the weight does not change since a H edge becomes a NE one. However, if $\ymp{l-1}=1$, the SE edge between $l-1$ and $ l$ is destroyed and get replaced by an H edge, so the weight is increased by $l-1$. On the other hand, if $\ymp{l+2}=-1$ the SE edge  linking the positions $l+1$ and $l+2$ is displaced by one unit toward the left, which increases the weight by 1. Otherwise, if $\ymp{l+2}=1$, a SE edge is created in between $l$ and $l+1$, decreasing the weight by $l$. Adding both contributions yields:
\begin{align}
	\Delta\wm(\beta_l)&=\frac{1}{2}(l-1)(\ymp{l-1}+1)+1 -\frac{1}{2}(l+1)(\ymp{l+2}+1)\nonumber\\
	&=\frac{1}{2}(l-1)\ymp{l-1}-\frac{1}{2}(l+1)\ymp{l+2}.
\end{align}
Using the bijection (\ref{bijection1}) to re-express this in terms of the $h$-path data, the above expression becomes
\begin{align}\label{ega1}
	\Delta\wm(\beta_l)=\frac{1}{2}(l-1)(\yap{l-1}-\yap{l-2})-\frac{1}{2}(l+1)(\yap{l+2}-\yap{l+1})=\Delta\wa(\beta_l),
\end{align}
giving the announced equality.

As $\beta_l^{-1}\beta_l$ is the identity operator {(assuming $\beta_lh\neq0$)}, it does not affect the weight, meaning that
\begin{align}\label{dinv}
	\Delta\wa(\beta_l^{-1})=-\Delta\wa(\beta_l)\qquad \Delta\wm(\beta_l^{-1})=-\Delta\wm(\beta_l).
\end{align}
It thus follows directly from (\ref{ega1}) that
\begin{align}
\Delta\wm(\beta_l^{-1})=\Delta\wa(\beta_l^{-1}).
\end{align}

These results show that $\beta_l$ and $\beta_l^{-1}$ change the weight by the same amount on two paths related by the bijection. As already said, starting from an arbitrary path, a combination of these operators can generate any other path with the same values of $s$ and $j$. Consequently, if we know the weight difference between a $\ak$ path and its $\mk$ image, it must be the same for every path having the same values of these parameters. This shows that this weight difference depends only on $\cm$ and $j$. 
By considering some simple paths, it will next  be shown 
that the correction $-(\ca+j)/2$ added to the weight $w$ nullifies this difference.

\begin{figure}[ht]\
\caption{The $\ak$ path $h^{(3)}$ for $L=10$. The straight-up vertices, those contributing to the sum part of the expression (\ref{wa}) are indicated by black dots. This path has $j=1/2, s=3/2$ (cf. eqs (\ref{jdehm})--(\ref{sdehm})) and $w=-\frac12(2)+\frac12(1+2+3)=2$.} 
\label{exh4}
\begin{center}
\begin{pspicture}(0,0)(6.,2.3)
{\psset{yunit=15pt,xunit=15pt,linewidth=.8pt}
\psline{->}(0,0)(0,4.5) \psline{->}(0,0)(12,0)
\rput(2,-0.4){{\scriptsize $2$}}
\rput(4,-0.4){{\scriptsize $4$}}
\rput(6,-0.4){{\scriptsize $6$}}
\rput(8,-0.4){{\scriptsize $8$}}
\rput(10,-0.4){{\scriptsize $10$}}

\psset{dotsize=3pt}\psset{dotstyle=*}
\psdots(1.0,1.0)(2.0,2.0)(3,3)

\rput[Br](-0.2,0){{\scriptsize $0$}}
\rput[Br](-0.2,1){{\scriptsize $1$}}
\rput[Br](-0.2,2){{\scriptsize $2$}}
\rput[Br](-0.2,3){{\scriptsize $3$}}
\rput[Br](-0.2,4){{\scriptsize $4$}}

\psset{linestyle=solid}
\psline{-}(1,-0.1)(1,0.1)\psline{-}(2,-0.1)(2,0.1)
\psline{-}(3,-0.1)(3,0.1)\psline{-}(4,-0.1)(4,0.1)
\psline{-}(5,-0.1)(5,0.1)\psline{-}(6,-0.1)(6,0.1)
\psline{-}(7,-0.1)(7,0.1)\psline{-}(8,-0.1)(8,0.1)
\psline{-}(9,-0.1)(9,0.1)\psline{-}(10,-0.1)(10,0.1)
\psline{-}(11,-0.1)(11,0.1)

\psline{-}(0,0)(0.1,0)
\psline{-}(0,1)(0.1,1)
\psline{-}(0,2)(0.1,2)
\psline{-}(0,3)(0.1,3)\psline{-}(0,4)(0.1,4)

\psline{-}(0,0)(4,4)
\psline{-}(4,4)(5,3)
\psline{-}(5,3)(6,4)
\psline{-}(6,4)(7,3)
\psline{-}(7,3)(8,4)
\psline{-}(8,4)(9,3)
\psline{-}(9,3)(10,4)
\psline{-}(10,4)(11,3)
}
\end{pspicture}
\end{center}
\end{figure}
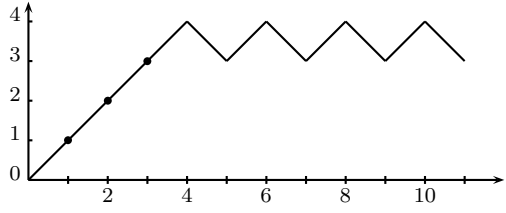

Consider the $\ak$ path $h^{(m)}$ composed of $m$ straight-up vertices when $m>0$, or $m$ straight-down vertices for $m<0$, followed by an oscillatory tail. 
In term of the sequence of vertices $(h_0,\cdots, h_{L+1})$,  $h^{(m)}$ reads:
\begin{align}
 h^{(m)}=(0,\pm1,\cdots,\pm |m|, \pm(|m|+1),\pm |m|,\cdots \pm(|m|+[L+m]_2),\pm (|m|+[L+1+m]_2))         ,
      \end{align} 
      where the signs are correlated with that of $m$. The path $h^{(3)}$ for the value $L=10$ is pictured in Fig. \ref{exh4}.

Let us determine  the characteristics of $h^{(m)}$ for the case $m>0$.
From the expression (\ref{ja}) and
\begin{equation}\label{jdehm}
h_{L+1}-h_L=[m+L+1]_2-[m+L]_2=(-1)^{L+m}\quad \Rightarrow\quad j=\frac12[m]_2.
\end{equation}
The sector is obtained  from (\ref{ca}):
\begin{equation}\label{sdehm}
2 \ca=m+ \text{min}([m+L+1]_2,[m+L]_2)\quad \Rightarrow\quad s=\frac12m.
\end{equation}
For the weight, since $|h_{l-1}-h_{l+1}|=2$ if $h_l$ is  a straight-up vertex, we have
\begin{align}
	\wa(h^{(m)}) &= 
	{-\frac{1}{2}(\ca+j)}+\frac{1}{2}\sum_{l=1}^{m}l=\frac{1}{4}(m(m+1)-m-{[m]_2})	\nonumber\\
	&=\frac{1}{4}\group{m^2-{[m]_2}}.
	\label{wam}
\end{align}
As it is easily checked, these results aslo hold for $m<0$.

\begin{figure}[ht]\
\caption{The path $\ga^{(-7)}$, with $L=10$. It has $j=1/2$ and $s=-7/2$ (these values are fixed by the bijection and eqs (\ref{jdehm}) and (\ref{sdehm})). The dotted lines indicate the path $\ga^{\gs(\frac12)}$, that is, the case $m=-1$, with respect to which the weight is calculated: $\wm(\ga^{(-7)})=-(-2-4-6)=12$ (in agreement with the formula (\ref{waa})).}
\label{ga-7}
\begin{center}
\begin{pspicture}(1,-0.5)(8,0.5)
{\psset{yunit=10pt,xunit=20pt,linewidth=.8pt}
\psline{->}(1,-1.5)(1,1.75) \psline{->}(1,0)(12,0)
\rput(2,-0.4){{\scriptsize $2$}}
\rput(4,-0.4){{\scriptsize $4$}}
\rput(6,-0.4){{\scriptsize $6$}}
\rput(8,-0.4){{\scriptsize $8$}}
\rput(10,-0.4){{\scriptsize $10$}}

\rput[Br](0.8,-1){{\scriptsize $-1$}}
\rput[Br](0.8,1){{\scriptsize $1$}}

\psset{linestyle=solid}
\psline{-}(1,-0.1)(1,0.1)\psline{-}(2,-0.1)(2,0.1)
\psline{-}(3,-0.1)(3,0.1)\psline{-}(4,-0.1)(4,0.1)
\psline{-}(5,-0.1)(5,0.1)\psline{-}(6,-0.1)(6,0.1)
\psline{-}(7,-0.1)(7,0.1)\psline{-}(8,-0.1)(8,0.1)
\psline{-}(9,-0.1)(9,0.1)\psline{-}(10,-0.1)(10,0.1)
\psline{-}(11,-0.1)(11,0.1)\psline{-}(12,-0.1)(12,0.1)

\psline{-}(1,-1)(1.1,-1)
\psline{-}(1,1)(1.1,1)

\psline{-}(1,-1)(7,-1)
\psline{-}(7,-1)(8,1)
\psline{-}(8,1)(9,-1)
\psline{-}(9,-1)(10,1)
\psline{-}(10,1)(11,-1)

\psset{linestyle=dotted}

\psline{-}(1,-1)(2,1)
\psline{-}(2,1)(3,-1)
\psline{-}(3,-1)(4,1)
\psline{-}(4,1)(5,-1)

\psline{-}(5,-1)(6,1)
\psline{-}(6,1)(7,-1)

\psline{-}(7,-1)(8,1)
\psline{-}(8,1)(9,-1)
}

\end{pspicture}
\end{center}
\end{figure}
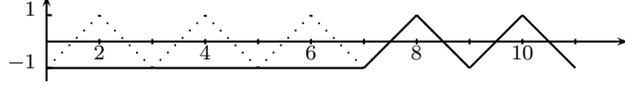

Consider now the related $\mk$ path $	\ga^{(m)}=(\ga_1,\cdots)$:
\begin{align}
	\ga^{(m)}=\underbrace{(\pm1,\cdots\pm1}_m,1,-1,1,\cdots),
\end{align}
Note that $\ga^{(-1)}=\ga^{\gs(\frac12)}$. $\ga^{(-7)}$ is displayed in Fig. \ref{ga-7}.
Recall that the weight of a $\mk_j$ path is always defined relative to the corresponding ground-state $\ga^{\gs(j)}$. Accordingly, we need to distinguish the cases $m$ even ($j=0)$ and $m$ odd $(j=1/2$). Let us start with $m$ even and $>0$. Recall that the path $\ga^{\gs(0)}$ has SE edges between $(2n-1,2n)$. Therefore, with respect to the ground-state weight, the plateau at height 1 in-between the positions $1$ and $m+1$  is equivalent to the removal of the first $m/2$ SE edges. The weight in this case is thus
\begin{align}\label{evenm}
	\wm(\ga^{(m)})=\sum_{n=1}^{m/2}(2n-1)=\frac{1}{4}m^2 \quad (m\,\text{even})
.\end{align}
For $m<0$ (and still even), the plateau at height $-1$ between 1 and $|m|$ is equivalent to the removal of the first $|m|/2$ SE edges; this leads again to the above expression.

For $m$ odd, the weight is evaluated with respect to $\ga^{\gs(\frac12)}$ ($m=-1$). In this case, the the plateau 
amounts to removing SE edges between $(2n,2n+1)$, with $n$ ranging from 1 to $(|m|-1)/2$. The weight of the path $\ga^{(m)}$ is thus
\begin{align}
	\wm(\ga^{(m)})=\sum_{n=1}^{(|m|-1)/2}2n=\group{\frac{|m|-1}{2}}\group{\frac{|m|+1}{2}}=\frac{1}{4}(m^2-1)
	\quad (m\,\text{odd}). \label{oddm}
\end{align}
Equations (\ref{evenm}) and (\ref{oddm}) can both be written in the single form:
\begin{align}\label{waa}
	\wm(\ga^{(m)})=\frac{1}{4}\group{m^2-{[m]_2}}.
\end{align}
Comparing (\ref{wam}) and{(\ref{waa})}, we see that
\begin{align}\label{wgh}
	\wa(h^{(m)})=\wm(\ga^{(m)}) .
\end{align}
Since the values of $j$ and $s$ for these two special paths are arbitrary (cf. eqs (\ref{jdehm}) and (\ref{sdehm})), the proof of the weight-preserving nature of the bijection is now complete.

As a consequence of this equivalence, both weight formulae  can be used for each path. In particular, $\wa(h)$ and $\wm(\ga)$
can be re-expressed, via the bijection, in terms of the $\mk$ and $\ak$ path data  respectively. In particular, we have
\begin{align}
	\wa(\ga)&=-\frac{1}{2}(\cm+j)+\sum_{l=1}^L \frac{l}{4}\,|\ymp{l+1}+\ymp{l}|  \nonumber\\ &=-\frac12(s+j)+\sum_{l=1}^Ll\, |j_{l+1}-j_l|,\label{wadm}
\end{align}
where the second form in 
 follows from (\ref{jlun}).

\subsubsection{Bijection $\ak\leftrightarrow\tl$}\label{bijt}

To get a $\tl$ path from a $\ak$ one, the first step is to adjust the boundary conditions to recast them within the allowed range of values for $\tl$ paths. Recall that for $\tl$ paths, we impose $\ytp{0}=1+\max(0,-2\ct)$, whereupon $\min(\ytp{0},\ytp{L},\ytp{L+1})=1$ (cf. eqs (\ref{3min}) -- (\ref{homin}) and the discussion thereafter). The boundaries adjustment is done by raising the whole $\ak$ path by $\Delta \ya=\ytp{0}$. This operation corresponds to a global vertical shift of all heights $\yap{l}$ by $\yap{l} \rightarrow \yap{l}+\ytp{0}$.

This translation is a mere adjustment of the vertical axis and does not change any property of the path. In particular, it directly relates the boundary conditions for $\ak$ and $\tl$ paths, respectively given by (\ref{hLa}) and (\ref{hLt}). The relation between $\ct$, $j$ and the boundary conditions in $\tl$ paths are just the ones for $\ak$ paths where the absolute heights $\yap{L}$ and $\yap{L+1}$ are replaced by the relative ones $ \ytp{L}-\ytp{0}$ and $\ytp{L+1}-\ytp{0}$ respectively. It does not however ensure the height of every point to be $\geq 1$. To remedy for this, we proceed via a succession of basic weight-preserving operations (defined precisely shortly) where at each  step, we replace a pair of SE and NE edges in down-most position by two H edges and raise the portion of the path in-between by one unit. This is pursued until all heights are $\geq 1$. Of course, if the $\ak$ path shifted upward by $\ytp{0}$ is everywhere $\geq1$, no further operation is needed. That would correspond to the case where the image $\tl$ path would have no H edge.

Let us now describe the procedure for transforming a $\ak$ path into a $\tl$ one for the case where $h_l+\ytp{0}$ is not $\geq 1$ for $1<l< L-1$.  The first step is to identify the points of lowest height $<1$. If this point is unique, say at $a$, it is raised it by one unit, thereby replacing a pair of adjacent SE-NE edges by two adjacent H edges. In this operation, two straight vertices are changed  into bends and one valley is transformed into a plateau. The weight change due to this operation is 
\begin{equation}
	\Delta w_{\text{shift}}=- \frac12(a-1)- \frac12(a+1)+ \frac14(a-1) +\frac{a}2+ \frac14(a+1) = 0.
\end{equation}
If there is more than one point at the lowest height, we identify among them the leftmost and the rightmost points, denoted respectively $a$ and $b$. We then raise these two points together with the portion of the path between them by 1. This creates two H edges -- linking $(a-1,a)$ and $(b,b+1)$ -- and leaves all other edges unchanged. In this way, the valleys at $a$ and $b$ (which did not contribute to the weight) are each replaced by two bends. The weight modification resulting from this operation is thus
\begin{equation}
	\Delta w_{\text{shift}}=- \frac12(a-1)+ \frac14(a-1)+ \frac{a}4 - \frac12(b+1)+ \frac14(b+1)+\frac{b}4 = 0.
\end{equation}
We repeat this procedure until the lowest point of the path has height 1. From the second step on,  a new possibility occurs which is that $a$ and $b$ might not be valleys but bends. A simple computation shows that the weight change due to these lifts is also 0 in that case.  Note that the number of H edges in the produced $\tl$ path is even, as they are generated in pairs. Since at every step, the weight is not modified, the resulting  $\tl$ path has the same weight as the starting $\ak$ path. 

Let us add a clarification concerning the correction term $-(j+s)/2$. Because the relative boundary conditions are not affected and that they determine $j$ and $s$, these two parameters $j$ and $s$ are thus preserved by the transformation. This completes the demonstration of the weight-preserving character of the map.

The inverse procedure, obtaining a $\ak$ path from a $\tl$ one, amounts to change the first half of the H edges, the $n$ leftmost ones if there is a total of  $2n$ H edges (recall  from the discussion following eq. (\ref{condtlpair}) that the relevant $\tl$ paths necessarily have an even number of H edges), into SE edges, and the second half into NE edges, then to translate the vertical axis to ensure that $\yap{0}=0$.

\begin{figure}[ht]\
\caption{The $\tl$ path corresponding to the $\ak$ path of Fig. \ref{exemplea1} via the bijection} 
\label{exbijection2}
\begin{center}
\begin{pspicture}(1,0.5)(7,2)
{\psset{yunit=15pt,xunit=15pt,linewidth=.8pt}
\psline{->}(0,1)(0,4) \psline{->}(0,1)(16,1)
\rput(2,0.6){{\scriptsize $2$}}
\rput(4,0.6){{\scriptsize $4$}}
\rput(6,0.6){{\scriptsize $6$}}
\rput(8,0.6){{\scriptsize $8$}}
\rput(10,0.6){{\scriptsize $10$}}
\rput(12,0.6){{\scriptsize $12$}}
\rput(14,0.6){{\scriptsize $14$}}

\rput[Br](-0.2,1){{\scriptsize $1$}}
\rput[Br](-0.2,2){{\scriptsize $2$}}
\rput[Br](-0.2,3){{\scriptsize $3$}}

\psset{linestyle=solid}
\psline{-}(1,0.9)(1,1.1)\psline{-}(2,0.9)(2,1.1)
\psline{-}(3,0.9)(3,1.1)\psline{-}(4,0.9)(4,1.1)
\psline{-}(5,0.9)(5,1.1)\psline{-}(6,0.9)(6,1.1)
\psline{-}(7,0.9)(7,1.1)\psline{-}(8,0.9)(8,1.1)
\psline{-}(9,0.9)(9,1.1)\psline{-}(10,0.9)(10,1.1)
\psline{-}(11,0.9)(11,1.1)\psline{-}(12,0.9)(12,1.1)
\psline{-}(13,0.9)(13,1.1)\psline{-}(14,0.9)(14,1.1)

\psline{-}(0,1)(0.1,1)
\psline{-}(0,2)(0.1,2)
\psline{-}(0,3)(0.1,3)

\psline{-}(0,1)(1,2)
\psline{-}(1,2)(2,1)
\psline{-}(2,1)(3,1)
\psline{-}(3,1)(4,2)
\psline{-}(4,2)(5,1)
\psline{-}(5,1)(6,1)
\psline{-}(6,1)(7,1)
\psline{-}(7,1)(8,2)
\psline{-}(8,2)(9,1)
\psline{-}(9,1)(10,1)
\psline{-}(10,1)(11,2)
\psline{-}(11,2)(12,3)
\psline{-}(12,3)(13,2)
\psline{-}(13,2)(14,1)
\psline{-}(14,1)(15,2)

}
\end{pspicture}
\end{center}
\end{figure}

\subsubsection{Summing up}\label{Suup}

We have thus achieved our first aim, which was to provide a weight-preserving bijection between $\mk$ and $\tl$ paths. In the process, we have found a useful tool, namely the equivalence between $\mk$ paths and the rather simple $\ak$ or SOS-type paths. Thanks to the bijection, the basic properties of a given path (namely, its values of $j,\,s$ and its weight) can be determined either from the $\ak$ or $\mk$ representation, using the one that makes the computation simpler. This is particularly useful for the evaluation of the weight. And actually, to get an alternative expression for the weight $\wm$ of a path $\ga$, there is no need to work out the corresponding path $h$ but simply to use the expression $w(h)$ rewritten in terms of $\ga$ data. 

The bottom line is that the weight of $\ga$ can be evaluated either with $\wm(\ga)$ given in (\ref{poidsre}) and (\ref{poids}) or from $\wa(\ga)$ given in (\ref{wadm}). This will be quite useful in the following construction of the generating functions for $\mk$ paths. But the real power of this observation will show up in the analysis of the higher level cases, for which it will not even be necessary to define the analogous $\ak$ paths: it suffices to guess the expression for $\wa(\ga)$ and demonstrate it \cite{LPM}. In brief, the immediate off-shoot of the present section is the gain of a new and simpler weighting of the $\mk$ paths.

In establishing this bijection $\mk\lrw \tl$, it proves convenient to consider finite paths, especially given the original definition of $\tl$ paths given in \cite{FPW}. However, for the remainder of this section, devoted to the combinatorial analysis of the $\mk$ paths, it is more natural
to consider infinite paths. And these are the genuine representatives of the $\su{1}$ states (cf. eq. (\ref{cara})).

\subsection{Description of the paths in terms of (nonlocal) operators}\label{b1}

Another implication of the bijection $\mk\lrw \ak$ is the following. There is a natural operator construction for the $\ak$ paths  -- introduced in Sect. \ref{opak} -- that is lifted (but in a slightly modified form) from \cite{JM}, applied there to unitary RSOS paths. The bijection implies a similar construction for the $\mk$ paths. It is presented in Sect. \ref{opmk}.  The path characteristics are readily expressed in terms of the operator data as shown in Sect. \ref{pacara}.

This operator description will be the starting point for the construction of the generating function in Sect. \ref{SGF}. The motivation for detailing this rather simple case is that the proper generalization of this approach to higher levels will be the cornerstone of our constructive method for obtaining the $\su{k}$ fermionic characters \cite{LPM}.

As indicated in Sect. \ref{Suup}, from now on we consider only infinite paths.

\subsubsection{Operators on $\ak$ paths}\label{opak}

We now introduce a description of the $\ak$  paths in terms of a sequence of nonlocal operators $\ab$ and $\ab^{-1}$ acting  on $h^{\gs(0)}$.\footnote{As just mentioned, these operators $b^{\pm 1}$ are closely related to the operators $\ab$ and $\ab^*$ introduced in \cite{JM} in the operator description of the finite RSOS paths representing the finitized version of the states in unitary minimal models \cite{ABF}.  In \cite{JM}, the operator $b_l$ acts on a vertex at $l$ by inserting a NE edge linking the positions $(l,l+1)$ and displacing the rest of the path by one unit {rightward} and eliminating the last edge. The operator $b^*$ acts similarly, but this time on a valley where  it creates a SE edge. Clearly, $b_lb^*_l=0$. In contrast here $b$ and $b^{-1}$ are genuine inverses, which modifies the repulsion condition (\ref{succ}).} Their action is defined in terms of the $\beta$ operators introduced in Sect. \ref{wpbeta}, by eqs (\ref{defbeta}) and (\ref{defbetai}): 
\begin{align}\label{bvsbe}
	\abp{l}{\pm1}\equiv \beta^{\pm1}_{l+1}\beta^{\pm1}_{l+3}\beta^{\pm1}_{l+5}\beta^{\pm1}_{l+7}....
\end{align}
In words, the action of $b_l^{\pm1}$ is equivalent to the application of a $\beta^{\pm1}$ at every position $l+1+2m$, $m\geq0$. As a result,  $b_l^{\pm1}$ increases/decreases by 2 the heights of the vertices $h_{l+1+2m}$, for $m\geq 0$ -- as long as the resulting path is well defined --, without affecting that of the vertices $h_{l+2m}$. These actions are illustrated in Fig. \ref{ex operation b1 p path}. Given that it modifies an infinite number of vertices, this action is clearly nonlocal.\footnote{Note that the relation (\ref{bvsbe}) can be inverted to give
\begin{equation*}\label{bevsb}
	\beta_{l}=(\abp{l-1}\,\abp{l+1}{-1})\qquad\text{and}\qquad \beta_{l}^{-1}=(\abp{l-1}{-1}\,\abp{l+1}{}),
\end{equation*}
where  the parentheses  are used to indicate a simultaneous action of the two operators as opposed to a successive one. Indeed, if we apply $\beta_{l}^{\pm1}$ on a generic path $h$, it could happen that $\abp{l+1}{\mp1}h=0$ (see below), while if the two operators $\abp{l-1}{\pm1}$ and $\abp{l+1}{\mp1}$ are applied simultaneously, the composite action might nevertheless be well-defined since the non-vanishing conditions are less stringent in the latter case. In addition, the composite action is local. Actually,  a sequence of the form ${(\abp{l}{\pm1}\abp{l+2n}{\mp1})}=\beta_{l+1}^{\pm1}\beta_{l+3}^{\pm1}...\beta_{l+2n-1}^{\pm1}$ affects a finite portion of the path (of length $2n$), without modifying the tail.}

\begin{figure}[ht]\
\caption{The action of the $\ab^{\pm1}$ operators on a $\ak$ path.}
\label{ex operation b1 p path}
\begin{center}
\begin{pspicture}(0,0)(7.5,1.7)
{\psset{yunit=15pt,xunit=15pt,linewidth=.8pt}

\psline{->}(0,-0.5)(0,2.5)
\rput[Br](-0.2,0){{\scriptsize $\yap{l}\!-\!1$}}
\rput[Br](-0.2,1){{\scriptsize $\yap{l}$}}
\rput[Br](-0.2,2){{\scriptsize $\yap{l}\!+\!1$}}
\psline{->}(0,-0.5)(7.5,-0.5)
\rput(1,-1){{\scriptsize $l$}}
\rput(3,-1){{\scriptsize $l\!+\!2$}}
\rput(5,-1){{\scriptsize $l\!+\!4$}}
\psline{-}(1,-0.5)(1,-0.35)
\psline{-}(2,-0.5)(2,-0.35)
\psline{-}(3,-0.5)(3,-0.35)
\psline{-}(4,-0.5)(4,-0.35)
\psline{-}(5,-0.5)(5,-0.35)
\psline{-}(6,-0.5)(6,-0.35)
\psline{-}(0,0)(0.1,0)
\psline{-}(0,1)(0.1,1)
\psline{-}(0,2)(0.1,2)
\rput(3.5,3){{\scriptsize $(a)=\abp{l}{-1}(b)$}}

\psline{-}(0,0)(1,1)
\psline{-}(1,1)(2,0)
\psline{-}(2,0)(3,1)
\psline{-}(3,1)(4,0)
\psline{-}(4,0)(5,1)
\psline{-}(5,1)(6,0)
\psline[linestyle=dotted]{-}(6,0)(7,1)

\psline{->}(9,-0.5)(9,2.5)
\psline{->}(9,-0.5)(16.5,-0.5)
\rput(10,-1){{\scriptsize $l$}}
\rput(12,-1){{\scriptsize $l\!+\!2$}}
\rput(14,-1){{\scriptsize $l\!+\!4$}}
\psline{-}(10,-0.5)(10,-0.35)
\psline{-}(11,-0.5)(11,-0.35)
\psline{-}(12,-0.5)(12,-0.35)
\psline{-}(13,-0.5)(13,-0.35)
\psline{-}(14,-0.5)(14,-0.35)
\psline{-}(15,-0.5)(15,-0.35)
\psline{-}(16,-0.5)(16,-0.35)
\rput(12.5,3){{\scriptsize $(b)=\abp{l}{}(a)$}}
\psline{-}(9,0)(9.1,0)
\psline{-}(9,1)(9.1,1)
\psline{-}(9,2)(9.1,2)

\psline{-}(9,0)(10,1)
\psline{-}(10,1)(11,2)
\psline{-}(11,2)(12,1)
\psline{-}(12,1)(13,2)
\psline{-}(13,2)(14,1)
\psline{-}(14,1)(15,2)
\psline[linestyle=dotted]{-}(15,2)(16,1)

}
\end{pspicture}
\end{center}
\end{figure}

The condition for the operator to be non-trivial are simply lifted from the defining restrictions of the $\beta^{\pm1}$ operators specified by (\ref{defbeta}) and (\ref{defbetai}), which are simply that $\beta$ ($\beta^{-1}$) can only act on a valley (peak). 
The action of $b^{\pm1}$, following (\ref{bvsbe}), is thus
\begin{align}\label{bsurha}
b_l^{\pm1}(h_0,h_1,\cdots &,h_l,h_{l+1},h_{l+2},\cdots)=0\nonumber\\
\text{unless} &\quad h_l=h_{l+2}=h_{l+4}=h_{l+6}=\cdots\nonumber\\
\text{and} &\quad h_l\mp1=h_{l+1}=h_{l+3}=h_{l+5}=\cdots,
\end{align}
(i.e., $j_{l+1}=j_{l+2}=j_{l+3}=\cdots$),
in which case it reads
\begin{align}\label{bsurh}
b_l^{\pm1}(h_0,h_1,\cdots ,h_l,h_{l}\mp1,h_{l},h_{l}\mp1,h_{l},\cdots)
=(h_0,h_1,\cdots ,h_l,h_{l}\pm1,h_{l},h_{l}\pm1,h_{l},\cdots).\end{align}
Since the action of $b_l^{\pm1}$ modifies the tail of the path, it modifies the value of $s$ by $\pm\tfrac12$ (the argument is detailed in Sect. \ref{pacara}.) and since $s+j\in\mathbb Z$, it also changes $j\rw \tfrac12-j$. 
In particular, $\abp{0}{-1}\ya^{\gs(0)}=\ya^{\gs(\frac12)}$,  the $j=\tfrac12$ ground-state path corresponding to $\ga^{\gs(\frac12)}$ defined in (\ref{gsgs'}). Similarly, the path $\ya^{\gs'(\frac12)}$, related to  $\ga^{\gs'(\frac12)}$, also given in (\ref{gsgs'}), is $\abp{1}{}\ya^{\gs(0)}$. That the two lowest-weight states in the module $j=\frac12$ are obtained in this way explains that it suffices to act with a string of operators on $\ya^{\gs(0)}$ to generate all paths. Actually, for a sequence of operators applied on $h^{\gs(0)}$, the non-vanishing conditions in (\ref{bsurha}) simply  means that 
the operators must be applied from left to right (i.e., in increasing value of their indices)
and must satisfy the repulsion conditions (cf. \cite{JM})
\begin{align}\label{succ}
	\abp{l'}{\pm1}\abp{l}{\mp1}\ne0\quad &\Leftrightarrow \quad l'-l\, \geq 0\quad  \text{and even}, \nonumber\\
	\abp{l'}{\pm1}\abp{l}{\pm1}\ne0\quad &\Leftrightarrow \quad l'-l{>0}\quad  \text{and odd}.
\end{align}
When describing a path by a sequence of operators, it is unnecessary to have two operators of different types acting at the same point since this is just the identity, as it follows directly from the definition that $\abp{l}{}\abp{l}{-1}=1$; the $\geq $ in (\ref{succ}) may then replaced by a strict inequality.

For the example of Fig. \ref{exbijection1}, the sequence of operators is
\begin{equation}\label{EX1}
	b_8^{-1} \, b_6\, b_5\,b_3^{-1}\, b_1\,\ya^{\gs(0)}, 
\end{equation}
while for the one in Fig. \ref{exemplea1}, it is
\begin{equation}\label{EX2}
	b_{13}^{-1} \, b_{11}\, b_{10}\,b_7\, b_5^{-1}\, b_2^{-1}\, \ya^{\gs(0)}.
\end{equation}

\subsubsection{Operators on $\mk$ paths}\label{opmk}

Let us see how the  action  of the operators $b^{\pm1}$ on $\ak$ paths is lifted onto $\mk$ paths. It it now understood that the string of operators acts on $\ga^{\gs(0)}$.  By means of the bijection (\ref{bijection2}), the action (\ref{bsurha})-(\ref{bsurh}) is transformed into:
\begin{align}\label{bsursa}
b_l^{\pm1}(\ga_1,\ga_2,\cdots &,\ga_l,\ga_{l+1},\ga_{l+2},\cdots)=0\nonumber\\
\text{unless} &\quad \mp1=\ga_{l+1}=-\ga_{l+2}=\ga_{l+3}=-\ga_{l+4}=\ga_{l+5}=\cdots,
\end{align}
where 
\begin{align}\label{bsurs}
b_l^{\pm1}(\ga_1,\ga_2,\cdots ,\ga_l,\ga_{l+1},\ga_{l+2},\ga_{l+3},
\cdots)
=(\ga_1,\ga_2,\cdots ,\ga_l,-\ga_{l+1},-\ga_{l+2},-\ga_{l+3},\cdots).\end{align}
 Here again, the non-vanishing condition
 forces an ordering on the sequential action of the operators. 
  As a consequence, every vertex at a position larger than $l-1$ must match the ground state $\ym^{\text{gs}(j_l)}$ (and in particular, there is no H edge at  the right of $l-1$). For the path of Fig. \ref{exemplem1}, we have $b_8^{-1} \, b_6\, b_5\,b_3^{-1}\, b_1\,\ga^{\gs(0)}$ (cf. (\ref{EX1})).

\begin{figure}[ht]\
\caption{The action of the $\ab^{\pm1}$ operators on a $\mk$ path.}
\label{ex operation b1 s path}
\begin{center}
\begin{pspicture}(1,-0.5)(9,0.9)
{\psset{yunit=8pt,xunit=16pt,linewidth=.8pt}

\psline{->}(1,-1.5)(1,1.5) \psline{->}(1,0)(8.5,0)
\rput(2,-1.4){{\scriptsize $l$}}
\rput(4,-1.4){{\scriptsize $l\!+\!2$}}
\rput(6,-1.4){{\scriptsize $l\!+\!4$}}
\rput(8,-1.4){{\scriptsize $l\!+\!6$}}
\rput[Br](0.8,-1){{\scriptsize $-1$}}
\rput[Br](0.8,1){{\scriptsize $1$}}
\psset{linestyle=solid}
\psline{-}(1,-0.1)(1,0.1)\psline{-}(2,-0.1)(2,0.1)
\psline{-}(3,-0.1)(3,0.1)\psline{-}(4,-0.1)(4,0.1)
\psline{-}(5,-0.1)(5,0.1)\psline{-}(6,-0.1)(6,0.1)
\psline{-}(7,-0.1)(7,0.1)\psline{-}(8,-0.1)(8,0.1)
\psline{-}(1,-1)(1.1,-1)
\psline{-}(1,1)(1.1,1)
\psline{-}(1,-1)(2,1)
\psline{-}(2,1)(3,-1)
\psline{-}(3,-1)(4,1)
\psline{-}(4,1)(5,-1)
\psline{-}(5,-1)(6,1)
\psline{-}(6,1)(7,-1)
\psline{-}(7,-1)(8,1)
\rput(4.5,2.5){{\scriptsize $(a)=\abp{l}{-1}(b)$}}

\psline{->}(10,-1.5)(10,1.5) \psline{->}(10,0)(17.5,0)
\rput(11,-1.4){{\scriptsize $l$}}
\rput(13,-1.4){{\scriptsize $l\!+\!2$}}
\rput(15,-1.4){{\scriptsize $l\!+\!4$}}
\rput(17,-1.4){{\scriptsize $l\!+\!6$}}
\rput[Br](9.8,-1){{\scriptsize $-1$}}
\rput[Br](9.8,1){{\scriptsize $1$}}
\psset{linestyle=solid}
\psline{-}(10,-0.1)(10,0.1)\psline{-}(11,-0.1)(11,0.1)
\psline{-}(12,-0.1)(12,0.1)\psline{-}(13,-0.1)(13,0.1)
\psline{-}(14,-0.1)(14,0.1)\psline{-}(15,-0.1)(15,0.1)
\psline{-}(16,-0.1)(16,0.1)\psline{-}(17,-0.1)(17,0.1)
\psline{-}(10,-1)(10.1,-1)
\psline{-}(10,1)(10.1,1)
\psline{-}(10,-1)(11,1)
\psline{-}(11,1)(12,1)
\psline{-}(12,1)(13,-1)
\psline{-}(13,-1)(14,1)
\psline{-}(14,1)(15,-1)
\psline{-}(15,-1)(16,1)
\psline{-}(16,1)(17,-1)
\rput(13.5,2.5){{\scriptsize $(b)=\abp{l}{}(a)$}}

}

\end{pspicture}
\end{center}
\end{figure}

More concretely,  with $\ga_l=\pm1$, the action specified by eq. (\ref{bsurs}) means that $b_l$ acts non-trivially when the path has  a SE edge linking $(l,l+1)$, which is then replaced by a plateau (H edge) at 1 in-between these positions, thereby displacing to the right the SE edge (and the rest of the path) by one unit. Similarly, $b_l^{-1}$ displaces by one the position of a NE edge starting at $l$ (and the rest of the path) by introducing an H edge linking $(l,l+1)$ at height $-1$.%
\footnote{This clearly disregards the situation where  $b_l^{-1}$ undoes the action of $b_l$ (or the inverse in describing the action of $b_l$), but, as already pointed out after eq. (\ref{succ}), this is actually ruled out in the operator description of paths.}

Finally, note that in terms of the variable $j_l$ defined in (\ref{jlun}), the effect of $\abp{l}{\pm}$ is to modify every $j_n$, with $n>l$, by $\pm\tfrac12(-1)^l$:
\begin{align}\label{bsurj}
	\abp{l}{\pm1}:\quad j_n\to j_n'= \frac12-j_n
	\qquad \forall n>l \qquad(\implies \; j'_n=j'_{n+1}).
\end{align}

\subsubsection{Operators vs particles}\label{oppart}

For $\mk$ paths, the above operator construction can be dressed with a natural particle interpretation. As just noted,  $\ab$ and $\ab^{-1}$ create H edges at respective heights $1$ and $-1$. It is natural to interpret these plateaux as particles, denoted $\ppp$ and $\ppm$ respectively. With this interpretation, $\abp{l}{\pm1}$ thus creates a particle $\pp{\pm}$ at position $l$, where the position of the particle is chosen to be that of the initial vertex of the H edge. Note however that for interpreting the action of $\abp{0}{-1}$ as creating a H edge between $\ymp{0}$ and $\ymp{1}$,  we need to add a path pre-segment linking  $\ymp{0}=-1$ to $\ymp{1}=-1$. The particle content of the path of  Fig. \ref{exemplem1} is presented in Fig. \ref{ex1parti}.

Specifying the sequence of operators  in a $\ym$ path (ruling out the identity combination $\abp{l}{\pm1}\abp{l}{\mp1}$) is equivalent to  specifying the position of the plateaux, or equivalently, that of the particles. The construction of the generating function can be phrased in either terms. Below, we use the operator point of view. 
In contrast, the analysis of the paths at level 2 presented in Sect. \ref{cask2} will be phrased in terms of the particle decomposition of the paths.

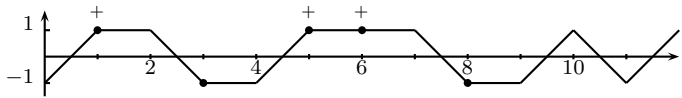
\begin{figure}[ht]\
\caption{Identification of the particles in the  $\mk_{1/2}$ path of Fig. \ref{exemplem1} augmented with a pre-segement linking the vertex $\ymp{0}=-1$. The particles correspond to H edges and their position,  indicated by a dot, is the leftmost  point of  the edge.}
\label{ex1parti}
\begin{center}
\begin{pspicture}(1,-0.5)(8,0.5)
{\psset{yunit=10pt,xunit=20pt,linewidth=.8pt}
\psline{->}(0,-1.5)(0,1.75) \psline{->}(0,0)(12,0)
\rput(2,-0.4){{\scriptsize $2$}}
\rput(4,-0.4){{\scriptsize $4$}}
\rput(6,-0.4){{\scriptsize $6$}}
\rput(8,-0.4){{\scriptsize $8$}}
\rput(10,-0.4){{\scriptsize $10$}}

\psset{dotsize=3pt}\psset{dotstyle=*}
\psdots(1.0,1.0)(3.0,-1)(5,1)(6,1)(8,-1)
\rput(1,1.7){{\scriptsize $_+$}}
\rput(3,-1.7){{\scriptsize $_-$}}
\rput(5,1.7){{\scriptsize $_+$}}
\rput(6,1.7){{\scriptsize $_+$}}
\rput(8,-1.7){{\scriptsize $_-$}}

\rput[Br](-0.2,-1){{\scriptsize $-1$}}
\rput[Br](-0.2,1){{\scriptsize $1$}}

\psset{linestyle=solid}
\psline{-}(1,-0.1)(1,0.1)\psline{-}(2,-0.1)(2,0.1)
\psline{-}(3,-0.1)(3,0.1)\psline{-}(4,-0.1)(4,0.1)
\psline{-}(5,-0.1)(5,0.1)\psline{-}(6,-0.1)(6,0.1)
\psline{-}(7,-0.1)(7,0.1)\psline{-}(8,-0.1)(8,0.1)
\psline{-}(9,-0.1)(9,0.1)\psline{-}(10,-0.1)(10,0.1)
\psline{-}(11,-0.1)(11,0.1)

\psline{-}(0,-1)(0.1,-1)
\psline{-}(0,1)(0.1,1)

\psline{-}(0,-1)(1,1)
\psline{-}(1,1)(2,1)
\psline{-}(2,1)(3,-1)
\psline{-}(3,-1)(4,-1)
\psline{-}(4,-1)(5,1)
\psline{-}(5,1)(6,1)
\psline{-}(6,1)(7,1)
\psline{-}(7,1)(8,-1)
\psline{-}(8,-1)(9,-1)
\psline{-}(9,-1)(10,1)
\psline{-}(10,1)(11,-1)
\psline{-}(11,-1)(12,1)

}

\end{pspicture}
\end{center}
\end{figure}

\subsubsection{Characteristics of the paths for a fixed operator content}\label{pacara}

Let us now consider those  $\mk$ paths obtained from the ground state $\ym^{\gs(0)}$ by the application of $n_+$ operators $b$ and $n_-$ operators $b^{-1}$, in any relative order but acting sequentially in strictly increasing values of $l$. The basic properties of a given path (namely, its values of $j,\,s$ and its weight) can be determined directly, or, thanks to the bijection,  from its $\ak$ image, choosing the one that makes the computation simpler.

We first calculate the value of $j(\ga)$ using directly the $\mk$ representation. The action of $b_l^{\pm1}$ modifies the tail, hence the value of $j$, as it is clear from (\ref{bsurj}). If $\ga\in\mk_j$, then $b_l^{\pm1}\ga \in\mk_{\frac12-j}$, irrespective of the nature of the operator (as long as $b_l^{\pm1}\ga\ne0)$. Since by construction we consider a path obtained from $\ga^{\gs(0)}$ (i.e., a path with $j=0$) by the application of a total of $n_++n_-$ operators, an inductive argument yields:
\begin{align}
	j=\frac{1}{2} [n_-+n_+]_2.\label{b j}
\end{align}

Let us now turn to the determination of $s(\ga)$ in terms of $n_+,\,n_-$.  As already stressed, the sector is a bulk property of $\mk$ paths but a boundary one for $\ak$ ones. Therefore, the value of $s(\ga)$ is most effectively computed from the $h$ image of $\ga$. Recall that $h$ oscillates between the values $2\ca$ and $2\ca+1$ (cf. the end of Sect. \ref{Sdefak}). The sector is thus given by 
\begin{align}
	 \ca=\frac12\liminf_{l\rw\infty}\{\yap{l}\}.
\end{align}
The action of $\ab_l$ has the following effect on the tail: it changes the valleys into peaks by increasing their heights by 2. Clearly, the value of the minimum is increased by one since the oscillation is now between $2\ca+2$ and $2\ca+1$ ($\ca$ being the sector before the application of the operator). As a result, the sector is increased by $1/2$. A similar analysis for $\abp{l}{-1}$ yields a sector modification of $-1/2$. It readily follows that
\begin{align}
	\cm={\frac{1}{2}(n_+-n_-)}.\label{b sector}
\end{align}
For the paths described by the sequence given in (\ref{EX1}) and (\ref{EX2}), we verify that $j=s=1/2$ and $j=s=0$ respectively, illustrating the formulae (\ref{b j}) and (\ref{b sector}).

With the values of $j$ and $s$ determined, the weight is now readily evaluated from (\ref{wa}), using again the image $h$ of $\ga$. Observe first that the action of $b_l^{\pm1}$ on $h$ -- by effectively creating a straight-up/-down vertex  at $l$ -- increases the sum-part of the weight (\ref{wa}) by $l/2$. If the operators act at positions $l_i$, it follows that
\begin{align}
	\wa&={-\frac{1}{2}(\ca+j)}+\frac{1}{2}\sum_{i=1}^{n_++n_-} l_i\nonumber\\
	&={-}{\frac{1}{4}(n_+-n_-{+}[n_++n_-]_2)}+\frac{1}{2}\sum_{i=1}^{n_++n_-} l_i.\label{wnpnm}
\end{align}.

\subsection{Determination of the $\mk$-paths generating function}\label{SGF}

In order to  construct the generating function for $\mk$ paths, we follow the procedure of \cite{OleJS} (see also \cite{PMjmp}). We first identify the configuration of minimal weight for fixed values of $(n_+,n_-)$, denoted $\gamma^{\text{mwc}}(n_+,n_-)$, and determine its weight (Sect. \ref{Smwc}). Then we indicate how the whole set of paths with same operator content can be obtained from $\gamma^{\text{mwc}}(n_+,n_-)$ by the displacements of the different operators and how these displacements modify the weight (Sect. \ref{Sodis}). Finally, the complete generating function is obtained by putting these pieces together and summing over all values of $n_+$ and $n_-$ (Sect. \ref{Sgf}).

\subsubsection{The minimal weight configuration}\label{Smwc}

The path $\gamma^{\text{mwc}}(n_+,n_-)$ is obtained from the ground-state path by the action of a sequence of $n_+$ operators  $b$ and $n_-$ operators $b^{-1}$ in positions $l_i$, with the $l_i$ arranged such that $\sum l_i$ is minimized. Because there must be a gap of two in-between the positions of two operators of different type, the weight is clearly minimized by acting first with all operators of one type -- in successive positions -- followed by the other sequence, also acting in successive positions.  By comparing the weight of the two possible configurations, it is then easy to see that the one of minimal weight is that with the $n_-$ operators $\ab^{-1}$ acting first (essentially because the minimum position at which a $b^{-1}$ can act is 0 but it is 1 for a $b$):
\begin{align}	
	\gamma^{\text{mwc}}(n_+,n_-)=
		{\abp{n_-+n_+}{}\abp{n_-+n_+-1}{}...\abp{n_-+1}{}\abp{n_--1}{-1}...\abp{1}{-1}\abp{0}{-1}} \gamma^{\text{gs(0)}}. 
\end{align}
The path $\gamma^{\text{mwc}}(3,4)$ is  illustrated in Fig. \ref{exmwc}.
The weight of $\gamma^{\text{mwc}}(n_+,n_-)$ is readily evaluated by means of  eq. (\ref{wnpnm}) :
\begin{align}
	\wm(\gamma^{\text{mwc}}(n_+,n_-))&={-\frac{1}{4}(n_+-n_-)}-
		{\frac{1}{2}j}+\frac{1}{2}\sum_{i=0}^{n_--1}i+\frac{1}{2}\sum_{i=n_-+1}^{n_++n_-}i\nonumber\\
	&={-\frac{1}{2}j}+\frac{1}{2}\sum_{i=0}^{n_--1}\left(i+\frac{1}{2}\right)+
		\frac{1}{2}\sum_{i=n_-+1}^{n_++n_-}\left(i-\frac{1}{2}\right)\nonumber\\
	&={-\frac{1}{2}j}+\frac{1}{4}\sum_{i=1}^{n_++n_-}(2i-1)\nonumber\\
	&=\frac{1}{4}(n_++n_-)^2{-\frac{1}{2}j}.
\end{align}

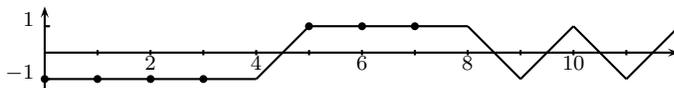
\begin{figure}[ht]\
\caption{The minimal weight configuration (augmented by a pre-segment linking the vertex $\ga_0=-1$)  with $n_-=4$ and $n_+=3$. The black dots indicate the operator application positions.}
\label{exmwc}
\begin{center}
\begin{pspicture}(1,-0.5)(8,0.5)
{\psset{yunit=10pt,xunit=20pt,linewidth=.8pt}
\psline{->}(0,-1.5)(0,1.75) \psline{->}(0,0)(12,0)
\rput(2,-0.4){{\scriptsize $2$}}
\rput(4,-0.4){{\scriptsize $4$}}
\rput(6,-0.4){{\scriptsize $6$}}
\rput(8,-0.4){{\scriptsize $8$}}
\rput(10,-0.4){{\scriptsize $10$}}

\psset{dotsize=3pt}\psset{dotstyle=*}
\psdots(0,-1)(1,-1)(2,-1)(3,-1)(5,1)(6,1)(7,1)

\rput[Br](-0.2,-1){{\scriptsize $-1$}}
\rput[Br](-0.2,1){{\scriptsize $1$}}

\psset{linestyle=solid}
\psline{-}(1,-0.1)(1,0.1)\psline{-}(2,-0.1)(2,0.1)
\psline{-}(3,-0.1)(3,0.1)\psline{-}(4,-0.1)(4,0.1)
\psline{-}(5,-0.1)(5,0.1)\psline{-}(6,-0.1)(6,0.1)
\psline{-}(7,-0.1)(7,0.1)\psline{-}(8,-0.1)(8,0.1)
\psline{-}(9,-0.1)(9,0.1)\psline{-}(10,-0.1)(10,0.1)
\psline{-}(11,-0.1)(11,0.1)

\psline{-}(0,-1)(0.1,-1)
\psline{-}(0,1)(0.1,1)

\psline{-}(0,-1)(4,-1)
\psline{-}(4,-1)(5,1)
\psline{-}(5,1)(8,1)
\psline{-}(8,1)(9,-1)
\psline{-}(9,-1)(10,1)
\psline{-}(10,1)(11,-1)
\psline{-}(11,-1)(12,1)

}

\end{pspicture}
\end{center}
\end{figure}

\subsubsection{The operator displacements}\label{Sodis} 

We next consider all {possible} displacements of the operators in order to generate all the paths with the same operator content. We first displace the $\ab$'s, by increasing the value of the application point. This is done by iterating the basic displacement  $b_l\to b_{l+2}.$ The elementary move is by two units since $b$ acts (in the tail)on a valley 
and two successive valleys 
are separated by at most two and precisely by two in the ground-state path. Every basic move  increases the weight by 1 according to eq (\ref{wnpnm}). To complete the description of the allowed displacements of the $b$  operators, we add the rule that an operator cannot pass over another one of the same type as this would generate configurations already taken into account. The displacements of the $n_+$ operators $b$ are thus is correspondence with partitions with at most $n_+$ parts. Their generating function  is the $q$-factorial $(q)_{n_+}^{-1}$ \cite[eq. 2.1.1]{Andr}, where
\begin{equation}
(q)_n=\prod_{i=1}^{n}(1-q^i).
\end{equation}

For the $\ab^{-1}$, the analogous basic displacement is $\abp{l}{-1}\rw\abp{l+2}{-1}$. However, a $b^{-1}$ can pass over a $b$ or a sequence of consecutive $b$'s according to the rule:
\begin{align}\label{rule}
	\abp{l+a+1}{0}\abp{l+a}{}\cdots\abp{l+3}{}\abp{l+2}{}\abp{l}{-1}\rw
	\abp{l+a+1}{-1}\abp{l+a}{0}\abp{l+a+1}{}\cdots{}\abp{l+2}{}\abp{l+1}{},
\end{align}
where $\abp{l}{0}$ indicates the absence of any operator at the position $l$ (see for instance Fig. \ref{ex1dis}). This operation also changes the weight by 1, irrespectively of the value of $a$. The generating function capturing all possible displacements of the $b^{-1}$ operators is $(q)_{n_-}^{-1}$.

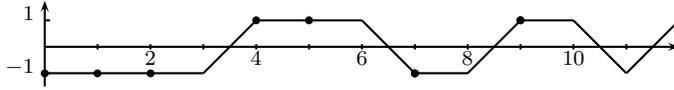
\begin{figure}[ht]\
\caption{The path obtained by two elementary  displacements from the minimal weight configuration of Fig. \ref{exmwc}, first the rightmost $b$ ($b_7\rw b_9$) and then the rightmost $b^{-1}$ ($b_6b_5b^{-1}_4\rw b^{-1}_7b_5b_4$), illustrating the rule (\ref{rule}).}
\label{ex1dis}
\begin{center}
\begin{pspicture}(1,-0.5)(8,0.5)
{\psset{yunit=10pt,xunit=20pt,linewidth=.8pt}
\psline{->}(0,-1.5)(0,1.75) \psline{->}(0,0)(12,0)
\rput(2,-0.4){{\scriptsize $2$}}
\rput(4,-0.4){{\scriptsize $4$}}
\rput(6,-0.4){{\scriptsize $6$}}
\rput(8,-0.4){{\scriptsize $8$}}
\rput(10,-0.4){{\scriptsize $10$}}

\psset{dotsize=3pt}\psset{dotstyle=*}
\psdots(0,-1)(1,-1)(2,-1)(4,1)(5,1)(7,-1)(9,1)
\psdots(7,-1)(9,1)

\rput[Br](-0.2,-1){{\scriptsize $-1$}}
\rput[Br](-0.2,1){{\scriptsize $1$}}

\psset{linestyle=solid}
\psline{-}(1,-0.1)(1,0.1)\psline{-}(2,-0.1)(2,0.1)
\psline{-}(3,-0.1)(3,0.1)\psline{-}(4,-0.1)(4,0.1)
\psline{-}(5,-0.1)(5,0.1)\psline{-}(6,-0.1)(6,0.1)
\psline{-}(7,-0.1)(7,0.1)\psline{-}(8,-0.1)(8,0.1)
\psline{-}(9,-0.1)(9,0.1)\psline{-}(10,-0.1)(10,0.1)
\psline{-}(11,-0.1)(11,0.1)

\psline{-}(0,-1)(0.1,-1)
\psline{-}(0,1)(0.1,1)

\psline{-}(0,-1)(3,-1)
\psline{-}(3,-1)(4,1)
\psline{-}(4,1)(6,1)
\psline{-}(6,1)(7,-1)
\psline{-}(7,-1)(8,-1)
\psline{-}(8,-1)(9,1)
\psline{-}(9,1)(10,1)
\psline{-}(10,1)(11,-1)
\psline{-}(11,-1)(12,1)

}

\end{pspicture}
\end{center}
\end{figure}

\subsubsection{Constructing the generating function}\label{Sgf}

Collecting these three components (the weight of $\gamma^{\text{mwc}}(n_+,n_-)$ and the generating factors for both types of displacements), we obtain the generating function for all paths with operator content ($n_+,n_-$):
\begin{align}\label{generfixe}
	f({n_+,n_-})=\frac{q^{\frac{1}{4}(n_++n_-)^2-{\frac{1}{2}j}}z^{{\frac{1}{2}(n_+-n_-)}}}{(q)_{n_+}(q)_{n_-}}.
\end{align}
Recall that $j$ is given by (\ref{b j}).

Summing over all values of $n_+, n_-$ gives the $\su{k}$ characters for both values of $j$ (cf. (\ref{cara}))
\begin{align}\label{char1}
	\chi_j^{(1)}(z;q)=\sum_{\substack{ n_+,n_-\geq 0 \\ n_++n_- = 2j\text{ mod }2}}
	\frac{q^{\frac{1}{4}(n_++n_-)^2-{\frac{1}{2}j}}z^{{\frac{1}{2}(n_+-n_-)}}}{(q)_{n_+}(q)_{n_-}}.
\end{align}
This is precisely the $k=1$ spinon formula \cite{BPS,BLS}.
Removing the parity constraint in the summation yields the direct sum of the two characters:
\begin{align}\label{char1b}
	\chi^{{(1)}}_{0\oplus\frac12}(z;q)& =	\chi^{{(1)}}_{0}(z;q)+q^{\frac14}	\, \chi^{{(1)}}_{\frac12}	(z;q) \nonumber \\ &
	=\sum_{n_+,n_-\geq 0}
	\frac{q^{\frac{1}{4}(n_++n_-)^2}z^{{\frac{1}{2}(n_+-n_-)}}}{(q)_{n_+}(q)_{n_-} }.
\end{align}

\section{Path description of the $\su{2}$ states and particle interpretation}\label{cask2}

\subsection{Introduction}

$\mk$ paths  have been shown to have a natural interpretation in terms of particles: these particles are nothing but the plateaux at $\pm1$ -- cf. Sect. \ref{oppart}. 
For $k=2$, there is also a natural way of decomposing the paths into interacting particles. 
In this section, we derive the fermionic form of the $\su{2}$ characters using the particle decomposition of the paths.

Before presenting the organization of this section, we need to extend the definition of the $\mk\equiv\mkp{1}$ paths to level 2, the set of which we call $\mkp{2}$. This is rather straightforward and only their infinite version will be considered.

\subsubsection{$\mkp{2}$ paths and their characteristics}

A $\mkp{2}$ path $\ga$ is a sequence of integers $(\ga_1,\ga_2,\cdots)$, such that $\ga_l\in\{-2,0,2\}.$ A path is characterized by: (1) its tail, which fixes the value of $j\in\{0,\frac12,1\}$ that specifies the integrable module, (2)  its sector $s$, and (3)  its weight $\wm$. These are expressed in terms of $\ga_l$ as \cite{JMMO,LPM}:
\begin{align}
&j= \lim_{l\rw \y}\left[\tfrac14(2-(-1)^l\ga_l\right],\qquad s= -j +\tfrac12\sum_{l=1}^\y(\ga_l-\ga^{\gs(j)}_l),
\qquad\ga_l^{\gs(j)}=(-1)^l(2-4j),	\label{jsgs}\\
	&\wm(\ga)=\wm^{\circ}(\ga)-\wm^{\circ}(\ga^{\gs(j)}),
	\quad \text{ with }\quad 
	\wm^{\circ}(\ga)= \tfrac12\sum_{l=1}^{\infty} l(2-\min(\ymp{l},-\ymp{l+1}))
	\label{w0}.
\end{align}
As for $k=1$, we also have the following alternative expression for the weight (whose proof is postponed to \cite{LPM}):
\begin{align}\label{weightako}
	\wa(\ga)&=-\tfrac12(\ca+j)+\tfrac14\sum_{l=1}^\y l|\ymp{l}+\ymp{l+1}|.
\end{align}

The nonlocal operator construction of the $\mkp{1}$ paths extends directly to the higher level \cite{LPM}.\footnote{For $k=2$, these operators are similar to the operators $a$ and $a^*$ introduced in \cite{JMsusy} for the paths representing states in superconformal unitary minimal models. These paths are similar to the direct $k=2$ generalization of the $\ak$ paths and therefore, via the bijection (\ref{bijection2}), related to the $\mkp{2}$ ones.} 
 In particular, any path  $\ga\in\mkp{2}$ can be written as follows
\begin{equation}\label{pirev}
\ga= \cdots b_{l}^{a_{l}}\cdots b_{2}^{a_{2}} b_{1}^{a_{1}} b_{0}^{a_{0}} \ga^{\gs(0)}\qquad \text{with}\quad a_{i}\in\{-2,-1,0,1,2\},
\end{equation}
with only finitely many $a_i\ne0$.
The action of $b_l^n$  on a  path $\ga$ is: 
\begin{align}\label{bundef}
	\abp{l}{n}&(\ymp{1}, \ymp{2}, ... , \ymp{l}, \ymp{l+1}, \ymp{l+2}, ...)=
		(\ymp{1}, \ymp{2}, ... , \ymp{l}, \ymp{l+1}', \ymp{l+2}', \ymp{l+3}', ...) \nonumber\\
	&	\quad \;\,\mbox{if} \;\; |\ymp{l+m}'|\leq 2\;\;\forall \; m\geq1,\quad\text{where}			
				\quad\ymp{l+m}'=\ymp{l+m}-2n(-1)^{m},
\end{align}
and it is 0 otherwise.
$\abp{0}{n}$ ($n<0$) is well defined and it shifts  the starting point of the path by $2n$.
In particular, the ground-state paths $\ym^{\gs(j)}$ for $j=\tfrac12,1$ are:
\begin{align}\label{gsmul}
	\ym^{\gs(\frac12)}=\abp{0}{-1}\ym^{\gs(0)}\qquad \text{and}\qquad\ym^{\gs(1)}=\abp{0}{-2}\ym^{\gs(0)}.
\end{align}
It is convenient to introduce a notation for a sequence of operators read in increasing values of the indices:
\begin{equation}
	\pi_b\equiv \abp{0}{a_0} \abp{1}{a_1}\abp{2}{a_2}\cdots \quad \implies \quad
	\ga=\pi_b^{\text{rev}}\ga^{\gs(0)}.
\end{equation}
where $\pi_b^{\text{rev}}$ means $\pi_b$ read in reversed order (cf. eq. (\ref{pirev})).
The weight $\wa$ is easily read off the operator construction: 
\begin{align}
	\frac14|\ga_l+\ga_{l+1}|=\frac{1}{2}|a_l|\quad \implies \quad w=-\frac12(s+j)+\frac12\sum_{l\geq1} l|a_l|\label{wb}.
\end{align}
Similarly, it is not difficult to express $s$  in terms of the $a_l$ \cite{LPM}: 
it takes the same form as for $k=1$ (eq. (\ref{b sector})):
\begin{align}
	s=\frac12\sum_{l\geq 0}a_l\label{cb}.
\end{align}
This operator construction will not be central in our subsequent analysis but it will facilitate the identification of the particles and some of their properties

\subsubsection{Organization of this section}

The particle decomposition of $\mkp{2}$ paths depends strongly upon the modules under consideration, or more precisely, upon the parity of $2j$. This explains the division  of this appendix in two parts. On the one hand, $\mkp{2}_{j=0,1}$ paths have their tail oscillating between $2$ and $-2$. Indeed, their tail must match that of their corresponding ground state and, from  eq. (\ref{jsgs}):  
\begin{equation}\label{jtail}
	\ymp{l}^{\gs(j)}=(-2)^{l+j+1}\quad\text{ for $j=0,1$.}
\end{equation}
Because they have similar tails, the particle decompositions of the $\mkp{2}_{j=0,1}$ paths have a common  description. On the other hand, the tail of $\mkp{2}_{1/2}$ paths consists in a straight line on the horizontal axis $( j=1/2\Rightarrow\ga_{l\gg1}=0)$; these paths require thus a separate treatment. Roughly, the tail constituents in one case acquire a particle interpretation in the other case.

Interestingly, the direct decomposition of the $\mkp{2}_{j=0,1}$ paths uses  at once four particles instead of the expected three independent fermions \cite{GodO} (see also Sect.\ 15.5.5 and Ex.\ 15.16 in \cite{CFT}).  However, closer inspection shows that two of these particles can be combined together. In a final step, the explicit summation over this unified  third type of particles is worked out exactly, resulting into a fermionic expression with only two summations. Quite remarkably, adding the resulting characters of the $j=0,1$ modules, gives, up to a simple pre-factor,  the sum of the $j=0,\tfrac12$ characters  at $k=1$ with $(z;q)\to(z^2;q^2)$ (cf. eq. (\ref{caun2})) -- an identity which appears to be new. 

The analysis of the $\mkp{2}_{1/2}$ paths is worked out directly in terms of three particles and leads to a fermionic expression that also looks to be new; however,  it takes a somewhat complicated form due to the presence of two $q$-binomial factors - cf. eq. (\ref{ca21/2}).

\subsection{The modules  $j=0$ and $j=1$}

\subsubsection{Identifying the four particles in $\mkp{2}_{0,1}$ paths}

Before we start our presentation of the particle decomposition of $\mkp{2}_{0,1}$ paths, we make a precision concerning the definition of the paths that simplifies considerably their particle description. It will be understood that all paths start at $l=0$ (instead of $l=1$) with the point $\ymp{0}=-2$. There is thus a pre-segment linking $\ymp{0}=-2$ to $\ymp{1}\in\{-2,0,2\}$, which is taken into account in the particle interpretation. This is akin to the introduction of a pre-segment starting from $\ga_0=-1$ in the $k=1$ case, introduced for the particle interpretation  to match  the operator construction (cf. Sect. \ref{oppart}). 

The first two types of particles, denoted  as  $\ppp$ and $\ppm$, are simply plateaux at $2$. Since these particles are represented by edges, a convention is needed in order to assign them a position: this position is understood to correspond to their initial (or leftmost) vertex. A third type of particle, denoted $\ppx$, corresponds to the vertex at $\ym=0$ in-between two NE or two SE edges (i.e, in $\diagup$ or $\diagdown$, hence the symbol $\t$). Finally,  the vertices at 0 other than those identified as  $\ppx$-type particles are taken to be particles of a fourth type, denoted $\ppo$. 

The $\ppo$ particles are distinguished by their property of always being attached to a particle of  another type, i.e., a $\ppo$ particle is never free. The structure of the particle, 
within which a number of $\ppo$ particles lie, is modified by this insertion. The different possibilities are illustrated in Fig. \ref{particules pmxo}. The figure also specifies the vertices that are identified as the $\ppo$ particles in a composite with a $\ppx$. The general rule is that the $\ppo$ are always at the right of the particle within which it lies. The identification of the particles within a generic path is illustrated in Fig. \ref{exempleS2}.

Note that the tail of $\mS^{(2)}_{j=0,1}$ paths is composed of a succession of NE$_2$ and SE$_2$ edges (where the subscript indicates that these edges are steps of 2 units, linking $\pm2$ to $\mp2$ respectively), and it is thus particle-free. This is also true for the whole $\ga^{\gs(0)}$ path.

\begin{figure}[htpb]\
\caption{The particles $\ppp$, $\ppm$ and $\ppx$ (the  later arising from two possible configurations), and their deformations caused by their interactions with one and two particles of type $\ppo$. The black dots indicate the position of the different particles.} 
\label{particules pmxo}
\begin{center}
\begin{pspicture}(0,1)(5.2,-6)
{\psset{yunit=12pt,xunit=12pt,linewidth=.8pt}
\psline{->}(0,0)(13,0)
\psline{->}(0,-1.2)(0,1.2)
\psline{*-}(1,1)(2,1)
\rput(1,-0.8){$_+$}
\psline{*-}(4,1)(5,0)
\psline{*-}(5,0)(6,1)
\rput(4,-0.8){$_+$}
\rput(5,-0.8){$_0$}
\psline{*-}(8,1)(9,0)
\psline{*-}(9,0)(10,0)
\psline{*-}(10,0)(11,1)
\rput(8,-0.8){$_+$}
\rput(9,-0.8){$_0$}
\rput(10,-0.8){$_0$}

\psline{->}(0,-4)(13,-4)
\psline{->}(0,-5.2)(0,-2.8)
\psline{*-}(1,-5)(2,-5)
\rput(1,-3.2){$_-$}
\psline{*-}(4,-5)(5,-4)
\psline{*-}(5,-4)(6,-5)
\rput(4,-3.2){$_-$}
\rput(5,-3.2){$_0$}
\psline{*-}(8,-5)(9,-4)
\psline{*-}(9,-4)(10,-4)
\psline{*-}(10,-4)(11,-5)
\rput(8,-3.2){$_-$}
\rput(9,-3.2){$_0$}
\rput(10,-3.2){$_0$}

\psline{->}(0,-8)(13,-8)
\psline{->}(0,-9.2)(0,-6.8)
\psline{-*}(1,-9)(2,-8)
\psline{-}(2,-8)(3,-7)
\rput(2,-8.8){$_\t$}
\psline{-*}(4,-9)(5,-8)
\psline{-*}(5,-8)(6,-8)
\psline{-}(6,-8)(7,-7)
\rput(5,-8.8){$_\t$}
\rput(6,-8.8){$_0$}
\psline{-*}(8,-9)(9,-8)
\psline{-*}(9,-8)(10,-8)
\psline{-*}(10,-8)(11,-8)
\psline{-}(11,-8)(12,-7)
\rput(9,-8.8){$_\t$}
\rput(10,-8.8){$_0$}
\rput(11,-8.8){$_0$}

\psline{->}(0,-12)(13,-12)
\psline{->}(0,-13.2)(0,-10.8)
\psline{-*}(1,-11)(2,-12)
\psline{-}(2,-12)(3,-13)
\rput(2,-12.8){$_\t$}
\psline{-*}(4,-11)(5,-12)
\psline{-*}(5,-12)(6,-12)
\psline{-}(6,-12)(7,-13)
\rput(5,-12.8){$_\t$}
\rput(6,-12.8){$_0$}
\psline{-*}(8,-11)(9,-12)
\psline{-*}(9,-12)(10,-12)
\psline{-*}(10,-12)(11,-12)
\psline{-}(11,-12)(12,-13)
\rput(9,-12.8){$_\t$}
\rput(10,-12.8){$_0$}
\rput(11,-12.8){$_0$}
}
\end{pspicture}
\end{center}
\end{figure}
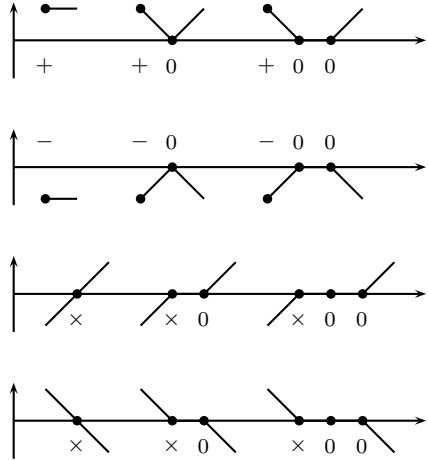

\subsubsection{Particles and operators}\label{partop2}

The different particles have a natural operator reinterpretation. For a $\pp{\pm}$ at position $l$, denoted $\pp{\pm}_l$, the equivalent operator description is $\abp{l}{\pm2}$. For $\ppx_l$, it is either $\abp{l-1}{}\abp{l}{-1}$ or $\abp{l-1}{-1}\abp{l}{}$ (note that we use subsequences of operators within $\pi_b$, that is, ordered in increasing values of $l$). The operator description of a $\ppo$ particle is more subtle since it is attached to a particle of another type and thereby deform it. Its operator content is identical to that of $\ppx_l$, i.e., $\abp{l-1}{\pm1}\abp{l}{\mp1}$. The difference is that the first operator of a $\ppo$  always cause an annihilation of the form $\abp{l}{\pm1}\abp{l}{\mp1}=1$ or $\abp{l}{\pm2}\abp{l}{\mp1}=\abp{l}{\pm1}$ with one of the operator of the host particle that precedes it (this is how the bound is created), while for $\ppx$ it does not. For instance, the pair $\pp{+}_{l-1}\ppo_l$ corresponds to $\abp{l-1}{}\abp{l}{}$, while an example of a $\ppx\ppo$ pair is $\abp{l-1}{}\abp{l+1}{-1}$.

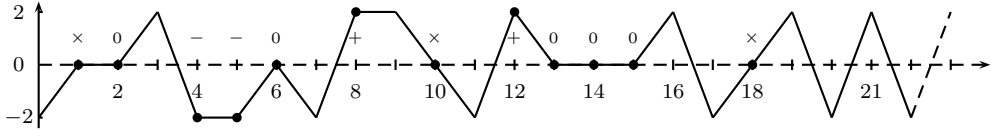
\begin{figure}[ht]\
\caption{The particle decomposition of a generic $\mk^{(2)}_{0}$ path (where the value of $j$ is read off eq. (\ref{jtail})) with a pre-segment originating from $\ga_0=-2$. The particle types are identified and their positions are indicated by black dots.
}
\label{exempleS2}
\begin{center}
\begin{pspicture}(1,0)(11,1.1)
{\psset{yunit=20pt,xunit=15pt,linewidth=.8pt}
\psline{->}(0,-1.2)(0,1.2) 
\psset{linestyle=dashed,dashadjust=false}
\psline{->}(0,0)(24,0)
\psset{linestyle=solid}

\psline{-}(0,-1)(1,0)
\psline{*-}(1,0)(2,0)
\psline{*-}(2,0)(3,1)
\psline{-}(3,1)(4,-1)
\psline{*-}(4,-1)(5,-1)
\psline{*-}(5,-1)(6,0)
\psline{*-}(6,0)(7,-1)
\psline{-}(7,-1)(8,1)
\psline{*-}(8,1)(9,1)
\psline{-}(9,1)(10,0)
\psline{*-}(10,0)(11,-1)
\psline{-}(11,-1)(12,1)
\psline{*-}(12,1)(13,0)
\psline{*-}(13,0)(14,0)
\psline{*-}(14,0)(15,0)
\psline{*-}(15,0)(16,1)
\psline{-}(16,1)(17,-1)
\psline{-}(17,-1)(18,0)
\psline{*-}(18,0)(19,1)
\psline{-}(19,1)(20,-1)
\psline{-}(20,-1)(21,1)
\psline{-}(21,1)(22,-1)
\psset{linestyle=dashed,dashadjust=false}
\psline{-}(22,-1)(23,1)

\psset{linestyle=solid}
\psline{-}(1,-0.1)(1,0.1)
\psline{-}(2,-0.1)(2,0.1)
\psline{-}(3,-0.1)(3,0.1)
\psline{-}(4,-0.1)(4,0.1)
\psline{-}(5,-0.1)(5,0.1)
\psline{-}(6,-0.1)(6,0.1)
\psline{-}(7,-0.1)(7,0.1)
\psline{-}(8,-0.1)(8,0.1)
\psline{-}(9,-0.1)(9,0.1)
\psline{-}(10,-0.1)(10,0.1)
\psline{-}(11,-0.1)(11,0.1)
\psline{-}(12,-0.1)(12,0.1)
\psline{-}(13,-0.1)(13,0.1)
\psline{-}(14,-0.1)(14,0.1)
\psline{-}(15,-0.1)(15,0.1)
\psline{-}(16,-0.1)(16,0.1)
\psline{-}(17,-0.1)(17,0.1)
\psline{-}(18,-0.1)(18,0.1)
\psline{-}(19,-0.1)(19,0.1)
\psline{-}(20,-0.1)(20,0.1)
\psline{-}(21,-0.1)(21,0.1)
\psline{-}(22,-0.1)(22,0.1)
\psline{-}(23,-0.1)(23,0.1)

\psline{-}(0,-1)(0.1,-1)
\psline{-}(0,1)(0.1,1)

\rput(2,-0.5){{\scriptsize $2$}}
\rput(4,-0.5){{\scriptsize $4$}}
\rput(6,-0.5){{\scriptsize $6$}}
\rput(8,-0.5){{\scriptsize $8$}}
\rput(10,-0.5){{\scriptsize $10$}}
\rput(12,-0.5){{\scriptsize $12$}}
\rput(14,-0.5){{\scriptsize $14$}}
\rput(16,-0.5){{\scriptsize $16$}}
\rput(18,-0.5){{\scriptsize $18$}}
\rput(21,-0.5){{\scriptsize $21$}}

\rput(-0.5,-1){{\scriptsize $-2$}}
\rput(-0.5,0){{\scriptsize $0$}}
\rput(-0.5,1){{\scriptsize $2$}}

\rput(1,0.5){{\scriptsize $_\t$}}
\rput(2,0.5){{\scriptsize $_0$}}
\rput(4,0.5){{\scriptsize $_-$}}
\rput(5,0.5){{\scriptsize $_-$}}
\rput(6,0.5){{\scriptsize $_0$}}
\rput(8,0.5){{\scriptsize $_+$}}
\rput(10,0.5){{\scriptsize $_\t$}}
\rput(12,0.5){{\scriptsize $_+$}}
\rput(13,0.5){{\scriptsize $_0$}}
\rput(14,0.5){{\scriptsize $_0$}}
\rput(15,0.5){{\scriptsize $_0$}}
\rput(18,0.5){{\scriptsize $_\t$}}
}
\end{pspicture}
\end{center}
\end{figure}
\vskip0.5cm

An interesting property of these sequences of operators describing the individual particles is that they form the basic blocks which change the value of $j$ by $1$, i.e., $j\rw[j+1]_2$.  For a string of $n_++n_-+n_0+n_\t$ basic blocks acting on the vacuum, it follows that 
\begin{equation}
j=[n_++n_-+n_0+n_\t]_2.\label{jvsN}
\end{equation}
This is easily justified. Indeed, it is clear from Fig. \ref{particules pmxo} that any particle has an effective diameter of 1. This implies that a finite-weight  path obtained from $\ga^{\gs(0)}$ by the adjunction of $N$ particles, has a tail of the form $\ga_l^{\text{mwc}}=(-2)^{l+N+1}$ for $l>N+1$. From (\ref{jtail}), the value of $j$ is found to be $j=[N]_2$.

We can use these operator--particle correspondences to deconstruct an arbitrary operator sequence $\pi_b$ into a sequence of  particles. Starting from the right (i.e., the largest position index), we identify the particle, which inserted at the position $l$ of the last operator, would be created by the operator $b_l^{a_l}$ (if $|a_l|=2$) or a pair containing it (if $|a_l|=1$). 
 In the latter case, we pair the operator at $l$ with an operator at $l-1$. The pairing procedure (i.e., the way this latter operator arises) determines the type of particle associated to the pair of operators. If the operator $b_{l-1}^{\pm1}$ is already present, the particle is a $\ppx$; if, on the other hand, it needs to be created via $1=b_{l-1}^{\pm1}b_{l-1}^{\mp1}$, it is a $\ppo$. For example, if the last operator is $\abp{l}{-1}$, we extract a $\ppx$ or $\ppo$ using the operation 
\begin{equation}
	\cdots\,\abp{l-1}{a_{l-1}}\abp{l}{-1}\rw\cdots\,\abp{l-1}{a_{l-1}-1}(\abp{l-1}{1}\abp{l}{-1})=\cdots\,\abp{l-1}{a_{l-1}-1}\pp{\epsilon}_l,
\end{equation}
where
\begin{equation}
	\epsilon=\left\{
	\begin{array}{ll}
		\t&\text{ if }a_{l-1}=1\\
		0&\text{ otherwise.}
	\end{array}\right.
\end{equation}
Doing this operation recursively reveals the particle content. For the example of Fig. \ref{exempleS2}, this gives:
\begin{align}
	\pi_b&=\abp{0}{-1}\,\abp{2}{1}\,\abp{4}{-2}\,\abp{5}{-1}\,\abp{6}{-1}\,\abp{8}{2}\,\abp{9}{1}\,
	\abp{10}{-1}\,\abp{12}{1}\,\abp{15}{1}\,\abp{17}{-1}\,\abp{18}{1}\nonumber\\
	&\rw(\abp{0}{-1}\abp{1}{1})(\abp{1}{-1}\abp{2}{1})(\abp{4}{-2})(\abp{5}{-2})(\abp{5}{1}\abp{6}{-1})(\abp{8}{2})(\abp{9}{1}\abp{10}{-1})(\abp{12}{2})
	(\abp{12}{-1}\abp{13}{1})(\abp{13}{-1}\abp{14}{1})(\abp{14}{-1}\abp{15}{1})(\abp{17}{-1}\abp{18}{1})\nonumber\\
	&\rw\;\ppx_{1}\,\ppo_{2}\,\ppm_{4}\,\ppm_{5}\,\ppo_{6}\,\ppp_{8}\,\ppx_{10}\,\ppp_{12}\,\ppo_{13}\,
	\ppo_{14}\,\ppo_{15}\,\ppx_{18}
\end{align}
which is indeed the path's particle sequence. 

\subsection{Path characteristics: sector and weight}

The evaluation of the sector and the weight of a path in terms of its particle content is most easily done via the corresponding operator sequence. Since $b_l^{a_l}$ contributes to $a_l/2$ to the sector (cf. eq. (\ref{cb})), it readily follows from the above operator description of the individual particles that
\begin{equation}\label{sec2}
	s=\nnp-\nnm.
\end{equation}

The operator representation is also useful for finding the weight of the particles using eq. (\ref{wb})
decomposed as follows
\begin{equation}\label{defwt}
w(\ga)=-\frac12j+\wt(\ga),\qquad 
	\wt(\ga)=-\frac12\ca+\frac12\sum_{l=1}^\y l|a_{l}|,
\end{equation}
and evaluating the contribution to the $-\cm/2$ part using (\ref{sec2}). Consider first the weight of an isolated particle at $l$. For a $\pp{\pm}$, the only contribution is from $\abp{l}{\pm2}$, for a weight of $\wt_{(\pm)}=l-\cm/2=l\mp1/2$. For a $\ppx$, there are two operators, at $l$ and $l+1$, so the weight is $\wt_{(\t)}=(l-1)/2+l/2-\cm/2=l-1/2$. When particles are not isolated, we add the contribution of the attached $\ppo$. The particle annihilates an operator at $l-1$, reducing the weight by $(l-1\pm1)/2$, but also adds one of the same type at $l$, increasing the weight by $(l\pm1)/2$. The net weight change associated to an attached $\ppo$ is thus always $1/2$, irrespectively of its position: $\wt_{(0)}=1/2$. Summing up, the contribution to the weight of a particle at position $l$ is
\begin{equation}\label{difw}
	\wt_{(+)}= \wt_{(\t)}=l-\frac12 ,\qquad\quad \wt_{(-)}=l+\frac12,\qquad \wt_{(0)}=\frac12.
\end{equation}

\subsubsection{The configuration of minimal weight}

Given a fixed particle content, that is, a fixed number $\nn{\epsilon}$ of particles  of type $\pp{\epsilon}$, with $\epsilon\in\{+,-,\t,0\}$, we look for the configuration of minimal weight  $\ga^{\text{mwc}}(\nnp,\nnm,\nnx,\nno)\in\mkp{2}_{j=0,1}$. In a first step, we identify  $\ga^{\text{mwc}}(\nnp,\nnm,0,0)$. This configuration is obtained by distributing the plateaux exactly as in the $k=1$ case: the $\nnm$ particles $\ppm$ are located at the beginning of the path (including the pre-segment) followed by the $\ppp$ particles. $\ga^{\text{mwc}}(\nnp,\nnm,\nnx,0)$ is obtained from this configuration by replacing the first $\nnx$ NE$_2$ and SE$_2$ edges, {including the one before the $\ppp$ if $\nnp>0$,} by $\nnx$ pairs of simple NE and SE edges ({explicitly} NE$_2\to$ NE,NE and SE$_2\to$ SE,SE), thereby inserting the $\nnx$ particles $\ppx$. Finally, the $\nno$ particles are inserted in successive positions within the rightmost particle. Indeed, their weight being independent of their position, it is energetically favorable to leave them at the far right of the sequence of particles since otherwise another type of particle would have to be moved toward the right, thereby increasing the weight. The configuration $\ga^{\text{mwc}}(4,2,4,4)$ is illustrated in Fig. \ref{exemplemwc}. We stress that to read off  the proper particle content, the pre-segment needs to be considered. As a further simple example, the three paths of weight 0 (where the weight is computed in the following paragraph) and respective sector $-1,0$ and 1, in the form $(\ga_0,\ga_1,\cdots)$, are:
\begin{equation}
	(-2,-2,2,-2,2,\cdots)\equiv\ga^{\gs(1)} ,\qquad (-2,0,2,-2,2,\cdots),\qquad (-2,2,2,-2,2,\cdots),
\end{equation}
with respective particle content: $\nnm=1,\,\nnx=1$ and $\nnp=1$.

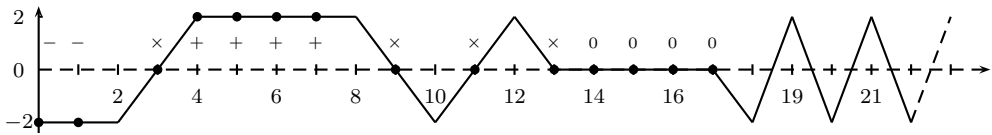
\begin{figure}[ht]\
\caption{The path $\ga^{\text{mwc}}(4,2,4,4)$ drawn  with a pre-segment originating from $\ga_0=-2$. The particle-type are identified and their positions are indicated by black dots.
}
\label{exemplemwc}
\begin{center}
\begin{pspicture}(1,-1)(11,1.1)
{\psset{yunit=20pt,xunit=15pt,linewidth=.8pt}
\psline{->}(0,-1.2)(0,1.2) 
\psset{linestyle=dashed,dashadjust=false}
\psline{->}(0,0)(24,0)
\psset{linestyle=solid}

\psline{*-}(0,-1)(1,-1)
\psline{*-}(1,-1)(2,-1)
\psline{-}(2,-1)(3,0)
\psline{*-}(3,0)(4,1)
\psline{*-}(4,1)(5,1)
\psline{*-}(5,1)(6,1)
\psline{*-}(6,1)(7,1)
\psline{*-}(7,1)(8,1)
\psline{-}(8,1)(9,0)
\psline{*-}(9,0)(10,-1)
\psline{-}(10,-1)(11,0)
\psline{*-}(11,0)(12,1)
\psline{-}(12,1)(13,0)
\psline{*-}(13,0)(14,0)
\psline{*-}(14,0)(15,0)
\psline{*-}(15,0)(16,0)
\psline{*-}(16,0)(17,0)
\psline{*-}(17,0)(18,-1)
\psline{-}(18,-1)(19,1)
\psline{-}(19,1)(20,-1)
\psline{-}(20,-1)(21,1)
\psline{-}(21,1)(22,-1)
\psset{linestyle=dashed,dashadjust=false}
\psline{-}(22,-1)(23,1)

\psset{linestyle=solid}
\psline{-}(1,-0.1)(1,0.1)
\psline{-}(2,-0.1)(2,0.1)
\psline{-}(3,-0.1)(3,0.1)
\psline{-}(4,-0.1)(4,0.1)
\psline{-}(5,-0.1)(5,0.1)
\psline{-}(6,-0.1)(6,0.1)
\psline{-}(7,-0.1)(7,0.1)
\psline{-}(8,-0.1)(8,0.1)
\psline{-}(9,-0.1)(9,0.1)
\psline{-}(10,-0.1)(10,0.1)
\psline{-}(11,-0.1)(11,0.1)
\psline{-}(12,-0.1)(12,0.1)
\psline{-}(13,-0.1)(13,0.1)
\psline{-}(14,-0.1)(14,0.1)
\psline{-}(15,-0.1)(15,0.1)
\psline{-}(16,-0.1)(16,0.1)
\psline{-}(17,-0.1)(17,0.1)
\psline{-}(18,-0.1)(18,0.1)
\psline{-}(19,-0.1)(19,0.1)
\psline{-}(20,-0.1)(20,0.1)
\psline{-}(21,-0.1)(21,0.1)
\psline{-}(22,-0.1)(22,0.1)
\psline{-}(23,-0.1)(23,0.1)

\psline{-}(0,-1)(0.1,-1)
\psline{-}(0,1)(0.1,1)

\rput(2,-0.5){{\scriptsize $2$}}
\rput(4,-0.5){{\scriptsize $4$}}
\rput(6,-0.5){{\scriptsize $6$}}
\rput(8,-0.5){{\scriptsize $8$}}
\rput(10,-0.5){{\scriptsize $10$}}
\rput(12,-0.5){{\scriptsize $12$}}
\rput(14,-0.5){{\scriptsize $14$}}
\rput(16,-0.5){{\scriptsize $16$}}
\rput(19,-0.5){{\scriptsize $19$}}
\rput(21,-0.5){{\scriptsize $21$}}

\rput(-0.5,-1){{\scriptsize $-2$}}
\rput(-0.5,0){{\scriptsize $0$}}
\rput(-0.5,1){{\scriptsize $2$}}

\rput(0.3,0.5){{\scriptsize $_-$}}
\rput(1,0.5){{\scriptsize $_-$}}
\rput(4,0.5){{\scriptsize $_+$}}
\rput(5,0.5){{\scriptsize $_+$}}
\rput(6,0.5){{\scriptsize $_+$}}
\rput(7,0.5){{\scriptsize $_+$}}
\rput(3,0.5){{\scriptsize $_\t$}}
\rput(9,0.5){{\scriptsize $_\t$}}
\rput(11,0.5){{\scriptsize $_\t$}}
\rput(13,0.5){{\scriptsize $_\t$}}
\rput(14,0.5){{\scriptsize $_0$}}
\rput(15,0.5){{\scriptsize $_0$}}
\rput(16,0.5){{\scriptsize $_0$}}
\rput(17,0.5){{\scriptsize $_0$}}
}
\end{pspicture}
\end{center}
\end{figure}

Let us now evaluate the weight of the configuration $\ga^{\text{mwc}}(\nnp,\nnm,\nnx,\nno)$. Again, it is simpler to proceed by steps and consider first the case where $\nnx=\nno=0$ 
(cf. eq. (\ref{defwt}) and (\ref{difw})):
\begin{align}
	\wa(\ga^{\text{mwc}}(\nnp,\nnm,0,0))=-\frac{j}2+\sum_{l=0}^{\nnm-1}\group{l+\frac{1}{2}}+\sum_{l=\nnm+1}^{\nnm+\nnp}\group{l-\frac{1}{2}}.
\end{align}
Consider the effect of introducing the $\nnx$ particles $\ppx$. The first $\ppx$ particle shifts the position of all the $\ppp$ particles, which increases the weight by  $\nnp$. Adding the weight of the particle itself, this gives a weight of $\nnm+\nnp+1/2$. This is equivalent to add a term in the second summation by increasing the upper bound by one unit. With $\nnx$ generic, we can shift this upper bound by $\nnx$, and correct for the fact that the particles $\ppx$ starting from the second one are separated from each other by two units instead of one, by introducing the correcting factor $\sum_{i=1}^{\nnx}(i-1)$:
\begin{align}
	\wa(\ga^{\text{mwc}}(\nnp,\nnm,\nnx,0))=&-\frac{j}2+\sum_{i=1}^{\nnx}(i-1)+\sum_{l=0}^{\nnm-1}\group{l+\frac{1}{2}}
		+\sum_{l=\nnm+1}^{\nnm+\nnp+\nnx}\group{l-\frac{1}{2}} \nonumber\\
	=&-\frac{j}2+\frac{1}{2}\nnx(\nnx-1)+\frac{1}{2}(\nnm+\nnp+\nnx)^2.
\end{align}
Finally, the addition of the $\nno$ particles $\ppo$ simply shifts the weight by $\nno/2$:
\begin{align}\label{wmwc}
	\wa(\ga^{\text{mwc}}(\nnp,\nnm,\nnx,\nno))=-\frac{j}2+\frac{1}{2}(\nnp+\nnm+\nnx)^2+\frac{1}{2}\nnx(\nnx-1)+\frac{1}{2}\nno.
\end{align}

\subsubsection{The particle displacements}

Having determined the minimal-weight configuration with a fixed  particle content, we now describe all paths containing the same particles in terms of the possible particle displacements.

Let us first describe the displacements of the particles of type $\ppo$ whose understanding is necessary for the analysis of the other particles. In $\ga^{\text{mwc}}(\nnp,\nnm,\nnx,\nno)$, the $\ppo$ particles are all attached to  the rightmost particle. Because they need to be attached to another particle, these $\ppo$ particles cannot be displaced toward the right. As usual, particles of a given type are viewed as being impenetrable to each other. The first move must then involve the leftmost $\ppo$ and is toward the left. Consider then the elementary displacement of this leftmost $\ppo$ particle which consists in leaving the rightmost host particle to become attached to the next to the last particle of type $\ne \ppo$. This displaces the (quitted) rightmost particle by one unit toward the right, augmenting the weight by 1 (cf. (\ref{difw}), and recall that the displacement of a $\ppo$, by itself, does not affect its weight). Two examples are presented in Fig. \ref{p0 move}. The maximal number of elementary displacements for the leftmost $\ppo$ particle is clearly $\nnp+\nnm+\nnx-1$: this is numbering of the most distant particle (counted from right to left)  to which the leftmost $\ppo$ can be attached following its displacement. The generating factor $G_{0}(q)$ describing all the possible displacements of  the $\ppo$ particles is clearly the number of partitions into at most $\nno$ parts, with the largest part $\leq \nnp+\nnm+\nnx-1$ \cite[Theo 3.1]{Andr}:
\begin{align}\label{Goq}
	G_{0}(q)=&\qbin{\nno+\nnp+\nnm+\nnx-1}{\nno}\equiv\qbin{N-1}{\nno},
\end{align}
where the $q$-binomial is defined as
\begin{equation}
\begin{bmatrix}
a\\ b\end{bmatrix} \equiv \begin{bmatrix}
a\\ b\end{bmatrix}_q =  \begin{cases} & \frac{(q)_a}{(q)_{a-b}(q)_b}  \quad \text{if}\quad 0\leq b\leq a,\\ &\qquad 0\qquad\qquad\text{otherwise}.
\end{cases}
\end{equation}
This definition is augmented by the condition
\begin{equation}\label{qbi-1}
\qbin{-1}{0}=1,
\end{equation}
required for the expression (\ref{Goq}) to make sense in absence of particles.

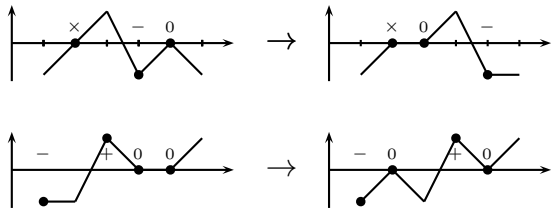
\begin{figure}[htpb]\
\caption{Two examples of elementary $\ppo$ displacements.}
\label{p0 move}
\begin{center}
\begin{pspicture}(0,1)(6.2,-2)
{\psset{yunit=12pt,xunit=12pt,linewidth=.8pt}

\psline{->}(0,0)(7,0)
\psline{->}(0,-1.2)(0,1.2)
\psline{-}(1,-1)(2,0)
\psline{*-}(2,0)(3,1)
\psline{-}(3,1)(4,-1)
\psline{*-}(4,-1)(5,0)
\psline{*-}(5,0)(6,-1)
\rput(2,0.5){{\scriptsize $_\t$}}
\rput(4,0.5){{\scriptsize $_-$}}
\rput(5,0.5){{\scriptsize $_0$}}
\psline{-}(1,-0.1)(1,0.1)
\psline{-}(2,-0.1)(2,0.1)
\psline{-}(3,-0.1)(3,0.1)
\psline{-}(4,-0.1)(4,0.1)
\psline{-}(5,-0.1)(5,0.1)
\psline{-}(6,-0.1)(6,0.1)
\rput(8.5,0){{\large$\rw$}}
\psline{->}(10,0)(17,0)
\psline{->}(10,-1.2)(10,1.2)
\psline{-}(11,-1)(12,0)
\psline{*-}(12,0)(13,0)
\psline{*-}(13,0)(14,1)
\psline{-}(14,1)(15,-1)
\psline{*-}(15,-1)(16,-1)
\rput(12,0.5){{\scriptsize $_\t$}}
\rput(15,0.5){{\scriptsize $_-$}}
\rput(13,0.5){{\scriptsize $_0$}}
\psline{-}(11,-0.1)(11,0.1)
\psline{-}(12,-0.1)(12,0.1)
\psline{-}(13,-0.1)(13,0.1)
\psline{-}(14,-0.1)(14,0.1)
\psline{-}(15,-0.1)(15,0.1)
\psline{-}(16,-0.1)(16,0.1)

\psline{->}(0,-4)(7,-4)
\psline{->}(0,-5.2)(0,-2.8)
\psline{*-}(1,-5)(2,-5)
\psline{-}(2,-5)(3,-3)
\psline{*-}(3,-3)(4,-4)
\psline{*-}(4,-4)(5,-4)
\psline{*-}(5,-4)(6,-3)
\rput(1,-3.5){{\scriptsize $_-$}}
\rput(3,-3.5){{\scriptsize $_+$}}
\rput(4,-3.5){{\scriptsize $_0$}}
\rput(5,-3.5){{\scriptsize $_0$}}
\psline{-}(1,-0.1)(1,0.1)
\psline{-}(2,-0.1)(2,0.1)
\psline{-}(3,-0.1)(3,0.1)
\psline{-}(4,-0.1)(4,0.1)
\psline{-}(5,-0.1)(5,0.1)
\psline{-}(6,-0.1)(6,0.1)
\rput(8.5,-4){{\large$\rw$}}
\psline{->}(10,-4)(17,-4)
\psline{->}(10,-5.2)(10,-2.8)
\psline{*-}(11,-5)(12,-4)
\psline{*-}(12,-4)(13,-5)
\psline{-}(13,-5)(14,-3)
\psline{*-}(14,-3)(15,-4)
\psline{*-}(15,-4)(16,-3)
\rput(11,-3.5){{\scriptsize $_-$}}
\rput(14,-3.5){{\scriptsize $_+$}}
\rput(12,-3.5){{\scriptsize $_0$}}
\rput(15,-3.5){{\scriptsize $_0$}}
\psline{-}(1,-0.1)(1,0.1)
\psline{-}(2,-0.1)(2,0.1)
\psline{-}(3,-0.1)(3,0.1)
\psline{-}(4,-0.1)(4,0.1)
\psline{-}(5,-0.1)(5,0.1)
\psline{-}(6,-0.1)(6,0.1)
\rput(8.5,0){{\large$\rw$}}
}
\end{pspicture}
\end{center}
\end{figure}

The basic data controlling the combinatorial description of these displacements is what we will call the {\it corank} of a $\ppo$ particle: this is the number of particles $\ne \ppo$ at the right of the $\ppo$ particle under consideration. In $\ga^{\text{mwc}}(\nnp,\nnm,\nnx,\nno)$, all $\ppo$ particles have corank zero; an elementary displacement increases the corank by one (and clearly, this displacement could be a step of more than one unit). In Fig. \ref{exempleS2}, there are 5 $\ppo$ and their coranks are respectively (from left to right): 6,4,1,1,1.
 
Consider next the displacements of the other types of particles. Basically, they all move in the same way, toward the right and at the next available position, with the ordering of the particles of a given type being always preserved. However, the weight modification depends upon the type of particle that is moved.
 
In $\ga^{\text{mwc}}(\nnp,\nnm,0,0)$, the rightmost $\ppp$ particle is followed by an oscillatory tail of type NE$_2$-SE$_2$; its elementary displacement  is by two units to reach the closest peak  (exactly as in the $k=1$ case) and the weight is clearly increased by 2 (cf. (\ref{difw})). For a generic configuration, there might be other particles of type $\ppx$ or $\ppm$ in between the next available position. If there are $m$ such particles, the displacement is by $2+m$ units. This is illustrated by the first set of pictures in Fig. \ref{particules move}. However, in the process, these $m$ intermediate particles are all displaced by one unit toward the left; the resulting weight shift is still $+2$. If, in addition, there are intermediate $\ppo$ particles, the prescription is that these should all be simultaneously moved (toward the right) by an elementary displacement that preserve their corank. 
In this way, the basic displacement of a $\ppp$ particle is always associated with a weight increase of 2 units.

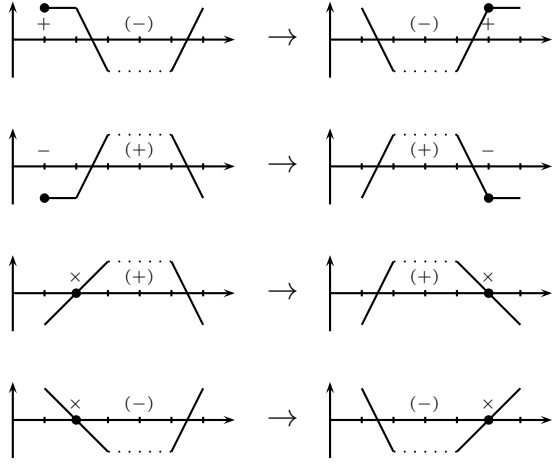
\begin{figure}[htpb]\
\caption{Elementary displacements of the particles $\ppp$, $\ppm$ and $\ppx$ in the absence of any $\ppo$. The dotted lines indicate an arbitrary number of H edges (possibly 0).} 
\label{particules move}
\begin{center}
\begin{pspicture}(0,1)(6.2,-6)
{\psset{yunit=12pt,xunit=12pt,linewidth=.8pt}

\psline{->}(0,0)(7,0)
\psline{->}(0,-1.2)(0,1.2)
\psline{*-}(1,1)(2,1)
\psline{-}(2,1)(3,-1)
\psline[linestyle=dotted]{-}(3,-1)(5,-1)
\psline{-}(5,-1)(6,1)
\rput(1,0.5){{\scriptsize $_+$}}
\rput(4,0.5){{\scriptsize $_{(-)}$}}
\psline{-}(1,-0.1)(1,0.1)
\psline{-}(2,-0.1)(2,0.1)
\psline{-}(3,-0.1)(3,0.1)
\psline{-}(4,-0.1)(4,0.1)
\psline{-}(5,-0.1)(5,0.1)
\psline{-}(6,-0.1)(6,0.1)
\rput(8.5,0){{\large$\rw$}}

\psline{->}(10,0)(17,0)
\psline{->}(10,-1.2)(10,1.2)
\psline{-}(11,1)(12,-1)
\psline[linestyle=dotted]{-}(12,-1)(14,-1)
\psline{-}(14,-1)(15,1)
\psline{*-}(15,1)(16,1)
\rput(15,0.5){{\scriptsize $_+$}}
\rput(13,0.5){{\scriptsize $_{(-)}$}}
\psline{-}(11,-0.1)(11,0.1)
\psline{-}(12,-0.1)(12,0.1)
\psline{-}(13,-0.1)(13,0.1)
\psline{-}(14,-0.1)(14,0.1)
\psline{-}(15,-0.1)(15,0.1)
\psline{-}(16,-0.1)(16,0.1)

\psline{->}(0,-4)(7,-4)
\psline{->}(0,-5.2)(0,-2.8)
\psline{*-}(1,-5)(2,-5)
\psline{-}(2,-5)(3,-3)
\psline[linestyle=dotted]{-}(3,-3)(5,-3)
\psline{-}(5,-3)(6,-5)
\rput(1,-3.5){{\scriptsize $_-$}}
\rput(4,-3.5){{\scriptsize $_{(+)}$}}
\psline{-}(1,-4.1)(1,-3.9)
\psline{-}(2,-4.1)(2,-3.9)
\psline{-}(3,-4.1)(3,-3.9)
\psline{-}(4,-4.1)(4,-3.9)
\psline{-}(5,-4.1)(5,-3.9)
\psline{-}(6,-4.1)(6,-3.9)
\rput(8.5,-4){{\large$\rw$}}

\psline{->}(10,-4)(17,-4)
\psline{->}(10,-5.2)(10,-2.8)
\psline{-}(11,-5)(12,-3)
\psline[linestyle=dotted]{-}(12,-3)(14,-3)
\psline{-}(14,-3)(15,-5)
\psline{*-}(15,-5)(16,-5)
\rput(15,-3.5){{\scriptsize $_-$}}
\rput(13,-3.5){{\scriptsize $_{(+)}$}}
\psline{-}(11,-4.1)(11,-3.9)
\psline{-}(12,-4.1)(12,-3.9)
\psline{-}(13,-4.1)(13,-3.9)
\psline{-}(14,-4.1)(14,-3.9)
\psline{-}(15,-4.1)(15,-3.9)
\psline{-}(16,-4.1)(16,-3.9)

\psline{->}(0,-8)(7,-8)
\psline{->}(0,-9.2)(0,-6.8)
\psline{-}(1,-9)(2,-8)
\psline{*-}(2,-8)(3,-7)
\psline[linestyle=dotted]{-}(3,-7)(5,-7)
\psline{-}(5,-7)(6,-9)
\rput(2,-7.5){{\scriptsize $_\t$}}
\rput(4,-7.5){{\scriptsize $_{(+)}$}}
\psline{-}(1,-8.1)(1,-7.9)
\psline{-}(2,-8.1)(2,-7.9)
\psline{-}(3,-8.1)(3,-7.9)
\psline{-}(4,-8.1)(4,-7.9)
\psline{-}(5,-8.1)(5,-7.9)
\psline{-}(6,-8.1)(6,-7.9)
\rput(8.5,-8){{\large$\rw$}}

\psline{->}(10,-8)(17,-8)
\psline{->}(10,-9.2)(10,-6.8)
\psline{-}(11,-9)(12,-7)
\psline[linestyle=dotted]{-}(12,-7)(14,-7)
\psline{-}(14,-7)(15,-8)
\psline{*-}(15,-8)(16,-9)
\rput(15,-7.5){{\scriptsize $_\t$}}
\rput(13,-7.5){{\scriptsize $_{(+)}$}}
\psline{-}(11,-8.1)(11,-7.9)
\psline{-}(12,-8.1)(12,-7.9)
\psline{-}(13,-8.1)(13,-7.9)
\psline{-}(14,-8.1)(14,-7.9)
\psline{-}(15,-8.1)(15,-7.9)
\psline{-}(16,-8.1)(16,-7.9)

\psline{->}(0,-12)(7,-12)
\psline{->}(0,-13.2)(0,-10.8)
\psline{-}(1,-11)(2,-12)
\psline{*-}(2,-12)(3,-13)
\psline[linestyle=dotted]{-}(3,-13)(5,-13)
\psline{-}(5,-13)(6,-11)
\rput(2,-11.5){{\scriptsize $_\t$}}
\rput(4,-11.5){{\scriptsize $_{(-)}$}}
\psline{-}(1,-12.1)(1,-11.9)
\psline{-}(2,-12.1)(2,-11.9)
\psline{-}(3,-12.1)(3,-11.9)
\psline{-}(4,-12.1)(4,-11.9)
\psline{-}(5,-12.1)(5,-11.9)
\psline{-}(6,-12.1)(6,-11.9)
\rput(8.5,-12){{\large$\rw$}}

\psline{->}(10,-12)(17,-12)
\psline{->}(10,-13.2)(10,-10.8)
\psline{-}(11,-11)(12,-13)
\psline[linestyle=dotted]{-}(12,-13)(14,-13)
\psline{-}(14,-13)(15,-12)
\psline{*-}(15,-12)(16,-11)
\rput(15,-11.5){{\scriptsize $_\t$}}
\rput(13,-11.5){{\scriptsize $_{(-)}$}}
\psline{-}(11,-12.1)(11,-11.9)
\psline{-}(12,-12.1)(12,-11.9)
\psline{-}(13,-12.1)(13,-11.9)
\psline{-}(14,-12.1)(14,-11.9)
\psline{-}(15,-12.1)(15,-11.9)
\psline{-}(16,-12.1)(16,-11.9)
}
\end{pspicture}
\end{center}
\end{figure}

The analysis of the displacements of the $\ppm$ particles is the same (cf. Fig. \ref{particules move}).  Both type of displacements are unbounded and their generating factor is the number of partitions into at most $\nn{\pm}$ parts, with $q\to q^2$ (since the weight increases in steps of 2). We thus conclude that the generating factor accounting for all possible displacements of the $\pp{\pm}$ particles is:
\begin{align}
	G_{}(q)=\frac{1}{(q^2;q^2)_{\nn{\pm}}},
\end{align}
where we used the notation
\begin{equation}
(a;q)_n=\prod_{i=0}^{n-1}(1-aq^i) \qquad [\text{with}\quad(a)_n\equiv (a;q)_n].
\end{equation}

Finally, for the $\ppx$ particles, the basic displacement transforms a straight-up (reps. down) vertex into a straight-down (resp. up) in the next available position. (More precisely, in the first case, NE, NE $\cdots$ SE$_2$ is replaced by NE$_2$ $\cdots$ SE, SE). This is exemplified in  Fig. \ref{particules move}. The weight increase is always 1, with the rule that the corank of eventual intermediate $\ppo$ particles is preserved. The generating function describing these moves is thus simply:
\begin{align}
	G_{\t}(q)=\frac{1}{(q)_{\nnx}}.
\end{align}

Since the different displacements are all independent, the full generating function describing all the particle displacements   is the product of the four separated generating factors:
\begin{align}
	G_{\text{disp.}}(q)=G_{0}G_{+}G_{-}G_{\t}=\frac{1}{(q^2;q^2)_{\nnp}(q^2;q^2)_{\nnm}(q)_{\nnx}}\qbin{\nno+\nnp+\nnm+\nnx-1}{\nno}.
\end{align}

\subsubsection{The fermionic character in terms four particles}
The generating function for all paths with a fixed particle content, $G(\nnp,\nnm,\nnx,\nno;q,{z})$, is obtained by multiplying $G_{\text{disp.}}(q)$ by $q^{\wa(\ga^{\text{mwc}}(\nnp,\nnm,\nnx,\nno))}$ and $z^s$, with $s=\nnp-\nnm$: 
\begin{align}\label{gfpar}
	G(\nnp,\nnm,\nnx,\nno;z;q)=\frac{q^{\frac{1}{2}(\nnp+\nnm+\nnx)^2+\frac{1}{2}\nnx(\nnx-1)+
		\frac{1}{2}\nno-\frac{j}2}z^{\nnp-\nnm}}{(q^2;q^2)_{\nnp}(q^2;q^2)_{\nnm}(q)_{\nnx}}\qbin{ N-1}		{\nno},
\end{align}
with $N=\nnp+\nnm+\nnx+\nno$.  Recall from (\ref{jvsN}) that $j$ is related to  the total number of particles $N$  
by $j=[N]_2$.

The full $\su{2}$ character for the modules $j=0$ and $j=1$
is obtained by summing over all possible particle contents:
\begin{align}
	\chi^{{(2)}}_{j\in\{0,1\}}(z;q)=\sum_{\substack{ \nnp,\nnm,\nnx,\nno \geq 0\\ N\equiv \nnp+\nnm+\nnx+\nno 
		= j\text{ mod }2}} \frac{q^{\frac{1}{2}(\nnp+\nnm+\nnx)^2+\frac{1}{2}\nnx(\nnx-1)
		+\frac{1}{2}\nno-\frac{j}2}z^{\nnp-\nnm}}{(q^2;q^2)_{\nnp}(q^2;q^2)_{\nnm}(q)_{\nnx}}\qbin{N-1}{\nno}.
\end{align}
\subsubsection{From four to three particles: a transmutation process}\label{secttransmut}

The fermionic character just obtained involves four summations. It turns out that it can be naturally re-expressed in terms of a triple summation by introducing a  transmutation process that changes a $\ppo$ into a $\ppx$. 

Consider the leftmost $\ppo$ and suppose that is has been displaced in a maximal way into the leftmost non-$\ppo$ particle. In our previous discussion, this particle would be frozen there. At this point, we introduce a further move for this particle, accompanied by its transmutation into a $\ppx$ particle. The actual materialization of the $\ppx$ particle in the path is on the first (i.e., leftmost) NE$_2$  edge, transforming it into a pair of unit NE edges; at the same time, all other $\ppo$ particles in-between the original leftmost one and the first NE$_2$ edge are displaced toward the right by one (inverse) elementary move, thereby preserving their corank. After the transmutation process, the particle can continue to move as a $\ppx$, of course without crossing another $\ppx$. This is illustrated in Fig. \ref{p0 transmut}.

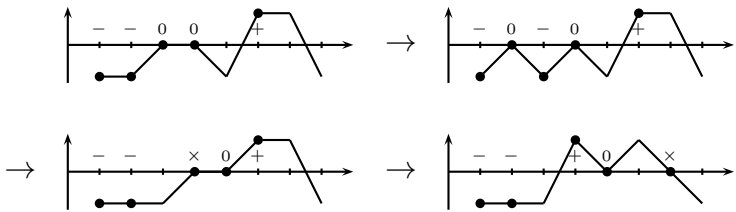
\begin{figure}[htpb]\
\caption{Transmutation of a $\ppo$ into a $\ppx$ and surrounding moves. As a result of the transmutation (which occurs at the third step), the second $\ppo$ moves to preserve its corank.  At each step the weight is increased by 1.}
\label{p0 transmut}
\begin{center}
\begin{pspicture}(0,1)(6.2,-2)
{\psset{yunit=12pt,xunit=12pt,linewidth=.8pt}

\psline{->}(0,0)(9,0)
\psline{->}(0,-1.2)(0,1.2)
\psline{*-}(1,-1)(2,-1)
\psline{*-}(2,-1)(3,0)
\psline{*-}(3,0)(4,0)
\psline{*-}(4,0)(5,-1)
\psline{-}(5,-1)(6,1)
\psline{*-}(6,1)(7,1)
\psline{-}(7,1)(8,-1)
\rput(1,0.5){{\scriptsize $_-$}}
\rput(2,0.5){{\scriptsize $_-$}}
\rput(3,0.5){{\scriptsize $_0$}}
\rput(4,0.5){{\scriptsize $_0$}}
\rput(6,0.5){{\scriptsize $_+$}}
\psline{-}(1,-0.1)(1,0.1)
\psline{-}(2,-0.1)(2,0.1)
\psline{-}(3,-0.1)(3,0.1)
\psline{-}(4,-0.1)(4,0.1)
\psline{-}(5,-0.1)(5,0.1)
\psline{-}(6,-0.1)(6,0.1)
\psline{-}(7,-0.1)(7,0.1)
\psline{-}(8,-0.1)(8,0.1)
\rput(10.5,0){{\large$\rw$}}

\psline{->}(12,0)(21,0)
\psline{->}(12,-1.2)(12,1.2)
\psline{*-}(13,-1)(14,0)
\psline{*-}(14,0)(15,-1)
\psline{*-}(15,-1)(16,0)
\psline{*-}(16,0)(17,-1)
\psline{-}(17,-1)(18,1)
\psline{*-}(18,1)(19,1)
\psline{-}(19,1)(20,-1)
\rput(13,0.5){{\scriptsize $_-$}}
\rput(14,0.5){{\scriptsize $_0$}}
\rput(15,0.5){{\scriptsize $_-$}}
\rput(16,0.5){{\scriptsize $_0$}}
\rput(18,0.5){{\scriptsize $_+$}}
\psline{-}(13,-0.1)(13,0.1)
\psline{-}(14,-0.1)(14,0.1)
\psline{-}(15,-0.1)(15,0.1)
\psline{-}(16,-0.1)(16,0.1)
\psline{-}(17,-0.1)(17,0.1)
\psline{-}(18,-0.1)(18,0.1)
\psline{-}(19,-0.1)(19,0.1)
\psline{-}(20,-0.1)(20,0.1)

\rput(-1.5,-4){{\large$\rw$}}
\psline{->}(0,-4)(9,-4)
\psline{->}(0,-5.2)(0,-2.8)
\psline{*-}(1,-5)(2,-5)
\psline{*-}(2,-5)(3,-5)
\psline{-}(3,-5)(4,-4)
\psline{*-}(4,-4)(5,-4)
\psline{*-}(5,-4)(6,-3)
\psline{*-}(6,-3)(7,-3)
\psline{-}(7,-3)(8,-5)
\rput(1,-3.5){{\scriptsize $_-$}}
\rput(2,-3.5){{\scriptsize $_-$}}
\rput(4,-3.5){{\scriptsize $_\t$}}
\rput(5,-3.5){{\scriptsize $_0$}}
\rput(6,-3.5){{\scriptsize $_+$}}
\psline{-}(1,-4.1)(1,-3.9)
\psline{-}(2,-4.1)(2,-3.9)
\psline{-}(3,-4.1)(3,-3.9)
\psline{-}(4,-4.1)(4,-3.9)
\psline{-}(5,-4.1)(5,-3.9)
\psline{-}(6,-4.1)(6,-3.9)
\psline{-}(7,-4.1)(7,-3.9)
\psline{-}(8,-4.1)(8,-3.9)
\rput(10.5,-4){{\large$\rw$}}

\psline{->}(12,-4)(21,-4)
\psline{->}(12,-5.2)(12,-2.8)
\psline{*-}(13,-5)(14,-5)
\psline{*-}(14,-5)(15,-5)
\psline{-}(15,-5)(16,-3)
\psline{*-}(16,-3)(17,-4)
\psline{*-}(17,-4)(18,-3)
\psline{-}(18,-3)(19,-4)
\psline{*-}(19,-4)(20,-5)
\rput(13,-3.5){{\scriptsize $_-$}}
\rput(14,-3.5){{\scriptsize $_-$}}
\rput(16,-3.5){{\scriptsize $_+$}}
\rput(17,-3.5){{\scriptsize $_0$}}
\rput(19,-3.5){{\scriptsize $_\t$}}
\psline{-}(13,-4.1)(13,-3.9)
\psline{-}(14,-4.1)(14,-3.9)
\psline{-}(15,-4.1)(15,-3.9)
\psline{-}(16,-4.1)(16,-3.9)
\psline{-}(17,-4.1)(17,-3.9)
\psline{-}(18,-4.1)(18,-3.9)
\psline{-}(19,-4.1)(19,-3.9)
\psline{-}(20,-4.1)(20,-3.9)

}
\end{pspicture}
\end{center}
\end{figure}
 
The interest in this new description is that the move associated to the transmutation process is always accompanied by a weight increase of 1. This is proved in the following paragraph. Note that when there is a newly created $\ppx$ before the first NE$_2$ edge, a further transmutation is impossible: this $\ppx$ blocks the eventual displacement (through which it would have been transformed) of the leftmost $\ppo$. Starting from the minimal weight configuration, this reduces by one the number of possible moves of the particles $\ppo$ at the right of the transmuted one. However, a $\ppo$ that would be in position for transmutation, as well as all the other $\ppo$ at the right of the transmuted particle, have now the possibility of increasing their corank by one unit resulting from the creation of a new non-$\ppo$-type particle. This increasing of their allowed displacements by 1 compensates the reduction by 1 caused the blocking process. There is thus a continuity in the number of unit displacements (all integer values are covered) taking together the $\ppo$ moves and that of its transmuted $\ppx$ form. 
This phenomenon is illustrated in Fig. \ref{p0 transmut block}. Since the elementary displacements before and after the transmutation point also come together with a unit weight increment, the generating factor of the $\pp{0/\t}$ moves is simply $(q)_{n^0}^{-1}$. This is, combinatorially,  a radical simplification.

\begin{figure}[htpb]\
\caption{A blocking situation for the transmutation process, obtained from the third configuration of Fig. \ref{p0 transmut} by moving the $p_0$ by two unit displacements. The blocking is compatible with the hard-core repulsion of identical particles (meaning that they cannot overpass each other): both particles ($p_0$ and $p_\t$) have been moved twice from the first configuration in Fig. \ref{p0 transmut}, in which they have the same corank.}
\label{p0 transmut block}
\begin{center}
\begin{pspicture}(0,1)(4,-1)
{\psset{yunit=16pt,xunit=16pt,linewidth=.8pt}

\psline{->}(0,0)(9,0)
\psline{->}(0,-1.2)(0,1.2)
\psline{*-}(1,-1)(2,0)
\psline{*-}(2,0)(3,-1)
\psline{*-}(3,-1)(4,-1)
\psline{-}(4,-1)(5,0)
\psline{*-}(5,0)(6,1)
\psline{*-}(6,1)(7,1)
\psline{-}(7,1)(8,-1)
\rput(1,0.5){{\scriptsize $_-$}}
\rput(2,0.5){{\scriptsize $_0$}}
\rput(3,0.5){{\scriptsize $_-$}}
\rput(5,0.5){{\scriptsize $_\t$}}
\rput(6,0.5){{\scriptsize $_+$}}
\psline{-}(1,-0.1)(1,0.1)
\psline{-}(2,-0.1)(2,0.1)
\psline{-}(3,-0.1)(3,0.1)
\psline{-}(4,-0.1)(4,0.1)
\psline{-}(5,-0.1)(5,0.1)
\psline{-}(6,-0.1)(6,0.1)
\psline{-}(7,-0.1)(7,0.1)
\psline{-}(8,-0.1)(8,0.1)

}
\end{pspicture}
\end{center}
\end{figure}
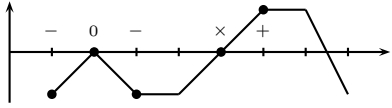

Let us now prove that the transformation just described of a $\ppo$ into a $\ppx$ increases the weight by one. Since $\ga_0=-2$, two situations need to be considered. Either the path starts with a number of $\ppm$ particles hosting some $\ppo$ followed by a NE$_2$ edge or it starts directly with a NE$_2$ edge. In the first case, suppose that there are $i_-$ $\ppm$ and $i_0$ $\ppo$, so that the NE$_2$ edge starts at position $i_0+i_-$. By assumption, one $\ppo$ is at  position 1: this is the particle that will be transformed. (See the second picture in Fig. \ref{p0 transmut}. For that example, $i_0=i_-=2$.) The transformation creates a $\ppx$ at position $i_0+i_-$, with weight $i_-+i_0-\tfrac12$. All the $\ppm$ particles but the first one are displaced toward the left, decreasing the weight by $i_--1$. Also, the remaining $i_0-1$ $\ppo$ are displaced toward the right by one elementary step each (to preserve their corank), which induces a unit move toward the left of $i_0-1$ non-$\ppo$ particles, lowering the weight by $i_0-1$.\footnote{These $i_0-1$ non-$\ppo$ particles are typically  all of $\ppm$-type but if the there is a $\ppo$ attached to the last $\ppm$ (the $i_-$th one from the left), then the newly created $\ppx$ is among the displaced particles. In that case, the $\ppx$ would then appear at $i_0+i_--1$ followed by a $\ppo$ at $i_0+i_-$. This is the case of the third step in Fig. \ref{p0 transmut}.} Finally, the annihilation of the $\ppo$ particle originally at position 1 decreases the weight by $\tfrac12$. The total weight difference is thus $i_-+i_0-\tfrac12-(i_-+i_0-2+\tfrac12)=1$. In the other case, where there is no $\ppm$ at the left of the first NE$_2$ edge, the situation is simpler in that after the transmutation the host particle is moved toward the right by 1, increasing the weight by 1. The annihilation of the $\ppo$ decreases the weight by $\tfrac12$, while the creation of the $\ppx$ at position 1 increases the weight by the same amount. Summing up, the total weight increase is again 1.

\subsubsection{The fermionic character in terms three particles}

The construction of the generating function proceeds as before, starting from the configuration $\ga^{\text{mwc}}(\nnp,\nnm,\nno)$, which is like the previous  $\ga^{\text{mwc}}(\nnp,\nnm,\nnx,\nno)$ but with $\nnx=0$.\footnote{Actually, $\ga^{\text{mwc}}(\nnp,\nnm,\nnx,\nno)$ can be regarded as obtained from the configuration  $\ga^{\text{mwc}}(\nnp,\nnm,\nnx+\nno)$ by displacing the first $\nnx$ $\ppo$ particles until they get transformed into $\ppx$ in a minimal way.} Hence, its weight is thus read off  $\wa(\ga^{\text{mwc}}(\nnp,\nnm,0,\nno))$ given in (\ref{wmwc}). The generating factors for the displacements of the $\pp{\pm}$ are unchanged and that for the $\ppo$ has just been determined. The character is thus simply: 
\begin{align}\label{caj01}
	\chi^{(2)}_{j\in\{0,1\}}(z;q)=\sum_{\substack{ \nnp,\nnm,\nno\geq 0 \\  \nnp+\nnm+\nno = j\text{ mod }2}}
		 \frac{q^{\frac{1}{2}(\nnp+\nnm)^2+\frac{1}{2}\nno-\frac12j}z^{\nnp-\nnm}}{(q^2;q^2)_{\nnp}(q^2;q^2)_{\nnm}(q)_{\nno}}.
\end{align}
This is the announced  fermionic expression involving three types of particles. The relation between the parity of $ \nnp+\nnm+\nno$ and $j$ is obtained as before (and it can be viewed as a consequence of the previous connection with $\nnx+\nno\to \nno$). The two characters can be combined together by removing the parity condition from the summation. The direct sum of the two characters reads then: 
\begin{align}\label{j01k2}
	\chi^{(2)}_{0\oplus1}(z;q)&=\chi^{(2)}_{0}(z;q)+q^{\frac12}\chi^{(2)}_{1}(z;q)\nonumber\\
	&=\sum_{ \nnp,\nnm,\nno\geq 0 } \frac{q^{\frac{1}{2}(\nnp+\nnm)^2
		+\frac{1}{2}\nno}z^{\nnp-\nnm}}{(q^2;q^2)_{\nnp}(q^2;q^2)_{\nnm}(q)_{\nno}}.
\end{align}

Interestingly, when the parity condition  is ignored, the $\nno$ terms decouple in (\ref{j01k2}) and the summation over $\nno$ is readily obtained by means of the identity \cite[eq. 2.2.5]{Andr}: 
\begin{align}
	\sum_{i=0}^{\infty}\frac{q^{\frac{1}{2}i}}{(q)_i}=\frac{1}{(q^{\frac{1}{2}})_{\infty}}.
\end{align}
The sum of the two characters reduces to
\begin{align}\label{chark2}
	\chi^{(2)}_{0\oplus1}(z;q)=\frac{1}{(q^{\frac{1}{2}})_{\infty}}\; \sum_{\nnp,\nnm\geq 0} 
		\frac{q^{\frac{1}{2}(\nnp+\nnm)^2}z^{\nnp-\nnm}}{(q^2;q^2)_{\nnp}(q^2;q^2)_{\nnm}}.
\end{align}
Note that this does not induces a genuine reduction to two particles: rather, one summation has been worked out explicitly.
By comparing this result with eq. (\ref{char1b}), one notices a somewhat surprising structural relationship between the $\su{2}$ and $\su{1}$ characters: 
\begin{align}\label{caun2}
	\chi^{{(2)}}_{0\oplus1}(z;q)=\frac{1}{(q^{\frac{1}{2}})_{\infty}}\;\chi^{{(1)}}_{0\oplus\tfrac12}(z^2;q^2).
\end{align}

The expression (\ref{caj01}) appears to be new. Its equivalent version (\ref{j01k2}), in its reduced form (\ref{chark2}), will be proved directly in the following subsection, by demonstrating its equality with another established fermionic form.

\subsubsection{Equivalence with the three-free-fermions character formula}

We now show that the character (\ref{chark2}) is equivalent to the following fermionic formula \cite[eq. (6.24)]{MR}:
\begin{align}\label{caMR}
	\chi^{(2)}_{0\oplus1}(z;q)=\sum_{ \ell_0,\ell_1,\ell_2\geq 0} \frac{q^{\frac{1}{2}(\ell_0^2+\ell_1^2+\ell_2^2)}
		z^{\ell_0-\ell_2}}{(q)_{\ell_0}(q)_{\ell_1}(q)_{\ell_2}},
\end{align}
obtained from an explicit construction of the basis of states. Notice that with $z=1$, the latter expression is simply the (expected) cube of the free-fermion vacuum character.

Our first step is to replace $n_+=n_-+S$ and $n_-\rw n$ in (\ref{chark2}):
\begin{align}
	\chi^{(2)}_{0\oplus1}(z;q)&=\frac{1}{(q^{\frac{1}{2}})_{\infty}}\; \sum_{n=0}^\y\sum_{S=-n}^\y 
		\frac{q^{2n(n+S)}q^{\frac12S^2}z^{S}} {(q^2;q^2)_{n}(q^2;q^2)_{n+S}}.
\end{align}
Let now change the lower bound of the $S$-summation from $-n$ to $-\y$, with the understanding that $(q)_n^{-1}=0$ if $n<0$. The order of two summations can then be interchanged.
Using the Durfee rectangle identity\footnote{This identity is a special case of $q$-Gauss sum. Combinatorially, it is easily proved: 
the generating function of partitions with Durfee rectangle
of size $m$ by $m+a$ is $q^{m(m+a)}/(q)_{m}(q)_{m+a}$;  its sum over $m$ gives the generating
function of all partitions, namely $(q)_\y^{-1}$.}
\begin{align}\label{durfee}
	\sum_{m=0}^{\infty}\frac{q^{m(m+a)}}{(q)_{m}(q)_{m+a}}=\frac{1}{(q)_{\infty}},
\end{align}
with $q\rw q^2$, to evaluate the sum over $n$ and get 
\begin{align}
	\chi^{(2)}_{0\oplus1}(z;q)&= \frac{1}{(q^{\frac{1}{2}})_{\infty}(q^2;q^2)_{\infty}}
	\sum_{S=-\infty}^{\infty}	q^{\frac{1}{2}S^2}z^S\nonumber\\
	&=
		\frac{(q;q^2)_{\infty}}{(q^{\frac{1}{2}})_{\infty} (q)_{\infty}}\sum_{S=-\infty}^{\infty} q^{\frac{1}{2}S^2}z^S\nonumber\\
	&=\frac{1}{(q)_{\infty} } \prod_{m=0}^{\infty}(1+q^{m+\frac12}) \sum_{S=-\infty}^{\infty}q^{\frac{1}{2}S^2}z^S.
\end{align}
 The second line follows from the simple relation $(q^2;q^2)_{\infty}=(q;q^2)_{\infty}/(q)_{\infty}$ and the last step relies on $(1-q^{2a})/(1-q^a)=(1+q^a)$. We then use the identity \cite[eq. 2.2.6]{Andr}
\begin{align}
	\prod_{m=0}^{\infty}(1+q^{m+\frac12})=\sum_{n=0}^{\infty}\frac{q^{\frac12n^2}}{(q)_n},
\end{align}
to rewrite the character as
\begin{align}
	\chi^{(2)}_{0\oplus1}(z;q)=\frac{1}{(q)_{\infty}} \sum_{S=-\infty}^{\infty}q^{\frac{1}{2}S^2}z^S
		\sum_{\ell_1=0}^{\infty}\frac{q^{\frac12\ell_1^2}}{(q)_{\ell_1}}.
\end{align}
Using the Durfee rectangle identity again, this time in order to decompose $(q)_\y^{-1}$ into a sum over $\ell_2$ say, with $a=S$ in (\ref{durfee}), we get (after interchanging the first two summations)
\begin{align}
	\chi^{(2)}_{0\oplus1}(z;q)&=\sum_{\ell_2=0}^\y\sum_{S=-\ell_2}^\y       
		\frac{q^{\ell_2(\ell_2+S)}q^{\frac12S^2}z^S}{(q)_{\ell_2}(q)_{\ell_2+S}}
		\sum_{\ell_1=0}^{\infty}\frac{q^{\frac12\ell_1^2}}{(q)_{\ell_1}}.
\end{align}
The change of variable $S=\ell_0-\ell_2$ gives 
directly (\ref{caMR}), as announced.

\subsection{The module $j=1/2$}

\subsubsection{Identifying the particles in $\mk^{(2)}_\frac12$ paths}

We now turn to the analysis of the particle decomposition of the $\mk^{(2)}$ paths with $j=1/2$. This decomposition preserves some features of the $\mk^{(2)}_{0,1}$ paths -- in particular, the definition of the particles $\ppp$ and $\ppm$ is unchanged --, but it also brings novel aspects. The main novelty is the tail shape which is now composed of an infinite sequence of $\ppo$ particles, that is, an infinite plateau at $\ym=0$. In other words, the paths are now viewed as finite deformations in a background of free $\ppo$ particles, by opposition with the $j=0,1$ paths which are defined in an infinite background of SE$_2$ and NE$_2$ edges. This means that the number $n_0$ of particles $\ppo$, being infinite, can no longer be used to characterize the particle content. We stress that such an infinite tail of $\ppo$ particles has zero weight, since a free $\ppo$ itself has zero weight. In the previous section a weight $\wa_{(0)}=\frac12 $ (cf. eq. (\ref{difw})) has been attributed to a $\ppo$, but it always occur there within an host particle: this weight is thus that of an attached $\ppo$. 

By a sort of duality, the role of the $\ppo$ particles in the $\mk^{(2)}_{0,1}$ paths  will now be played by the SE$_2$ and NE$_2$ edges, which are thereby lifted to the status of particles; these will be labeled $\ppi$ indifferently of their type (SE$_2$ or NE$_2$, hence the notation $\vee$). Although the number of $\ppo$ is no longer a relevant parameter, the $\ppo$ can be displaced and get attached to a particle of a different type. These displacements will be taken into account. In that vein, a $\ppx$ can be seen as a $\ppo$ bound to a $\ppi$, and therefore it does not have to be considered as an independent particle (i.e.,. we directly use the results of Sect. \ref{secttransmut}).

Summing up, the particles entering in the description of the $j=1/2$ paths are $\ppp$, $\ppm$, $\ppi$ and $\ppo$ and  the particle content is specified by the three parameters, $\nnp$, $\nnm$ and $\nni$. 

A further new issue must be considered. Observe that a path portion composed by a number of particles $\ppp$, $\ppm$ and $\ppi$, including possibly a number of attached  $\ppo$, cannot be  linked to a tail of free $\ppo$. In other words, such a portion cannot end up at position $l$ with $j_l=1/2$. This is even the case in absence of all non-$\ppo$ particles -- recall that the path starts with the pre-segment $\ymp{0}=-2$, i.e., with $j_0=0$. The missing piece is a SE or NE edge that needs to be added  just before the infinite sequence of $\ppo$, after all other particles. This edge can be dressed with a natural  particle interpretation. Indeed, since this edge is described by the operator $\abp{l}{\pm1}$, depending on its exact nature (NE or SE), it is conveniently viewed as a half-$\pp{\pm}$, to be denoted $\pp{\pm/2}$. An example of a particle decomposition of a $\mk^{(2)}_{\frac12}$ path is illustrated in Fig. \ref{exemple j1s2}.

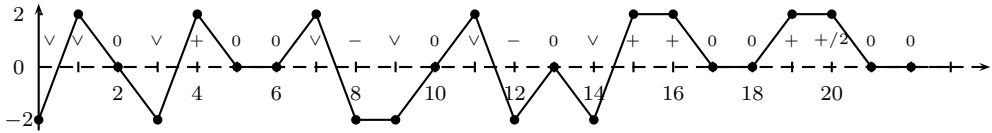
\begin{figure}[ht]\
\caption{The particle decomposition of a generic $\mk^{(2)}_{1/2}$ path with a pre-segment originating from $\ga_0=-2$. The particle types are identified and their positions are indicated by black dots. Note that to each vertex corresponds a particle and that a $p^0$ attached  to a $p^\vee$ is equivalent to a $p^\t$ (cf. at position 10). Here $n_\vee(=7)$ is odd so that the last non-$p^0$ particle is $+/2$.}
\label{exemple j1s2}
\begin{center}
\begin{pspicture}(1,-1)(11,1.1)
{\psset{yunit=20pt,xunit=15pt,linewidth=.8pt}
\psline{->}(0,-1.2)(0,1.2) 
\psset{linestyle=dashed,dashadjust=false}
\psline{->}(0,0)(24,0)
\psset{linestyle=solid}

\psline{*-}(0,-1)(1,1)
\psline{*-}(1,1)(2,0)
\psline{*-}(2,0)(3,-1)
\psline{*-}(3,-1)(4,1)
\psline{*-}(4,1)(5,0)
\psline{*-}(5,0)(6,0)
\psline{*-}(6,0)(7,1)
\psline{*-}(7,1)(8,-1)
\psline{*-}(8,-1)(9,-1)
\psline{*-}(9,-1)(10,0)
\psline{*-}(10,0)(11,1)
\psline{*-}(11,1)(12,-1)
\psline{*-}(12,-1)(13,0)
\psline{*-}(13,0)(14,-1)
\psline{*-}(14,-1)(15,1)
\psline{*-}(15,1)(16,1)
\psline{*-}(16,1)(17,0)
\psline{*-}(17,0)(18,0)
\psline{*-}(18,0)(19,1)
\psline{*-}(19,1)(20,1)
\psline{*-}(20,1)(21,0)
\psline{*-}(21,0)(22,0)
\psset{linestyle=dashed,dashadjust=false}
\psline{*-}(22,0)(23,0)

\psset{linestyle=solid}
\psline{-}(1,-0.1)(1,0.1)
\psline{-}(2,-0.1)(2,0.1)
\psline{-}(3,-0.1)(3,0.1)
\psline{-}(4,-0.1)(4,0.1)
\psline{-}(5,-0.1)(5,0.1)
\psline{-}(6,-0.1)(6,0.1)
\psline{-}(7,-0.1)(7,0.1)
\psline{-}(8,-0.1)(8,0.1)
\psline{-}(9,-0.1)(9,0.1)
\psline{-}(10,-0.1)(10,0.1)
\psline{-}(11,-0.1)(11,0.1)
\psline{-}(12,-0.1)(12,0.1)
\psline{-}(13,-0.1)(13,0.1)
\psline{-}(14,-0.1)(14,0.1)
\psline{-}(15,-0.1)(15,0.1)
\psline{-}(16,-0.1)(16,0.1)
\psline{-}(17,-0.1)(17,0.1)
\psline{-}(18,-0.1)(18,0.1)
\psline{-}(19,-0.1)(19,0.1)
\psline{-}(20,-0.1)(20,0.1)
\psline{-}(21,-0.1)(21,0.1)
\psline{-}(22,-0.1)(22,0.1)
\psline{-}(23,-0.1)(23,0.1)

\psline{-}(0,-1)(0.1,-1)
\psline{-}(0,1)(0.1,1)

\rput(2,-0.5){{\scriptsize $2$}}
\rput(4,-0.5){{\scriptsize $4$}}
\rput(6,-0.5){{\scriptsize $6$}}
\rput(8,-0.5){{\scriptsize $8$}}
\rput(10,-0.5){{\scriptsize $10$}}
\rput(12,-0.5){{\scriptsize $12$}}
\rput(14,-0.5){{\scriptsize $14$}}
\rput(16,-0.5){{\scriptsize $16$}}
\rput(18,-0.5){{\scriptsize $18$}}
\rput(20,-0.5){{\scriptsize $20$}}

\rput(-0.5,-1){{\scriptsize $-2$}}
\rput(-0.5,0){{\scriptsize $0$}}
\rput(-0.5,1){{\scriptsize $2$}}

\rput(0.3,0.5){{\scriptsize $_\vee$}}
\rput(1,0.5){{\scriptsize $_\vee$}}
\rput(2,0.5){{\scriptsize $_{0}$}}
\rput(3,0.5){{\scriptsize $_\vee$}}
\rput(4,0.5){{\scriptsize $_+$}}
\rput(5,0.5){{\scriptsize $_0$}}
\rput(6,0.5){{\scriptsize $_0$}}
\rput(7,0.5){{\scriptsize $_\vee$}}
\rput(8,0.5){{\scriptsize $_-$}}
\rput(9,0.5){{\scriptsize $_\vee$}}
\rput(10,0.5){{\scriptsize $_{0}$}}
\rput(11,0.5){{\scriptsize $_\vee$}}
\rput(12,0.5){{\scriptsize $_-$}}
\rput(13,0.5){{\scriptsize $_0$}}
\rput(14,0.5){{\scriptsize $_\vee$}}
\rput(15,0.5){{\scriptsize $_+$}}
\rput(16,0.5){{\scriptsize $_+$}}
\rput(17,0.5){{\scriptsize $_0$}}
\rput(18,0.5){{\scriptsize $_0$}}
\rput(19,0.5){{\scriptsize $_+$}}
\rput(20,0.5){{\scriptsize $_{+/2}$}}
\rput(21,0.5){{\scriptsize $_0$}}
\rput(22,0.5){{\scriptsize $_0$}}
}
\end{pspicture}
\end{center}
\end{figure}

We now determine the weight of the different particles. The $\pp{\pm}$ weight is the same as for $j=0,1$ since these particles are unchanged and it is given in (\ref{difw}). This weighting is compatible with the interpretation of $\abp{l}{\pm1}$ as creating $\pp{\pm/2}_l$:
\begin{equation}\label{difwd}
	\wa_{(\pm/2)}=\frac12\wa_{(\pm)}= \frac12\left(l\mp\frac12\right).
\end{equation}
As noted before, the weight of a $p^0$ is $1/2$ if it is bound and $0$ otherwise. Finally, the $\ppi$ have zero weight since NE$_2$ and SE$_2$ edges do not contribute to the  weight.

\subsubsection{The configuration of minimal weight }

The minimal weight configuration $\ym^{\text{mwc}}(\nnp,\nnm,\nni)$ is obtained as in the $j=0,1$ case. Consider first $\ym^{\text{mwc}}(0,0,0)$, which is just $\ym^{\text{gs}(\frac12)}=\abp{0}{-1}\ym^{\text{gs}(0)}$.\footnote{We do not write this as $\ym^{\text{mwc}}(0,\frac12,0)$: the half particle, being always there and its nature being position dependent, will not be counted in $n_\pm$.} Equivalently, there is the single $p^{-/2}$ at position 0. We then add the $\ppm$ at the beginning (one at each $l$ from $0$ to $\nnm-1$) to form $\ym^{\text{mwc}}(0,\nnm,0)$. This moves the $p^{-/2}$ by $\nnm$ units. In the next step, we add the $\ppi$ particles. Since they have zero weight, we can place them after the sequence of $\ppm$. They must however remain before the  $p^{-/2}$, which as already said must always stay at the right of every particle other than $\ppo$. In the case where $\nni$ is even, we can just place them (after the $p_-$ as said) in the order NE$_2$, SE$_2$,..., SE$_2$ and move the $p^{-/2}$ by $\nni$ units. However, if $\nni$ is odd, the sequence ends with NE$_2$, which means the $p^{-/2}$ is forbidden (it must start at $-2$). To solve this problem, we change the half particle to a $p^{+/2}$. This is how the edge before the tail can become a SE. The $\ym^{\text{mwc}}(0,\nnm,\nni)$ configuration is associated to the particle content:
\begin{align}
	\ym^{\text{mwc}}(0,\nnm,\nni)\lrw\ppm_0\ppm_1...\ppm_{\nnm-1}\ppi_{\nnm}\ppi_{\nnm+1}...\ppi_{\nnm+\nni-1} p^{-(-1)^{\nni}/2}_{\nnm+\nni}.
\end{align}
We then add the $\ppp$. Since these have non-zero weight, they must be placed at the leftmost possible position. This is just after the first $\ppi$, in a way similar to $j=0,1$, of course assuming $\nni>0$. In the case $\nni=0$, there is no way to insert any $\ppp$ since there is no vertex at $\ym=2$. Since this argument applies to any path and not just to the minimal weight path, the configurations $(\nnp,\nnm,0)$ are strictly forbidden if $\nnp>0$. The resulting minimal weight configuration corresponds thus to: 
\begin{align}
	\ym^{\text{mwc}}(\nnp,\nnm,\nni)\lrw\ppm_0\ppm_1...\ppm_{\nnm-1}\ppi_{\nnm}\ppp_{\nnm+1}...\ppp_{\nnm+\nnp}\ppi_{\nnm+\nnp+1}...\ppi_{\nnm+\nnp+\nni}p_{\nnm+\nnp+\nni}^{-(-1)^{\nni}/2}.
\end{align}
The example $\ym^{\text{mwc}}(5,3,8)$ is displayed in Fig. \ref{exemplemwc j1s2}.

The weight of this path is
\begin{align}
	\wa(\ym^{\text{mwc}}(\nnp,\nnm,\nni))&=-\frac12j+\sum_{i=0}^{\nnm-1}\left(i+\frac12\right)+\sum_{i=\nnm+1}^{\nnm+\nnp}\left(i-\frac12\right)+\frac12(\nnm+\nnp+\nni)+\frac14(-1)^{\nni}\nonumber\\
	&=-\frac14+\frac12(\nnp+\nnm)^2+\frac12(\nnm+\nnp+\nni)+\frac14(-1)^{\nni},
\end{align}
where the last two terms are the contribution of the half-particle at the end of the sequence. In particular, this accounts for the $\nni$-dependence of the weight, the $\ppi$ particles being themselves weightless.

\begin{figure}[ht]\
\caption{The path $\ym^{\text{mwc}}(5,3,8)$ is pictured, with its pre-segment. The particle types are identified, with their positions indicated by black dots. Since in this example $n_\vee$ is  even, the last particle before the sea of $p^0$ ones is of type $-/2$. }
\label{exemplemwc j1s2}
\begin{center}
\begin{pspicture}(1,-1)(11,1.1)
{\psset{yunit=20pt,xunit=15pt,linewidth=.8pt}
\psline{->}(0,-1.2)(0,1.2) 
\psset{linestyle=dashed,dashadjust=false}
\psline{->}(0,0)(24,0)
\psset{linestyle=solid}

\psline{*-}(0,-1)(1,-1)
\psline{*-}(1,-1)(2,-1)
\psline{*-}(2,-1)(3,-1)
\psline{*-}(3,-1)(4,1)
\psline{*-}(4,1)(5,1)
\psline{*-}(5,1)(6,1)
\psline{*-}(6,1)(7,1)
\psline{*-}(7,1)(8,1)
\psline{*-}(8,1)(9,1)
\psline{*-}(9,1)(10,-1)
\psline{*-}(10,-1)(11,1)
\psline{*-}(11,1)(12,-1)
\psline{*-}(12,-1)(13,1)
\psline{*-}(13,1)(14,-1)
\psline{*-}(14,-1)(15,1)
\psline{*-}(15,1)(16,-1)
\psline{*-}(16,-1)(17,0)
\psline{*-}(17,0)(18,0)
\psline{*-}(18,0)(19,0)
\psline{*-}(19,0)(20,0)
\psline{*-}(20,0)(21,0)
\psline{*-}(21,0)(22,0)
\psset{linestyle=dashed,dashadjust=false}
\psline{*-}(22,0)(23,0)

\psset{linestyle=solid}
\psline{-}(1,-0.1)(1,0.1)
\psline{-}(2,-0.1)(2,0.1)
\psline{-}(3,-0.1)(3,0.1)
\psline{-}(4,-0.1)(4,0.1)
\psline{-}(5,-0.1)(5,0.1)
\psline{-}(6,-0.1)(6,0.1)
\psline{-}(7,-0.1)(7,0.1)
\psline{-}(8,-0.1)(8,0.1)
\psline{-}(9,-0.1)(9,0.1)
\psline{-}(10,-0.1)(10,0.1)
\psline{-}(11,-0.1)(11,0.1)
\psline{-}(12,-0.1)(12,0.1)
\psline{-}(13,-0.1)(13,0.1)
\psline{-}(14,-0.1)(14,0.1)
\psline{-}(15,-0.1)(15,0.1)
\psline{-}(16,-0.1)(16,0.1)
\psline{-}(17,-0.1)(17,0.1)
\psline{-}(18,-0.1)(18,0.1)
\psline{-}(19,-0.1)(19,0.1)
\psline{-}(20,-0.1)(20,0.1)
\psline{-}(21,-0.1)(21,0.1)
\psline{-}(22,-0.1)(22,0.1)
\psline{-}(23,-0.1)(23,0.1)

\psline{-}(0,-1)(0.1,-1)
\psline{-}(0,1)(0.1,1)

\rput(2,-0.5){{\scriptsize $2$}}
\rput(4,-0.5){{\scriptsize $4$}}
\rput(6,-0.5){{\scriptsize $6$}}
\rput(8,-0.5){{\scriptsize $8$}}
\rput(11,-0.5){{\scriptsize $11$}}
\rput(13,-0.5){{\scriptsize $13$}}
\rput(15,-0.5){{\scriptsize $15$}}
\rput(18,-0.5){{\scriptsize $18$}}
\rput(20,-0.5){{\scriptsize $20$}}

\rput(-0.5,-1){{\scriptsize $-2$}}
\rput(-0.5,0){{\scriptsize $0$}}
\rput(-0.5,1){{\scriptsize $2$}}

\rput(0.3,0.5){{\scriptsize $_-$}}
\rput(1,0.5){{\scriptsize $_-$}}
\rput(2,0.5){{\scriptsize $_-$}}
\rput(3,0.5){{\scriptsize $_\vee$}}
\rput(4,0.5){{\scriptsize $_+$}}
\rput(5,0.5){{\scriptsize $_+$}}
\rput(6,0.5){{\scriptsize $_+$}}
\rput(7,0.5){{\scriptsize $_+$}}
\rput(8,0.5){{\scriptsize $_+$}}
\rput(9,0.5){{\scriptsize $_\vee$}}
\rput(10,0.5){{\scriptsize $_\vee$}}
\rput(11,0.5){{\scriptsize $_\vee$}}
\rput(12,0.5){{\scriptsize $_\vee$}}
\rput(13,0.5){{\scriptsize $_\vee$}}
\rput(14,0.5){{\scriptsize $_\vee$}}
\rput(15,0.5){{\scriptsize $_\vee$}}
\rput(16,0.5){{\scriptsize $_{-/2}$}}
\rput(17,0.5){{\scriptsize $_0$}}
\rput(18,0.5){{\scriptsize $_0$}}
\rput(19,0.5){{\scriptsize $_0$}}
\rput(20,0.5){{\scriptsize $_0$}}
\rput(21,0.5){{\scriptsize $_0$}}
\rput(22,0.5){{\scriptsize $_0$}}
}
\end{pspicture}
\end{center}
\end{figure}
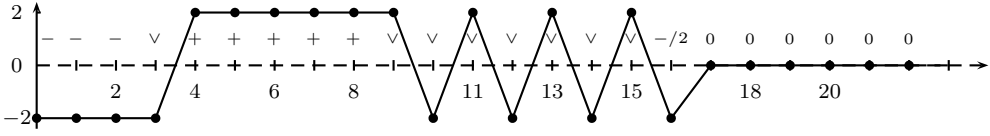

\subsubsection{The particle displacements}

The  displacements of the particles within $j=1/2$ paths are similar to the $j=0,1$ case. For instance, every elementary displacement of a $p^\pm$ particle is by two units toward the right and it increases the weight by 2. Similarly, the $\ppo$ can be displaced toward the left to be attached to another type of particle, increasing their corank and the weight by 1. Note that for the first move, the binding adds $1/2$ to the weight, the other $1/2$ comes from the displacement of the $\pp{\pm/2}$.

However there are two differences with the $j=0,1$ situation: the displacements of the $\pp{\pm}$ are now limited and there are infinitely many particles $\ppo$ to be considered. Note that these displacements of the $p^\pm$ and $\pp0$ particles indirectly take care  of the $\ppi$ ones.
 
Consider first the $p^\pm$.  The maximal displacement of the rightmost  $\ppm$ is $\lfloor\nni/2\rfloor$ (the integer part of $\nni/2$). This is easily seen from the fact that a $\ppm$ displacement is equivalent to a jump over a pair of $\ppi$ and there are  $\lfloor\nni/2\rfloor$ such pairs. The same argument applies for the $\ppp$, but since these are all initially positioned after the first $\ppi$, the maximal number of moves of the rightmost $p^+$ is $\lf(\nni-1)/2\rf$. Consequently, the generating factor for the displacements of the $p^\pm$ is given by the number of partitions into $n_\pm$ parts with the first part at most equal to $\lf(\nni-1)/2\rf$ or $\lf\nni/2\rf$ respectively, with $q\to q^2$, that is,
\begin{align}
	G_{+}(q)&=\qbin{\nnp+\lf\frac{\nni-1}{2}\rf}{\nnp}_{q^2}\qquad\text{and}\qquad
	G_{-}(q)=\qbin{\nnm+\lf\frac{\nni}{2}\rf}{\nnm}_{q^2}.
\end{align}
We recall the condition (\ref{qbi-1}):
\begin{align}\qbin{-1}{0}=1 \quad \Rightarrow \quad \qbin{n-1}{n}=\delta_{n,0}
\end{align}
which takes directly into account the constraint that $\nni=0\Rightarrow\nnp=0$.

Consider next the $\ppo$.  A $\ppo$ can be attached to every $\pp{\pm}$ or $\ppi$; in the latter case, it forms a $\ppx$. There can be only one $\ppx$ formed from a single $\ppi$, but recall that each $\ppx$ adds one possible displacement to the other $\ppo$; the number of possible moves is thus $\nnp+\nnm+\nni$ for every $\ppo$. All these displacements are generated by the function:
\begin{align}
	G_{0}(q)=\lim_{\nno\rw\infty}\qbin{\nno+\nnp+\nnm+\nni}{\nno}
	=\frac{1}{(q)_{\nnp+\nnm+\nni}}.
\end{align}
The generating function for all possible displacements is thus
\begin{align}
	G_{\text{disp}}(q)=\qbin{\nnp+\lf\frac{\nni-1}{2}\rf}{\nnp}_{q^2}\qbin{\nnm+\lf\frac{\nni}{2}\rf}{\nnm}_{q^2}\frac{1}{(q)_{\nnp+\nnm+\nni}}.
\end{align}

\subsubsection{The fermionic character}

Adding the weight of the minimal weight configuration and the sector to $G_{\text{disp}}(q)$, we obtain the generating function for all $\mk^{(2)}_\frac12$ paths with a fixed operator content:
\begin{align}
	G(\nnp,\nnm,\nni;z;q)=&\frac{q^{-\frac14+\frac12(\nnp+\nnm)^2+\frac12(\nnm+\nnp+\nni)+\frac14(-1)^{\nni}}z^{\nnp-\nnm-\frac12(-1)^{\nni}}}{(q)_{\nnp+\nnm+\nni}}\nonumber\\
		&\times\qbin{\nnp+\lf\frac{\nni-1}{2}\rf}{\nnp}_{q^2}\qbin{\nnm+\lf\frac{\nni}{2}\rf}{\nnm}_{q^2}.
\end{align}
Summing over all possible configurations yields the character for the $j=1/2$ module at level 2 in a fermionic form: 
\begin{align}\label{ca21/2}
	\chi^{(2)}_{\frac12}(z;q)=\sum_{\substack{ \nnp,\nnm,\nni\geq 0 \\   }}&
	\frac{q^{-\frac14+\frac12(\nnp+\nnm)^2+\frac12(\nnm+\nnp+\nni)+\frac14(-1)^{\nni}}z^{{\nnp-\nnm}-\frac12(-1)^{\nni}}}{(q)_{\nnp+\nnm+\nni}}\nonumber\\
		&\times\qbin{\nnp+\lf\frac{\nni-1}{2}\rf}{\nnp}_{q^2}\qbin{\nnm+\lf\frac{\nni}{2}\rf}{\nnm}_{q^2}.
\end{align}
This is, at least superficially, a new form of the $\chi^{(2)}_{\frac12}$ character. Curiously, it is more complicated than the $j=0,1$ characters obtained from the particle decomposition of the paths (containing two $q$-binomial factors vs none).
The correctness of the above expression has been checked by $q$-expansion.
The corresponding three-free-fermions version of this fermionic character reads \cite[eq. (6.26)]{MR}\footnote{To recover the $j=1/2$ character, the expression \cite[eq. (6.26)]{MR} must be divided by 2, or equivalently, the sign of the linear term $\ell_1$ in the exponent must be reversed, which is the version given here.}
\begin{align}\label{formeMR}
	\chi^{(2)}_{\frac12}(z;q)=\sum_{ \ell_0,\ell_1,\ell_2\geq 0} \frac{q^{\frac{1}{2}(\ell_0^2+\ell_1^2+\ell_2^2)+(\ell_0+\ell_1-\ell_2)/2}
		z^{\ell_0-\ell_2+\frac12}}{(q)_{\ell_0}(q)_{\ell_1}(q)_{\ell_2}}.
\end{align}
The direct equivalence between the forms (\ref{ca21/2}) and (\ref{formeMR}) has not been demonstrated.

\section{Conclusion}

Our first result was to establish the equivalence of the $\mkp{1}$ paths with those paths introduced in \cite{FPW} in the special case $j=s=0$ and extended here to $j=0,1/2$ and all values of $s$. This analysis went through an intermediate step where both paths are put in a bijective relationship with SOS-type paths. This step had two offshoots: it suggests the new weight expression (\ref{weightako}) valid to all levels $k$ -- which turns out to be  a disguised form of the standard RSOS weight, up to the term $-(j+s)/2$ -- and, in the light of \cite{JM,JMsusy}, a formal operator construction for the $\mkp{1}$ paths.

Our second result concerning $\mkp{1}$ paths  is a constructive derivation of their generating function along the lines of  \cite{OleJS}, which results in a new derivation of the spinon character formula of \cite{BPS,BLS} (see also \cite{MR}). This derivation relies on the operator method just mentioned. This, as we have indicated, is equivalent (for $k=1$) to the decomposition of a path into its basic constituents which we call the particles, the path representation of the spinons.

The particle description
  is more difficult to implement at higher levels. However, for $k=2$, this is still manageable. The analysis of $\mkp{2}$ paths from this point of view as been worked out in detail. Interestingly, this description leads to novel forms of the fermionic characters; we point out that the one pertaining to the $j=1/2$ module has not been proved to be directly reducible to the standard form.

\vskip0.3cm
\noindent {\bf ACKNOWLEDGMENTS}

We thank T. Welsh and O. Blondeau-Fournier for useful discussions. We also thank M. Okado and K. Misra for pointing out \cite{JMMO} and T. Welsh for explanations in relation with this work, 
his critical reading of the article and, his numerous suggestions.
This work is supported  by NSERC.

\end{document}